\documentclass[acmsmall,screen]{acmartRev}
\usepackage{url}
\AtBeginDocument{%
  \providecommand\BibTeX{{%
    \normalfont B\kern-0.5em{\scshape i\kern-0.25em b}\kern-0.8em\TeX}}}

\newcommand{\ismain}{1}
\usepackage{docmute}
\usepackage{proof}

\usepackage{xifthen}
\ifthenelse{\isundefined{\ismain}}{
  \newcommand{\ismain}{0}
}{}

\usepackage{enumitem}

\usepackage[draft]{tikzit}
\input{hypergraph.tikzdefs}

\tikzstyle{hedge}=[fill=white, draw=black, shape=rectangle, rounded corners=2mm, inner sep=0.2mm, outer sep=-2mm, scale=0.8, minimum height=8mm, minimum width=8mm, tikzit category=hypergraph]
\tikzstyle{hedge blue}=[hedge, fill={rgb,255: red,102; green,204; blue,255}, draw=black, shape=rectangle, tikzit category=hypergraph]
\tikzstyle{node}=[fill=black, draw=black, shape=circle, minimum size=1.5mm, inner sep=0mm, tikzit category=hypergraph]
\tikzstyle{red node}=[fill=red, draw=black, shape=circle, minimum size=1.5mm, inner sep=0mm, tikzit category=hypergraph]
\tikzstyle{node highlight}=[fill=black, draw=blue, thick, shape=circle, minimum size=1.5mm, inner sep=0mm, tikzit category=hypergraph]
\tikzstyle{red node highlight}=[fill=red, draw=blue, thick, shape=circle, minimum size=1.5mm, inner sep=0mm, tikzit category=hypergraph]
\tikzstyle{yellow hedge}=[hedge, fill=yellow, draw=black, shape=rectangle, tikzit category=hypergraph]
\tikzstyle{green hedge}=[hedge, fill=green, draw=black, shape=rectangle, tikzit category=hypergraph]
\tikzstyle{small box}=[fill=white, draw=black, shape=rectangle, minimum height=6mm, minimum width=6mm, tikzit category=string diagram]
\tikzstyle{vsmall box}=[fill=black, draw=black, shape=rectangle, minimum height=4mm, minimum width=1mm, tikzit category=string diagram, inner sep=0]
\tikzstyle{medium box}=[fill=white, draw=black, shape=rectangle, minimum height=11mm, minimum width=6mm, tikzit category=string diagram]
\tikzstyle{semilarge box}=[fill=white, draw=black, shape=rectangle, minimum height=16mm, minimum width=6mm, tikzit category=string diagram]
\tikzstyle{large box}=[fill=white, draw=black, shape=rectangle, minimum height=21mm, minimum width=6mm, tikzit category=string diagram]
\tikzstyle{black dot}=[fill=black, draw=black, shape=circle, minimum size=2mm, inner sep=0mm, tikzit category=string diagram]
\tikzstyle{white dot}=[fill=white, draw=black, shape=circle, minimum size=2mm, inner sep=0mm, tikzit category=string diagram]
\tikzstyle{red dot}=[fill=red, draw=black, shape=circle, minimum size=2mm, inner sep=0mm, tikzit category=string diagram]
\tikzstyle{wlabel}=[fill=none, draw=none, shape=rectangle, tikzit category=string diagram, font={\footnotesize}, inner sep=0pt, tikzit fill={rgb,255: red,102; green,204; blue,255}, tikzit draw={rgb,255: red,102; green,204; blue,255}, yshift=0.3mm]
\tikzstyle{BRchange}=[draw=black, shape=diamond, tikzit shape=circle, tikzit fill={rgb,255: red,96; green,0; blue,0}, diamond split part fill={black,red}, inner sep=-5mm, minimum width=2.7mm, minimum height=1.7mm]
\tikzstyle{RBchange}=[draw=black, shape=diamond, tikzit shape=circle, tikzit fill={rgb,255: red,165; green,0; blue,0}, diamond split part fill={red,black}, inner sep=0, minimum width=2.7mm, minimum height=1.7mm]
\tikzstyle{dummy}=[fill=none, draw=none, shape=circle, font={\small}, inner sep=1pt, tikzit draw=blue, tikzit fill=white]
\tikzstyle{node label}=[fill=none, draw=none, shape=rectangle, tikzit fill=cyan, tikzit draw=cyan, font={\scriptsize}, tikzit shape=circle, inner sep=0pt]
\tikzstyle{empty diag}=[fill=white, draw={rgb,255: red,165; green,165; blue,165}, shape=rectangle, minimum size=1.2 cm, dashed, thick]

\tikzstyle{dashed edge}=[-, dashed, very thick]
\tikzstyle{alt sort}=[-, dashed, dash pattern=on 2pt off 0.5pt, thick, draw=red]
\tikzstyle{diredge}=[->, >={Latex[length=1.5mm]}]
\tikzstyle{diredge highlight}=[->, >={Latex[length=1.5mm]}, draw=blue, thick]
\tikzstyle{boundary frame}=[-, draw={rgb,255: red,170; green,170; blue,255}, dashed, fill={rgb,255: red,238; green,238; blue,255}, thick, dash pattern=on 2pt off 0.5pt]
\tikzstyle{graph frame}=[-, draw={rgb,255: red,191; green,191; blue,191}, dashed, fill={rgb,255: red,238; green,238; blue,238}, thick, dash pattern=on 2pt off 0.5pt]
\tikzstyle{def sort}=[-]
\tikzstyle{component}=[-, draw=red, thick]
\tikzstyle{map edge}=[{|->}, >=latex, shorten <=0.5mm, shorten >=0.5mm]
\tikzstyle{hypergraph map edge}=[{|->}, draw=red, shorten <=1mm, shorten >=1mm]
\tikzstyle{cdedge}=[->]
\tikzstyle{big cdedge}=[->, very thick, >=latex]
\tikzstyle{pointer edge}=[->, draw=gray, thick]

\usepackage{amsthm,amsmath}
 
\newcommand{\ignora}[1]{ }

\ignora{
\usepackage{aliascnt}
\newaliascnt{definition}{thm}
\newaliascnt{proposition}{thm} 
\newaliascnt{lemma}{thm}
\newaliascnt{corollary}{thm}
\newaliascnt{conjecture}{thm}
\newaliascnt{remark}{thm}
\newaliascnt{example}{thm}
\newaliascnt{examples}{thm}
\newaliascnt{assumption}{thm}
\newaliascnt{construction}{thm}
\newaliascnt{claim}{thm}

\theoremstyle{definition}

\newtheorem{theorem}[thm]{Theorem}
\newtheorem{proposition}[proposition]{Proposition} 
\aliascntresetthe{proposition}
\newtheorem{lemma}[lemma]{Lemma}
\aliascntresetthe{lemma}
\newtheorem{corollary}[corollary]{Corollary}
\aliascntresetthe{corollary}

\aliascntresetthe{conjecture}
\newtheorem{remark}[remark]{Remark}
\aliascntresetthe{remark}

\aliascntresetthe{assumption}

\aliascntresetthe{construction}

\aliascntresetthe{claim}

\theoremstyle{definition}
\newtheorem{definition}[definition]{Definition}
\aliascntresetthe{definition}
\newtheorem{example}[example]{Example}
\aliascntresetthe{example}
\newtheorem{examples}[examples]{Examples}
\aliascntresetthe{examples}
}

\usepackage{cleveref} 


\usepackage{microtype}
\setlist[itemize]{noitemsep, topsep=0pt}
\usepackage{xspace}
\usepackage{mathtools}
\usepackage{multicol}
\usepackage[many]{tcolorbox}
\usepackage[all,2cell]{xy}
\CompileMatrices
\UseAllTwocells
\SilentMatrices
\usepackage{graphicx}
\usepackage{float}
\usepackage{fancybox}
\setenumerate{noitemsep,topsep=0pt,parsep=0pt,partopsep=0pt}
\usepackage{comment}
\usepackage{anyfontsize}
 \usepackage{xcolor,colortbl}
\usepackage{multirow}

\newcommand{\col}{C}
\def \catC {\mathbb{C}}
\def \catD {\mathbb{D}}

\def \catDE {\mathbb{D}_{\scriptscriptstyle \Upsilon}}
\def \catA {\mathbb{A}}

\def \catC {\mathbb{C}}

\def \PROP {\mathsf{PROP}} 
\def \CPROP {\mathsf{CPROP}} 

\newcommand{\sg}{\!\lower1pt\hbox{$\includegraphics[width=8pt]{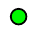}$}\!} 
\newcommand{\sr}{\!\lower1pt\hbox{$\includegraphics[width=8pt]{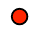}$}\!} 
\newcommand{\sbl}{\!\lower1pt\hbox{$\includegraphics[width=8pt]{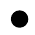}$}\!} 

\newcommand{\BRinput}{\itikzfig{BRinput}}
\newcommand{\RBoutput}{\itikzfig{RBoutput}}
\newcommand{\BRBio}{\itikzfig{BRBio}}
\newcommand{\RBRio}{\itikzfig{RBRio}}
\newcommand{\Bid}{\itikzfig{Bid}}
\newcommand{\Rid}{\itikzfig{Rid}}

\newcommand{\RsRmultIO}{\itikzfig{RsRmultIO}}
\newcommand{\RsRcounitIO}{\itikzfig{RsRcounitIO}}
\newcommand{\RsRcomultIO}{\itikzfig{RsRcomultIO}}
\newcommand{\RsRunitIO}{\itikzfig{RsRunitIO}}

\newcommand{\CFrob}[1]{\ensuremath{\mathbf{Frob}_{#1}}\xspace}
\newcommand{\frob}{\ensuremath{\mathbf{Frob}}\xspace}

\newcommand{\perm}[1]{\mathbf{P}_{\scriptscriptstyle #1}}

\newcommand{\old}[1]{}

\newcommand{\tr}[1]{\xrightarrow{#1}}    
\newcommand{\tl}[1]{\xleftarrow{#1}}    

\newcommand{\dlcorner}{{\ar@{}[dl]|(.8){\text{\large $\urcorner$}}}}
\newcommand{\drcorner}{{\ar@{}[dr]|(.8){\text{\large $\ulcorner$}}}}

\newcommand{\synTosem}[1]{[\! [ #1 ]\! ]}
\newcommand{\frobTosem}[1]{[ #1 ]}
\newcommand{\allTosem}[1]{\langle\! \langle #1 \rangle \! \rangle}

\newcommand{\freehyp}[1]{\mathbf{H}_{\scriptscriptstyle #1}}

\newcommand{\copair}[1]{({#1})}    

\newcommand{\Cospan}[1]{\mathsf{Csp}(#1)}
\def \FINSET {\mathbf{FinSet}} 
\def \SMC{\mathbf{SymCat}} 
\def \CAT{\mathbf{Cat}} 

\newcommand\symNet{\lower3pt\hbox{$\includegraphics[width=20pt]{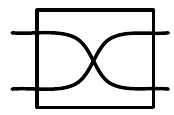}$}}
\newcommand\Idnet{\lower3pt\hbox{$\includegraphics[width=20pt]{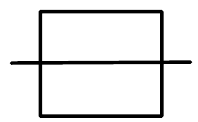}$}}

\newcommand\lccB{\lower5pt\hbox{$\includegraphics[width=25pt]{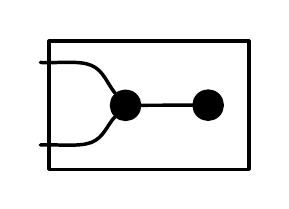}$}}
\newcommand\rccB{\lower5pt\hbox{$\includegraphics[width=25pt]{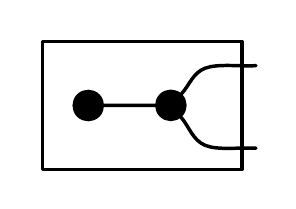}$}}
\newcommand\lccn{\lower5pt\hbox{$\includegraphics[width=20pt]{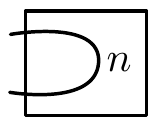}$}}
\newcommand\rccn{\lower5pt\hbox{$\includegraphics[width=20pt]{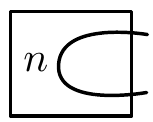}$}}
\def \df {\ \ensuremath{:\!\!=}\ }

\DeclareMathOperator{\id}{id}

\newcommand{\Defeq}
 {\stackrel{{def}}{=}}

\newcommand{\stran}{\raise1pt\hbox{$\centerdot$}}

\newcommand{\rring}[1]{\ensuremath{\mathbb{#1}}}

\newcommand{\N}{\rring{N}}

\newcommand{\ra}{\to}

\newcommand{\la}{\gets}

\newcommand{\Ra}{\Rightarrow}

\newcommand{\Ab}{\cat{Ab}}

\newcommand{\Set}{\cat{Set}\xspace}
\newcommand{\Vect}{\cat{Vect}\xspace}

\renewcommand{\emptyset}{\varnothing}

\newcommand{\ladj}[2]{\ar@/^/[#1]^-{#2} \ar@{}[#1]|-%

{\ifthenelse{\equal{#1}{r}}{\bot}{%

{\ifthenelse{\equal{#1}{rr}}{\bot}{%

{\ifthenelse{\equal{#1}{l}}{\top}{%

{\ifthenelse{\equal{#1}{u}}{\dashv}{%

{\vdash}}}}}}}}}}

\newcommand{\radj}[2]{\ar@/^/[#1]^-{#2}}

\newcommand{\radjff}[2]{\ar@{_{(}->}[#1]^{#2}}

\newcommand{\pullbacktop}[4]{%

{#1} \ar@/_/[ddr]_{#4} \ar@/^/[drr]^{#2}%

\ar@{.>}[dr]|-{#3} \\}

\newcommand{\cat}[1]{\ensuremath{\mathbf{#1}}}

\newcommand{\xra}[1]{\stackrel{#1}{\longrightarrow}}

\newcommand{\xla}[1]{\stackrel{#1}{\longleftarrow}}

\let\from\colon

\newcommand{\ltsred}[1]
{ \setbox0=\hbox{$\ {}^{#1}\ $}
  \setbox1=\hbox{$\longrightarrow$}
  \loop\setbox1=\hbox{$-$\kern-0.3em\unhbox1}\ifdim\wd1<\wd0\repeat
  \hbox{$\ \ \mathop{\box1}\limits^{#1}\ \ $}
}

\newcommand{\arx}[2]{\!\xymatrix@=15pt{\ar[r]^{{#1}}_{{#2}}&}\!}

\newlength{\mylength}

{\setlength{\fboxsep}{15pt}
\setlength{\mylength}{\linewidth}%
\addtolength{\mylength}{-2\fboxsep}%
\addtolength{\mylength}{-2\fboxrule}%
\Sbox
\minipage{\mylength}%
\setlength{\abovedisplayskip}{0pt}%
\setlength{\belowdisplayskip}{0pt}%
}%
{\endminipage\endSbox
\[\fbox{\TheSbox}\]}

\newcommand{\DCospan}[2]{\mathsf{Csp}_{#1}(#2)}

\newcommand{\syntax}[1]{\mathbf{S}_{#1}}

\newcommand{\FTerm}[1]{\DCospan{D}{\Hyp{{\scriptscriptstyle #1}}}}
\newcommand{\Hyp}[1]{\mathbf{Hyp}_{#1}}

\usepackage{color}
\def\bR{\begin{color}{red}}
\def\bB{\begin{color}{blue}}
\def\bM{\begin{color}{magenta}}
\def\bC{\begin{color}{cyan}}
\def\bW{\begin{color}{white}}
\def\bBl{\begin{color}{black}}
\def\bG{\begin{color}{green}}
\def\bY{\begin{color}{yellow}}
\def\e{\end{color}\xspace}

\newcommand{\cgr}[2][scale=0.45]{\raisebox{0.1em}{\begingroup
\setbox0=\hbox{\includegraphics[#1]{graffles/#2}}%
\parbox{\wd0}{\box0}\endgroup}}

\def \poi {\,\ensuremath{;}\,}
\def \df {\ensuremath{:=}}
\def \tns {\ensuremath{\oplus}}
\def \: {\colon}

\newcommand\Bmult{\itikzfig{Bmult}\xspace}
\newcommand\Bcomult{\itikzfig{Bcomult}\xspace}
\newcommand\Bunit{\itikzfig{Bunit}\xspace}
\newcommand\Bcounit{\itikzfig{Bcounit}\xspace}
\newcommand\Rmult{\itikzfig{Rmult}\xspace}
\newcommand\Rcomult{\itikzfig{Rcomult}\xspace}
\newcommand\Runit{\itikzfig{Runit}\xspace}
\newcommand\Rcounit{\itikzfig{Rcounit}\xspace}
\newcommand\BsRmult{\itikzfig{BsRmult}\xspace}
\newcommand\BsRcomult{\itikzfig{BsRcomult}\xspace}
\newcommand\BsRunit{\itikzfig{BsRunit}\xspace}
\newcommand\BsRcounit{\itikzfig{BsRcounit}\xspace}

\newcommand\BRchange{\itikzfig{BRchange}\xspace}
\newcommand\RBchange{\itikzfig{RBchange}\xspace}

\newcommand{\rrule}[2]{\ensuremath{\left\langle #1,#2 \right\rangle}}

\newcommand{\node}{\lower0pt\hbox{$\includegraphics[width=6pt]{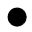}$}}
\newcommand{\hyperedge}{\lower2pt\hbox{$\includegraphics[width=25pt]{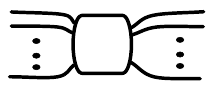}$}}
\newcommand{\ZeronetT}{\lower4pt\hbox{$\includegraphics[width=14pt]{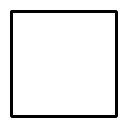}$}}
\def \Mon {\mathbf{Mon}}

\newcommand\idncircuit{\lower4pt\hbox{$\includegraphics[width=18pt]{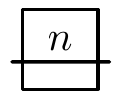}$}}
\newcommand{\rigidDPOstep}[1]{\Rrightarrow_{#1}}

\newcommand{\TODO}[1]{}

\begin{document}
\theoremstyle{acmdefinition}
\newtheorem{remark}[theorem]{Remark}
\newtheorem{examples}[theorem]{Examples}


\title[String Diagram Rewrite Theory I: Rewriting with Frobenius Structure]{String Diagram Rewrite Theory I:\\ Rewriting with Frobenius Structure}

\author{Filippo Bonchi}
\email{filippo.bonchi@unipi.it}
\author{Fabio Gadducci}
\email{fabio.gadducci@unipi.it}
\affiliation{%
  \institution{University of Pisa}
  \city{Pisa}
  \country{Italy}
 }

\author{Aleks Kissinger}
\affiliation{%
  \institution{University of Oxford}
  \city{Oxford}
  \country{United Kingdom}}
\email{aleks.kissinger@cs.ox.ac.uk}

\author{Pawel Sobocinski}
\affiliation{%
  \institution{Tallinn University of Technology}
  \city{Tallinn}
  \country{Estonia}}
\email{sobocinski@gmail.com}

\author{Fabio Zanasi}
\affiliation{%
  \institution{University College London}
  \city{London}
  \country{United Kingdom}}
\email{f.zanasi@ucl.ac.uk}

\renewcommand{\shortauthors}{Bonchi, Gadducci, Kissinger, Sobocinski and Zanasi}

\begin{abstract}
String diagrams are a powerful and intuitive graphical syntax, originating in theoretical physics and later formalised in the context of symmetric monoidal categories. In recent years, they have found application in the modelling of various computational structures, in fields as diverse as Computer Science, Physics, Control Theory, Linguistics, and Biology.

In several of these proposals, transformations of systems are modelled as rewriting rules of diagrams. These developments require a mathematical foundation for string diagram rewriting: whereas rewriting theory for terms is well-understood, the two-dimensional nature of string diagrams poses quite a few additional challenges.

This work systematises and expands a series of recent conference papers, laying down such a foundation. As first step, we focus on the case of rewriting systems for string diagrammatic theories that feature a Frobenius algebra. This common structure provides a more permissive notion of composition than the usual one available in monoidal categories, and has found many applications in areas such as concurrency, quantum theory, and electrical circuits. Notably, this structure provides an exact correspondence between the syntactic notion of string diagrams modulo Frobenius structure and the combinatorial structure of hypergraphs.

Our work introduces a combinatorial interpretation of string diagram rewriting modulo Frobenius structures in terms of double-pushout hypergraph rewriting. We prove this interpretation to be sound and complete and we also show that the  approach can be generalised to rewriting modulo multiple Frobenius structures. As a proof of concept, we show how to derive from these results a termination strategy for Interacting Bialgebras, an important rewriting theory in the study of quantum circuits and signal flow graphs.
\end{abstract}



\maketitle

\if\ismain0 

\setcounter{section}{0} 
\tableofcontents

\fi 

\section{Introduction}\label{sec:intro}

This is the first of a series of papers~\cite{BGKSZ-parttwo,BGKSZ-partthree} giving a comprehensive foundation for the rewriting theory of string diagrams. String diagrams are a graphical syntax that is particularly well-suited for capturing the behaviour of structures whose basic operations take many inputs to many outputs. This can be contrasted with term syntax, which is best suited for algebraic structures, whose basic operations take many inputs to a single output. In recent years, structures that mix algebraic (i.e. many-to-1 operations) and coalgebraic (i.e. 1-to-many) operations have increasingly found applications in a variety of fields, such as concurrency theory, control theory, quantum physics, biology, computational linguistics, and even cognition and consciousness.

Notable features of string diagrams include their flexibility and intuitive aspects. Two diagrams that can be topologically deformed into each other without cutting or joining wires must necessarily describe the same map. This makes string diagrams into a powerful language for reasoning about interacting processes, but also introduces unique challenges when it comes to formalising and implementing string diagrammatic equational reasoning and rewriting theory.

The formal basis for string diagrammatic syntax is the theory of symmetric monoidal categories (SMCs). SMCs provide a minimal setting for reasoning about processes that can compose in parallel or in sequence. String diagrams describe compositions of morphisms in an SMC, where boxes represent the morphisms themselves and wires represent objects, which serve as inputs and outputs
\[
f : A \tns B \tns C \to D \tns E \qquad\Longrightarrow\qquad
\input{./figures/basic-maps.tikz}
\]
Composition in parallel (``$\tns$'') and in sequence (``$\poi$'') are then depicted diagrammatically. For example
\[
(f \tns g) \poi h \poi (1 \tns f) \qquad\Longrightarrow\qquad
\input{./figures/arbitrary-sd-types.tikz}
\]
The relevant data of such a composition is the connectivity, rather than the
physical layout of the boxes and wires. Hence, it makes sense to consider `crossing'
of wires, and the SMC axioms make the interpretation invariant under deformation, e.g.
\[
\input{./figures/arbitrary-sd.tikz} \ \ =\ \ \input{./figures/deformation.tikz}
\]

There are many contexts where it furthermore makes sense to consider
`wires' that do not have just two endpoints, but many. Or, put another way, we may wish to allow wires to branch and merge in string diagrams, like in this example
\ctikzfig{splitting}
Rather than representing a path that information flows along, a branching wire represents a more general kind of interface between components. For example, branching wires in a digital circuit can be used to represent a physical wire, where junctions are used to join multiple wires into one. Systems whose primitive pieces compose via shared names (e.g. indices in tensor networks, variables in programming languages, or channels in process calculi) also naturally have such splitting structures, unless we specifically impose that names only occur in matched pairs. For non-deterministic, probabilistic, or quantum processes, branching wires represent the appropriate notion of correlation between their endpoints, e.g. statistical perfect correlations or (GHZ-like) quantum entanglement.

These branching wires can be accommodated categorically by requiring that
each object in the category comes equipped with certain basic operations for
`splitting' and `merging' wires, as well as `initialising' and `terminating' them
\[
\delta \ =\  \input{./figures/comult.tikz}
\qquad
\mu \ =\ \input{./figures/mult.tikz}
\qquad
\eta \ =\ \begin{tikzpicture}
	\begin{pgfonlayer}{nodelayer}
		\node [style=black dot] (0) at (-1, 0) {};
		\node [style=none] (3) at (0, 0) {};
	\end{pgfonlayer}
	\begin{pgfonlayer}{edgelayer}
		\draw (0) to (3.center);
	\end{pgfonlayer}
\end{tikzpicture}

\qquad
\epsilon \ = \ \begin{tikzpicture}
	\begin{pgfonlayer}{nodelayer}
		\node [style=black dot] (0) at (0, 0) {};
		\node [style=none] (3) at (-1, 0) {};
	\end{pgfonlayer}
	\begin{pgfonlayer}{edgelayer}
		\draw (0) to (3.center);
	\end{pgfonlayer}
\end{tikzpicture}

\]
satisfying some rules that again ensure that only the connectivity (i.e. the set of endpoints of a connected component) matters. These generators and rules are known in the literature as a \textit{special commutative Frobenius algebra} (SCFA). A category where every object is equipped with an SCFA is called a \textit{hypergraph category}.

In this paper, we will focus on the study of hypergraph categories, and their associated string diagrams. While this may seem counter-intuitive at first, the rewriting theory for hypergraph categories is actually simpler than that for generic symmetric monoidal categories. Hence, it will serve as a stepping stone toward doing rewriting for string diagrams with non-branching wires, which is the topic of~\cite{BGKSZ-parttwo}.

One can model composition and interaction between processes in hypergraph categories using equational reasoning. That is, we can introduce some primitive `boxes' into our theory and impose a set $\mathcal E$ of (diagram) equations that those boxes satisfy. For example, we may introduce a box $g$ and requires that it satisfies
\[
\mathcal E := \left\{ \ \input{./figures/ex-eq1.tikz} \ , \quad \input{./figures/ex-eq2.tikz} \  \right\}
\]
Just like in the case of equational reasoning with terms, when it comes to automated equational reasoning with diagrams, it becomes more useful to consider equations as rewriting rules, with a preferred orientation from left-to-right
\[
\mathcal R := \left\{ \ \input{./figures/ex-rw1.tikz} \ , \quad \input{./figures/ex-rw2.tikz} \  \right\}
\]
It is then natural to ask questions about how well-behaved such a rewriting system is, for example whether it terminates and in that case, whether it yields unique normal forms. However, more fundamentally, one can ask what rewriting in the context of hypergraph categories means. A simple answer is that diagram rewriting is simply term rewriting, performed modulo the axioms of a hypergraph category. Indeed all of the diagram equations above can be represented as terms over the basic generators of the theory, combined via two connectives $\tns$ and $\poi$, representing parallel and sequential composition, respectively. For example, adding boxes $f$ and $h$, we may have
\[
\input{./figures/splitting.tikz} \ \ := \ \ (1 \tns g) \poi (f \tns \delta) \poi (h \tns 1) \poi (\mu \tns 1)
\]
Then, we can apply the axioms of a hypergraph category in order to produce the LHS of one of our rewriting rules (i.e. a reducible expression, or redex) as a subterm. For example, the \textit{interchange law}
\begin{equation}\label{eq:interchange}
    (a \tns b) \poi (c \tns d) = (a \poi c) \tns (b \poi d)
\end{equation}
allows us to produce the subterm $g ; \delta$, which is the LHS of the rule $\mathcal R_1$ above, at which point we can replace the LHS of $\mathcal R_1$ with the RHS
\begin{align*}
(1 \tns g) \poi (f \tns \delta) \poi (h \tns 1) \poi (\mu \tns 1)
= & \,
((1 \poi f) \tns {\color{blue} (g \poi \delta)}) \poi (h \tns 1) \poi (\mu \tns 1) \\
\overset{\mathcal R_1}{\rewr} & \,
((1 \poi f) \tns {\color{green!50!black} ((\delta \tns \delta) \poi (1 \tns \sigma \tns 1) \poi (g \tns g))}) \poi (h \tns 1) \poi (\mu \tns 1)
\end{align*}
In other words, we do term rewriting in the usual way, modulo the SMC and Frobenius equations. This rewriting-modulo step can be seen as the formal, syntactic underpinning to the intuitive notion of rewriting defined directly on string diagrams. For example, the term rewrite above can be depicted diagrammatically as
\ctikzfig{diagram-rewrite}
However, there are numerous drawbacks to this rewriting-modulo approach. First, some care must be taken in applying the axioms of a hypergraph category to terms. For example, depending on the types of $a, b, c, d$ it is possible for the LHS of the interchange law~\eqref{eq:interchange} to be well-defined while the RHS is not. Second, and more importantly, even well-behaved rewriting systems quickly become intractable when doing rewriting-modulo. Even a very simple rewriting system $\mathcal R$ becomes very difficult to work with when taken modulo the 12 additional equations implied by the axioms of a hypergraph category. Working purely with terms, this requires a great deal of careful book-keeping to `reshuffle' a term in such a way that it contains an exact copy of the LHS of a given rule as a subterm, then applying it. This complexity then transfers into all the aspects of the rewriting theory, making it difficult to define what it means, e.g. to enumerate all of the distinct places that a rule \textit{matches} or to determine whether the application of two rules overlap.

Computationally, a much better approach is to take seriously the notion that `only connectivity matters', and represent a morphism in a hypergraph category as a combinatorial object capturing exactly this connectivity data. Indeed, as the name would suggest, the best choice for that object is a hypergraph. In this paper (and in~\cite{BGKSZ-parttwo})
we combine and integrate the results of a series of recent
works~\cite{BonchiGKSZ_lics16, BGKSZ-esop17,BonchiGKSZ18}
to develop a complete rewriting theory for hypergraph categories based on hypergraphs with interfaces. In particular, we show the free hypergraph category over a collection of generators can be represented combinatorially in terms of hypergraphs and show that diagram substitution can be performed via double-pushout hypergraph rewriting. This corresponds exactly to rewriting modulo the axioms of a hypergraph category, hence all of the hypergraph category axioms are subsumed by hypergraph isomorphism.


Often it is interesting to consider rewriting modulo in systems where each object is equipped with more than one Frobenius algebra. For example, the ZX-calculus used in quantum computation, as well as the system $\mathbb{IB}$ and its derivatives used to model linear systems, signal-flow graphs, and concurrency, involve two interacting Frobenius algebras at their heart. We will show that such systems can be accommodated within this framework while still preserving its good computational properties. As a demonstration, we define a theory of two Frobenius algebras interacting as a bialgebra and illustrate a simple, terminating strategy for transforming diagrams to a pseudo-normal form using hypergraph rewriting.

%

In Section~\ref{sec:background} we provide an introduction to the rationale
of string diagram rewriting, while at the same time recalling the preliminary notions of symmetric monoidal categories
and PROPs, as well as their respective extensions incorporating a Frobenius structure, with a particular
attention for the multi-sorted case.
In Section~\ref{sec:isosyntaxgraphs} we introduce hypergraphs with interfaces, and the main tool for manipulating
them, i.e., \emph{double pushout} (shortly, DPO) rewriting.
The section also provides a purely combinatorial interpretation for string diagrams as cospans of hypergraphs,
and this result in exploited in Section~\ref{sec:equiv} to prove the completeness of such interpretation and
the precise correspondence between string diagram rewriting and DPO rewriting on hypergraphs with interfaces.
Section~\ref{sec:example-group-alg} offers a first proof-of-concept for our approach, showing that
the equational reasoning on the algebra of groups can be modelled via string diagrams and then
by DPO rewriting with a terminating reduction strategy.
The main formalisation in this paper captures rewriting modulo a single Frobenius algebra per sort, but we show in Section~\ref{sec:multi-frob} that this can be easily extended to multiple Frobenius algebras per sort.
This latter result is then put to use in Section~\ref{sec:example} by tackling a more comprehensive example, which
is represented by interacting bialgebras, one of the  most-often studied structures in graphical quantum theory. As for the
group algebra, our graphical interpretation of such diagrams allows for recasting the associated equational reasoning
in term of DPO rewriting with a terminating reduction strategy.
A concluding section finally wraps up the paper.
\bigskip

\noindent \textbf{Related work.} There is a long tradition of works, both in
mathematics and computer science, exploring the link between syntactic
and combinatorial representations of formal systems. In our work, we
make the following two contributions:
\begin{enumerate}[label=(\roman*)]
	\item a correspondence between symmetric monoidal categories with Frobenius structure, and cospans of hypergraphs.
	\item a correspondence between rewriting of string diagrams modulo Frobenius, and double pushout rewriting of hypergraphs with interfaces.
\end{enumerate}
Regarding (i), the research on the precise correspondence between visual
languages for monoidal categories and combinatorial structures
witnessed a renewed interest in the
1990s~\cite{Joyal1991,Joyal_tracedcategories}, and we refer
to \cite{Selinger2009} for an extensive survey. The case of monoidal categories with
Frobenius structures is particularly important: their role in visually modelling circuit-like systems, and possibly their manipulation, has been recognised early on
 \cite{Carboni1987,Gadducci1998},
and more recently it has appeared in categorical models of quantum
processes \cite{CoeckeKissinger_book}, signal flow graphs
\cite{BonchiSZ17,BaezErbele-CategoriesInControl}, electrical circuits
\cite{BPSZ-lics19,baez2017props}, Petri nets \cite{BonchiHPS17} and more.

Also the formal link between Frobenius monoids and cospans
(both of sets and of graphs)
have been observed, with some variations, in several
works~\cite{BruniG01,Lack2004a,Rosebrugh2005,albasini2009cospans},
and most recently in~\cite{abs-1806-08304}.
Besides its fully worked out connection with PROPs, what is definitely
new of our formulation is the presentation of the multi-sorted version, and
especially its lifting to the ``multi-Frobenius''  case (Section~\ref{sec:multi-frob}).

Regarding (ii) above, the study of string diagram rewriting dates back at least to
Burroni's work on polygraphs~\cite{Burroni1993}. In that tradition, the driving
motivation is to generalise term rewriting to higher dimensions, including the
three-dimensional case of string diagram rewriting: see e.g.~\cite{Mimram14}
for a survey.
The approach does not rely on graph rewriting, and goes instead via a completely
``syntactic'' route: the laws of SMCs are  considered as explicit rewriting rules, resulting
in rather elaborate rewriting systems,  whose analysis is often challenging (see e.g.
\cite{Lafont2003}). Our approach is instead to ``absorb'' the structural equations of string
diagrams into the graph interpretation, and study as rewriting rules only the
``domain-specific'' equations of the theory under consideration.

A first attempt at modelling string diagram rewriting as graph rewriting was proposed in~\cite{dixon2010open1,DixonK13}, and was subsequently used in the Quantomatic proof assistant~\cite{quanto-cade}. That formalism differs from ours in that it does not capture rewriting modulo Frobenius structure, and instead assumes the presence of a categorical trace. Moreover, we directly work in an \emph{adhesive category}~\cite{Lack2005}, while the category of ``open graphs'' used in~\cite{DixonK13} is not adhesive, but instead inherits its good rewriting properties from an embedding into a larger adhesive category. 
An important aspect of our rewriting is the requirement that DPO rewriting respects a fixed mapping into a graph that serves as an interface to a larger, possibly unknown context.
As we will see in Section~\ref{sec:dpoi}, the presence of an interface has implications on which rewrites are applicable: only those that respect the interface will be liftable to a larger context. This has been studied in the DPO literature, most notably within the \textit{rewriting with borrowed contexts} approach~\cite{Ehrig2004}, and it will play an important role in our study of confluence in a sequel to this paper~\cite{BGKSZ-parttwo}.
The construction in~\cite{DixonK13} also considers rewriting relative to an interface, but here we consider much more general kinds of interfaces, as afforded by rewriting modulo Frobenius structure. This is explored in Section~\ref{sec:contexts}.

\if\ismain0 

\bibliographystyle{plain}
\bibliography{catBib3Rev}

\fi 

\if\ismain0 

\setcounter{section}{1} 
\tableofcontents

\fi 

\section{Syntactic foundations of string diagram rewriting}\label{sec:background}

We will develop two interlinked perspectives on string diagrams and the (co)algebraic structures they express: a \textit{syntactic} perspective based on symmetric monoidal theories and PROPs and a \textit{combinatorial} perspective based on hypergraphs. We will assume the reader is familiar with the basics of category theory in general, and symmetric monoidal categories (of which cartesian monoidal categories are a special case) in particular. The standard reference is~\cite{mclane}.

To begin, one can ask simply: why is string diagram rewriting interesting? One could give many different answers to this question, but perhaps the one that fits most naturally into a story about rewriting is the following

\begin{quote}
  \it String diagram rewriting is the natural extension of term rewriting for operations with many outputs.
\end{quote}

Term rewriting allows us to recast (universal) algebra as \textit{computation}. That is, it translates an algebraic structure -- like a monoid, a ring, or some more exotic formalism -- into a system for performing `computations' in a very general sense.

An \textit{algebraic theory} consists of a signature, i.e. a collection of symbols with arities, as well as a collection of equations between terms built from those symbols and some free variables. Arities are natural numbers that tell us how many inputs a symbol should take, where an arity of $0$ indicates something has no inputs, i.e. it is a constant. A simple example is the theory of monoids, consisting of the signature $\Sigma_{\Mon} = \{ \cdot : 2, \epsilon : 0 \}$ and the equations $E_{\Mon} = \{ (x \cdot y) \cdot z = x \cdot (y \cdot z), x \cdot \epsilon = x, \epsilon \cdot x = x \}$.

By making a tiny tweak to an algebraic theory, one obtains a \textit{term rewriting system}. The equations of an algebraic theory are effectively unordered pairs of terms that share some variables. A term rewriting system consists of a signature and a collection of \textit{ordered} pairs of terms called \textit{rewriting rules}. A possible rewriting system for monoids is the following: $R_{\Mon} = \{ (x \cdot y) \cdot z \Rew{} x \cdot (y \cdot z), x \cdot \epsilon \Rew{} x, \epsilon \cdot x \Rew{} x \}$.

Unlike the equational theory, the rewriting system has some computational content. Namely, if we apply the rules in a rewriting system from left-to-right in an arbitrary order until no rule applies any more, we obtain (non-deterministic, possibly non-terminating) computations on terms. In the case of $R_{\Mon}$, a particularly nice rewriting theory, the computation in fact always terminates with a unique answer: the term obtained from removing any extra units and bracketing to the right. For example:
\[
  ((a \cdot b) \cdot (c \cdot \epsilon)) \cdot d
  \Rew{}
  ((a \cdot b) \cdot c) \cdot d
  \Rew{}
  (a \cdot b) \cdot (c \cdot d)
  \Rew{}
  a \cdot (b \cdot (c \cdot d))
\]
One could also refine this notion of computation, e.g. by considering different rewriting strategies and/or termination criteria. Classical term rewriting, as a discipline, studies the properties of such theories and their associated computations. Namely, it gives techniques for proving a rewriting system is well-behaved in various ways, like being terminating or admitting unique normal forms. It also provides techniques such as Knuth-Bendix completion~\cite{knuth1970simple} that turn ill-behaved rewriting systems into well-behaved ones. Such techniques have proved to have far-reaching applications in programming languages, computer algebra, and automated theorem proving, see e.g. \cite{Dershowitz1990RewriteS,Visser_rewritingsurvey} for a survey.

However, algebraic theories, and hence term rewriting systems, can only handle signatures where every operation produces exactly one output. This is so fundamental to the concept of what a `formula' or a `term' is, that this limitation may even go unnoticed. Returning to the monoid example, we could write the arities (i.e. numbers of inputs) as well as the \textit{co-arities} (i.e. number of outputs) explicitly as: $\Sigma_{\Mon} = \{ \mu : 2 \to 1, \epsilon : 0 \to 1 \}$. Note that we have switched from `$\cdot$' to $\mu$, which we will shortly find more convenient.

Suppose we wanted to consider an operation with the signature $\delta : 1 \to 2$, i.e. something which takes 1 input, but produces 2 outputs. If this symbol is meant to represent a function of some kind, this is not a problem. Instead of using just one operation, we can simply give two: $\delta_1 : 1 \to 1$, which gives the first output of $\delta$, and $\delta_2 : 1 \to 1$, which gives the second output. However, this translation makes use of a fundamental property of functions (or more generally, of morphisms in a cartesian category) that the two maps $\langle \delta_1, \delta_2 \rangle$ contain exactly the same data as the overall map $\delta$. If we wish to generalise from functions to other kinds of maps (e.g. non-deterministic operations, probabilistic or quantum processes, etc.), this may not be true any more. For example, if we interpret symbols probabilistically, a constant of the form $p : 0 \to 1$ could represent a fixed probability distribution for a single value, whereas $q : 0 \to n$ could represent a joint distribution over $n$ values. In this case, we would clearly lose some information if we attempt to factorise $q$ as $q_1, q_2, \ldots, q_n$. Indeed, probabilistic maps are most naturally expressed in a (non-cartesian) symmetric monoidal category.

Even in cartesian categories, there may be good computational reasons for wanting to consider operations with multiple outputs. For example, suppose $f : 1 \to 2$ is a function that does some really difficult computation, then returns two copies of the output. While it is surely possible to represent the same function as a pair of functions $f_1 : 1 \to 1$ and $f_2 : 1 \to 1$, each doing the same difficult computation, in some contexts it might be incredibly wasteful. Following this idea to its natural conclusion leads to generalisation from terms to termgraphs~\cite{Plump-termgraph,Barendregt-termgraph}.

We argue that the appropriate generalisation of algebraic theory that allows operations with multiple outputs is a \textit{symmetric monoidal theory} (SMT). Indeed, while the syntax of algebraic theories is terms, we will see in the following section that the syntax of SMTs is \textit{string diagrams}.

\subsection{Symmetric Monoidal Theories and PROPs}\label{sec:smt}

Much like an algebraic theory axiomatises a structure in a cartesian category (typically the category of sets), a symmetric monoidal theory axiomatises a structure in a more general symmetric monoidal category. It consists of two parts: a signature, which we will call a \textit{monoidal signature} to avoid confusion with the notion of signature in an algebraic theory, and a set of equations.

A monoidal signature consists of a set $\Sigma$ of operations $f \: n \to m$ with a fixed \emph{arity} $n$ and \textit{coarity} $m$, for $n,m \in \N$.
Symmetric monoidal theories are defined using \textit{$\Sigma$-terms}. The set of $\Sigma$-terms is obtained by combining the operations in $\Sigma$, \emph{identities} $\id_n \colon n\to n$ and \emph{symmetries} $\sigma_{m,n} \colon m+n\to n+m$ for each $m,n\in \mathbb N$, by sequential ($;$) and parallel ($\tns$) composition. This is a purely formal process: given $\Sigma$-terms $r\colon m\to n$, $s\colon n\to o$, and $t\colon m' \to n'$, we construct new $\Sigma$-terms $r\poi s \colon m \to o$ and $r \tns t \colon m+m' \to n+n'$.

\begin{definition}
A \emph{symmetric monoidal theory} is a pair $(\Sigma, \mathcal E)$ where $\Sigma$ is a monoidal signature and $\mathcal E$ is a set of equations, namely pairs $\langle l, r \rangle$ of $\Sigma$-terms $l,r \colon v \to w$ with the same arity and coarity.
\end{definition}

Typically we depict $\Sigma$-terms (or more accurately: equivalence classes of $\Sigma$-terms) graphically as \textit{string diagrams}. The operation $f : n \to m$ is represented as a box with $n$ wires in a $m$ wires out
\ctikzfig{f-nm}
We sometimes also write $f$ as a box with single wires in and out, labelled by their (co)arities
\ctikzfig{f-nm-label}
Sequential and parallel composition are depicted as one would expect
\[
  f \poi g \ =\ \input{./figures/seq.tikz}
  \qquad\qquad
  f \tns h \ =\ \input{./figures/par.tikz}
\]
Identities are (sets of) blank wires
\[
  \id_1 \ = \ \begin{tikzpicture}
	\begin{pgfonlayer}{nodelayer}
		\node [style=none] (0) at (-0.75, 0) {};
		\node [style=none] (1) at (0.75, 0) {};
	\end{pgfonlayer}
	\begin{pgfonlayer}{edgelayer}
		\draw (0.center) to (1.center);
	\end{pgfonlayer}
\end{tikzpicture}

  \qquad\qquad
  \id_n \ = \ \input{./figures/id-n.tikz}
\]
and the symmetries are represented by wire-crossings
\[
  \sigma_{1,1} \ = \ \input{./figures/sigma-11.tikz}
  \qquad\qquad
  \sigma_{n,m} \ = \ \input{./figures/sigma-nm.tikz}
\]

However this notation is ambiguous. For example
\ctikzfig{abcd}
could represent either the $\Sigma$-term $(a \tns b) \poi (c \tns d)$ or the $\Sigma$-term $(a \poi c) \tns (b \poi d)$. Hence string diagram notation does not represent just one $\Sigma$-term, but instead an equivalence class of $\Sigma$-terms modulo some equations. It is noteworthy that the appropriate set of equations for this are exactly the axioms of a symmetric monoidal category~\cite{Selinger2009}. Figure~\ref{fig:axSMC} shows these equations, adapted to the special case of $\Sigma$-terms. This allows us to give a fully syntactic definition of string diagram.

\begin{definition}\label{def:string-diagram-syn}
  A \textit{string diagram} is an equivalence class of $\Sigma$-terms, taken modulo the equations in Figure~\ref{fig:axSMC}.
\end{definition}

\begin{figure}[t]
\[
  \begin{array}{rcl}
    (s \poi t) \poi u \equiv s \poi (t \poi u) & &
    \id_n \poi s \equiv s \equiv s\poi \id_m\\
    (s \tns t) \tns u \equiv s \tns (t \tns u) & &
    \id_0 \tns\ s \equiv s  \equiv s \tns \id_0 \\
  \end{array}
\]
\[
  \begin{array}{c}
    (s \poi u) \tns (t \poi v) \equiv (s \tns t) \poi (u \tns v)\\
    \id_m \tns \id_n \equiv \id_{m+n} \\
    (\sigma_{m,n} \tns \id_o) \poi (\id_n \tns\ \sigma_{m,o}) \equiv \sigma_{m,n+o} \\
    \sigma_{m,n}\poi\sigma_{n,m}\equiv\id_{m+n} \\
    (s\tns \id_m) \poi \sigma_{m,n} \equiv \sigma_{m,o} \poi (\id_m \tns\ s) 
  \end{array}
\]
\caption{Equivalences of $\Sigma$-terms that generate the same string diagram.}\label{fig:axSMC}
\end{figure}

It is often useful to consider not just the SMT itself, but a particularly simple kind of symmetric monoidal category that it presents, called a PROP. This somewhat cryptic name was coined by MacLane~\cite{MacLane1965} as shorthand for `(monoidal) \textbf{PRO}duct and \textbf{P}ermutation category'.

\begin{definition}
A PROP is a symmetric strict monoidal category with objects the natural numbers, where $\tns$ on objects is addition. PROP-morphisms are identity-on-objects symmetric strict monoidal functors. PROPs and their morphisms form a category $\PROP$.
\end{definition}

In particular, no restriction is made on how either the morphisms of a PROP or the tensor product of morphisms are defined.

\begin{definition}\label{def:syn-prop}
  Let $\syntax{\Sigma, \mathcal E}$ be the free (i.e. `syntactic') PROP presented by the SMT $(\Sigma, \mathcal E)$. Namely, arrows $u\to v$ are $\Sigma$-terms $u\to v$ modulo the laws of symmetric monoidal categories given in Figure~\ref{fig:axSMC} and the smallest congruence containing the equations $t=t'$ for any $\langle t,t' \rangle \in \mathcal E$. When $\mathcal E$ is empty, we will denote $\syntax{\Sigma, \emptyset}$ as $\syntax{\Sigma}$.
\end{definition}




\begin{example}\label{eq:cmon}
Consider the SMT $(\Sigma_{\textbf{CMon}}, {\mathcal E}_{\textbf{CMon}})$, where
\[ \Sigma_{\textbf{CMon}} := \left\{\, \Bmult\colon 2 \to 1,\, \Bunit\colon 0 \to 1 \, \right\} \]
and ${\mathcal E}_{\textbf{CMon}}$ is the set consisting of the following three equations
\begin{equation}\label{eq:monoid-laws}
\input{./figures/monoid-laws.tikz}
\end{equation}
Intuitively, the leftmost and the rightmost equations state associativity and commutativity of  $\Bmult$, while the central equation that $\Bunit$ is the unit of $\Bmult$. For this reason, the  PROP freely generated  by $(\Sigma_{\textbf{CMon}}, {\mathcal E}_{\textbf{CMon}})$ is called the PROP of commutative monoids, denoted by
$\textbf{CMon}$.
\end{example}

Just as an algebraic structure can have many different presentations via generators and equations, many different SMTs can generate the same PROP. For example, we could (redundantly) add a second unit law to \eqref{eq:monoid-laws} and the PROP \textbf{CMon} would be unchanged. In that sense, a PROP captures the `essence' of an algebraic structure in a presentation independent way, much like monads or Lawvere theories do for algebraic theories~\cite{hyland2007category}.

\begin{remark}
  Note that we have abused notation in Example~\ref{eq:cmon} by giving the equations of the SMT directly as string diagrams, which are technically equivalence classes of $\Sigma$-terms, not $\Sigma$-terms themselves. Since we form the PROP associated with commutative monoids by quotienting over the axioms of a symmetric monoidal category, this distinction becomes irrelevant. In other words, choosing any $\Sigma$-term to represent the lefthand-side and righthand-sides of the equations \eqref{eq:monoid-laws} will yield the same PROP.
\end{remark}

%

It is often interesting to not only consider PROPs freely generated from an SMT, but also more concrete PROPs, built out of more combinatorial structures. A common theme in the study of PROPs is to initially define a PROP syntactically, then give a more convenient concrete characterisation. A canonical example is the following.


\begin{proposition}\label{thm:prop-monoids}
  $\textbf{CMon} \cong \mathbb F$, where $\mathbb F$ is the PROP whose morphisms $f : m \to n$ are functions from the finite set $[m] := \{0, \ldots, m-1\}$ to $[n] := \{0, \ldots, n-1\}$ and $[m] \oplus [n] := [m] + [n]$ is given by the disjoint union of finite cardinals.
\end{proposition}


The formal proof of Proposition~\ref{thm:prop-monoids} can be found e.g.
in~\cite{Lack2004a}.
However, for our purposes, it will suffice to give some intuition as to why $\mathbb F$ is the PROP for commutative monoids. A morphism $d : m \to n$ in \textbf{CMon} can be described as an equivalence class of string diagrams with generators $\Bmult$ and $\Bunit$, modulo the equations~\eqref{eq:monoid-laws}. Such an equivalence class is precisely identified by specifying, for each input $i \in [m]$ the associated output $j \in [n]$, connected to $i$ by means of multiplication, identity, and/or swap maps. For example, consider this equivalence class of string diagrams, modulo the monoid laws
\[
\left\{ \ \ \input{./figures/cmon-morphisms.tikz} \ \ \right\}
\]
Every diagram in this class has the property that the inputs $\{0,1,2,3\}$ connect to output $1$ and the inputs $\{4,5\}$ connect to output $0$. Hence, we can identify the above equivalence class of diagrams with the function $f : [6] \to [2]$ given by
\[
f ::
\{
\ 0 \mapsto 1 ,
\ 1 \mapsto 1 ,
\ 2 \mapsto 1 ,
\ 3 \mapsto 1 ,
\ 4 \mapsto 0 ,
\ 5 \mapsto 0
\ \}
\]

Of course, commutative monoids can also be presented as an algebraic theory and reasoned about using the usual machinery of terms and term rewriting. A simple example of an SMT that does not have an evident presentation as an algebraic theory is a \textit{comonoid}, which simply turns all of the generators around.

\begin{example}\label{eq:ccomon}
  The SMT $(\Sigma_{\textbf{CComon}}, {\mathcal E}_{\textbf{CComon}})$ of \textit{cocommutative comonoids} has a monoidal signature
\[ \Sigma_{\textbf{CComon}} := \left\{\, \Bcomult\colon 1 \to 2,\, \Bunit\colon 1 \to 0 \, \right\} \]
with equations ${\mathcal E}_{\textbf{CComon}}$ given by
\begin{equation}\label{eq:comonoid-laws}
\input{./figures/comonoid-laws.tikz}
\end{equation}
\end{example}

By essentially the same argument we gave before, one can see that $\textbf{CComon} \cong \mathbb F^{op}$.

As algebraic structures are presented by SMTs where all of the generators have co-arity 1, it is natural to think of \textit{coalgebraic} structures as SMTs where all of the generators have arity 1. Cocommutative comonoids, as described above, are a simple example of such a coalgebraic structure.

\subsection{Frobenius algebras and cospans}\label{sec:frob}

In section~\ref{sec:smt}, we saw an SMT for an algebraic structure (commutative monoids) as well as an SMT for a coalgebraic structure (cocommutative comonoids).  The most interesting SMTs are the ones that have both algebraic \textit{and} coalgebraic parts interacting with each other. One such SMT, and its associated PROP, will play a central role throughout this paper.

\begin{example}
Consider the SMT $(\Sigma_{\frob}, {\mathcal E}_{\frob})$, where
\[ \Sigma_{\frob} := \left\{\, \Bmult,\, \Bunit,\, \Bcomult,\, \Bcounit \, \right\} \]
and ${\mathcal E}_{\frob}$ contains the equations in \eqref{eq:monoid-laws}-\eqref{eq:comonoid-laws} and the following ones
\begin{equation} \label{eq:frob-laws}
\input{./figures/frobenius-law.tikz} \qquad
\input{./figures/special-law.tikz}
\end{equation}

The PROP freely generated by $(\Sigma_{\frob}, {\mathcal E}_{\frob})$ is called the PROP of \textit{special commutative Frobenius algebras} and denoted by $\frob$.
\end{example}

\noindent
Note that we will often refer to special commutative Frobenius algebras simply
as Frobenius algebras.
Just like we could give convenient alternative characterisations for PROPs of monoids and comonoids, there is a concrete characterisation of the PROP of Frobenius algebras. To obtain this characterisation, we need to generalise from functions to cospans.

\begin{definition}[$\Cospan{\catC}$] \label{defn:cospan}
  Let $\catC$ be a category with all finite colimits. A \emph{cospan} from $X$ to $Y$ is a pair of arrows $X \to A \leftarrow Y$ in $\catC$.
A morphism $\alpha \from (X \to A \leftarrow Y) \Rightarrow (X \to B \leftarrow Y)$ is an arrow $\alpha \from A \to B$ in $\catC$ such that the diagram below  commutes
\begin{equation*}\label{eq:preorder}
\xymatrix@R=2pt@C=10pt{
& A \ar[dd]^{\alpha} & \\
X \ar@/_/[dr] \ar@/^/[ur] & & Y \ar@/_/[ul] \ar@/^/[dl] \\
& B & \\
}
\end{equation*}
 Cospans $X \to A \leftarrow Y$ and $X \to B \leftarrow Y$ are \emph{isomorphic} if
 there exists a morphism of cospans as above, where $\alpha: A\to B$ is an isomorphism.
 For $X \in \catC$, the \emph{identity cospan} is $X\xrightarrow{\id_X} X \xleftarrow{\id_X} X$.
 The composition of $X\to  A \xleftarrow{f} Y$
and $Y\xrightarrow{g} B \leftarrow Z$ is $X\to A+_{f,g}B \leftarrow Z$, obtained by taking the pushout of $f$ and $g$. 
This data is the category $\Cospan{\catC}$: the objects are those of $\catC$ and the arrows are isomorphism classes of cospans. Finally, $\Cospan{\catC}$ has a monoidal product given by the coproduct in $\catC$, with unit the initial object $0\in \catC$.
\end{definition}

\begin{remark}
  It is natural to consider the bicategory of cospans~\cite{BenabouBicategories}, where cospans form the 1-cells and cospan morphisms form the 2-cells. In this case, composition is only associative up to isomorphism. However, taking isomorphism classes makes composition associative on-the-nose, which allows us to define $\Cospan{\catC}$ simply as a category.
\end{remark}

Note that, when $\catC = \mathbb F$, $\Cospan{\catC}$ is a PROP, and furthermore
the following holds.

\begin{proposition}[\cite{BruniG01,Lack2004a}]\label{prop:cospanfrob}
There is an isomorphism of PROPs $\frob \cong \Cospan{\mathbb F}$.
\end{proposition}

Again we refer to \cite{Lack2004a} for a formal proof and give some intuition of why this is the case. As with monoids, the relevant data of an equivalence class of morphisms in \frob is the connectivity of inputs and outputs. However, since we have both algebraic and coalgebraic generators, it is possible for many inputs to be connected to many outputs
\begin{equation}\label{eq:frob-morphism}
\input{./figures/frob-morphism.tikz}
\end{equation}

To represent a morphism in \frob as a cospan, the inputs become the set on the left, the outputs the set on the right, and the connected components of generators in the diagram form the middle set. The two functions in the cospan then respectively map each input or output to its associated connected component
\[
\input{./figures/frob-morphism-components.tikz}
\qquad
\mapsto
\qquad
\scalebox{0.75}{\input{./figures/frob-morphism-cospan.tikz}} \]

For any cospan, we can construct a diagram using the generators \Bmult, \Bunit, \Bcomult, and \Bcounit whose components induce that cospan. Conversely, any two $\Sigma_{\frob}$ terms inducing the same cospan are equivalent by the Frobenius equations--- see e.g. Example~\ref{ex:frobeniuscospans} below.

For $\Sigma_{\frob}$ terms $s, t$, the connected components of $s \tns t$ are the disjoint union of the components of $s$ and $t$ respectively, which matches the tensor product in $\Cospan{\mathbb F}$. The composition $s \poi t$ acts on connected components by first taking the disjoint union, then amalgamating together those components that become connected, which is exactly what happens when taking the pushout. This is easiest to see by means of an example.

\begin{example}\label{ex:frobeniuscospans}
Consider the following composition of \frob-morphisms, where we label the connected components on the LHS and RHS
\[
\input{./figures/frob-morphism-components.tikz} \ \ \textrm{\Huge $\poi$} \ \
\input{./figures/frob-morphism-components-2.tikz}
\ \ =\ \ \input{./figures/frob-morphism-components-3.tikz}
\]
The composition of the associated cospans in $\Cospan{\mathbb F}$ is computed as follows
\[
\scalebox{0.75}{\input{./figures/frob-morphism-cospan.tikz}}
\ \ \textrm{\Huge $\poi$} \ \
\scalebox{0.75}{\input{./figures/frob-morphism-cospan-2.tikz}}
\ \ =\ \
\scalebox{0.75}{\input{./figures/frob-morphism-cospan-3.tikz}}
\]
That is, the pushout on the RHS is computed as the disjoint union $\{ 0, 1, 2 \} + \{ 3, 4, 5, 6 \}$ modulo the equivalence relation generated by $\{ j(a) \sim k(a) \,|\, a \in [5] \} = \{ 0 \sim 3, 0 \sim 4, 2 \sim 4, 2 \sim 5, 2 \sim 6 \}$. The $\sim$-equivalence classes are given by $\textbf{x} = \{ 0, 2, 3, 4, 5, 6 \}$ and $\textbf y = \{ 1 \}$ and the new cospan maps $i', l'$ on the RHS are induced by composing $i$ and $j$ with the pushout injections.
\end{example}


\subsection{Models of SMTs and PROPs}

For algebraic theories, one often wishes to study not the theory in isolation, but its concrete realisation in a category. A model of an SMT is similar in spirit to a model of an algebraic theory.

\begin{definition}\label{def:model-smt}
  A \textit{model} of an SMT $(\Sigma, \mathcal E)$ in a symmetric monoidal category $(\mathcal C, \otimes, I)$ consists of an object $A \in \mathcal C$ and a morphism $\hat f : A^{\otimes n} \to A^{\otimes m}$ for each $f : n \to m \in \Sigma$ such that the equations in $\mathcal E$, interpreted as equations between compositions of $\mathcal C$-morphisms, are all satisfied.
\end{definition}

Note that $A^{\otimes n}$ is  shorthand for the $n$-fold monoidal product of $A$ with itself, where $A^{\otimes 0} := I$. When we refer to a structure defined by an SMT \textit{in} a monoidal category, we really mean a model of that SMT in that category.

A nice feature of PROPs is they allow us to give a presentation-independent notion of model.

\begin{definition}\label{def:model-prop}
  A \textit{model} of a PROP $\mathbb A$ in a symmetric monoidal category $(\mathcal C, \otimes, I)$ is a strong symmetric monoidal functor $F : \mathbb A \to \mathcal C$.
\end{definition}

For a PROP $\syntax{\Sigma, \mathcal E}$ presented by an SMT $(\Sigma, \mathcal E)$, these two notions of model coincide. Namely, the chosen `carrier' object $A$ from Definition~\ref{def:model-smt} is $F(1)$ and since each generator of $\Sigma$ can be regarded as a morphism in the associated PROP, we can let $\hat f := F(f)$. The fact that $F$ is a strong symmetric monoidal functor forces all of the equations $\mathcal E$ to hold in $\mathcal C$, since they hold, by definition, in $\syntax{\Sigma, \mathcal E}$.

Since our primary interest will be in theories, rather than their models, we will not go into further details here, and conclude our discussion with some examples.

\begin{examples}
  For the following symmetric monoidal categories
  \begin{itemize}
    \item $(\Set, \times, 1)$ the category of sets with cartesian product,
    \item $(\Ab, \otimes, \mathbb N)$ the category of abelian groups with the tensor product, and
    \item $(\Vect_k, \otimes, k)$ the category of vector spaces over a field $k$ with the tensor product
  \end{itemize}
  the models of $\textbf{CMon}$ are commutative monoids, commutative rings, and associative, commutative $k$-algebras, respectively.
  The models of $\textbf{CComon}$ in $\Set$ are just sets, since the only comonoids in a cartesian category come from the diagonal map $\Delta : A \to A \times A$ on an object $A$.

  The only model of $\textbf{Frob}$ in $\Set$ (or any cartesian category) is the trivial one on the 1-element set $1$. The models of $\textbf{Frob}$ in $\Vect_k$ are in 1-to-1 correspondence with bases~\cite{CPV}.
\end{examples}

\subsection{Syntactic rewriting for PROPs}

Equational reasoning on algebraic theories can be mechanised by means of term rewriting. Rewriting plays the same role for SMTs, but here terms are more sophisticated structures than trees.

\begin{definition} \label{defn:rewprop}
A \emph{rewriting system} $\RS$ in a PROP $\catA$ consists of a set of \emph{rewriting rules}, i.e. pairs $\rrule{l}{r}$ of morphisms $l, r \: i \to j$ in $\catA$ with the same arities and coarities. Given $a,b \: m \to n$ in $\catA$, $a$ rewrites into $b$ via $\RS$, written $a \Rew{\RS} b$, if they are decomposable as follows, for some rule $\rrule{l}{r} \in \RS$
\begin{equation}
\label{eq:rewpropmatch}
\input{./figures/lhs-ctx.tikz} \qquad\qquad
\input{./figures/rhs-ctx.tikz}
\end{equation}
In this case, we say that $a$ contains a \emph{redex} for $\rrule{l}{r}$.
\end{definition}

When the PROP under consideration $\catA$ is $\syntax{\Sigma}$ for some signature $\Sigma$, the notion of rewriting step can be reformulated as follows: in order to apply a rewriting rule $\langle l,r \rangle$ for $l,r : i \to j$ to a diagram $d$ in $\syntax{\Sigma}$
we need to find a redex of $l$ in $d$. This means finding a \emph{context} $C$: A term in $\syntax{\Sigma + \{\star \: i \to j\}}$ with exactly one occurrence of $\star$, such that $d = C[l/\star]$. The rewrite then takes $d\Ra_\mathcal{R}  C[r/\star]$.

\medskip

With these definitions, diagrammatic reasoning can now be seen as a special case of rewriting. Given an arbitrary SMT $(\Sigma, {\mathcal E})$
we can obtain a rewriting system $\mathcal{R}_{\mathcal E}$ as
\[
\mathcal{R}_{\mathcal E} = \{\, \langle t,t'\rangle \,|\, (t,t') \in {\mathcal E} \} \cup \{\, \langle t',t\rangle  \,|\, (t,t')\in {\mathcal E} \,\}.
\]
\begin{proposition}\label{pro:diagrammaticreasoning}
Let $c, d$ be two diagrams in $\syntax{\Sigma}$. Then $c = d$ in the PROP freely generated by $(\Sigma,{\mathcal E})$ iff
$c \Ra_{\mathcal{R}_{\mathcal E}}^* d$.
\end{proposition}

In order to find a redex in the string diagram $a$ as in \eqref{eq:rewpropmatch} one needs to transform $a$ according to the laws of SMCs (Figure~\ref{fig:axSMC}). Thus rewriting in a PROP always happens modulo these laws.

%


\subsection{Frobenius theories and FROPs}\label{sec:frop}

In this section, we will specialise the notions of symmetric monoidal theory and PROP to the situation where there is a fixed, `default' Frobenius algebra. There are two reasons we might want to do this. The first, as explained in the introduction, is to allow structures that naturally admit ``many-to-many'' wiring between boxes. For example, wires might capture names or variables that can be shared across many processes, or data that can be copied and deleted.

The second, formal reason that it is useful to consider theories with a fixed Frobenius algebra is that, as we'll see in the following two sections, such structures correspond exactly to hypergraphs. This enables us to get a direct combinatoric handle on symmetric monoidal theories and PROPs, and conversely, to capture the structure of hypergraphs in a purely syntactic way.

We will call the `Frobeniated' versions of SMTs and PROPs \textit{Frobenius theories} and \textit{FROPs}, respectively.

\begin{definition}
A \textit{Frobenius theory} is a pair $(\Sigma, \mathcal{E})$ of a symmetric monoidal signature $\Sigma$ and a set $\mathcal{E}$ of well-typed equations over $\Sigma'$-terms, where $\Sigma' := \Sigma + \Sigma_\frob$.
\end{definition}

\begin{definition}\label{def:FROP} A \emph{FROP} is a PROP $\catA$ equipped
with a fixed special commutative Frobenius algebra
$(\Bmult, \Bunit, \Bcomult, \Bcounit)$ on the object $1 \in \textrm{ob}(\catA)$.
\end{definition}

Note that we only assume that the object $1$ carries a Frobenius algebra.
However, this extends in the obvious way to a Frobenius algebra on every
object $n \in \textrm{ob}(\catA)$
\begin{equation}\label{eq:n-frob}
\input{./figures/n-mult.tikz} \qquad
\input{./figures/n-unit.tikz} \qquad
\input{./figures/n-comult.tikz} \qquad
\input{./figures/n-counit.tikz}
\end{equation}

Just as SMTs present PROPs, Frobenius theories present FROPs.  The FROP  $\freehyp{\Sigma, \mathcal E}$ presented by a Frobenius theory $(\Sigma, \mathcal{E})$ has morphisms $\Sigma'$-terms over $\Sigma' := \Sigma + \Sigma_\frob$, modulo the SMC equations in Figure~\ref{fig:axSMC} as well as the equations $\mathcal{E}' := \mathcal{E} + \mathcal{E}_\frob$. The chosen Frobenius algebra on the object $1 \in \mathbb N$ of the presented FROP is then the one arising from the generators in $\Sigma_\frob$.

\begin{remark}
  While it should be clear why we used the notation $\syntax{\Sigma, \mathcal E}$ to represent the `syntactic' PROP, defined by an SMT, it is not yet obvious why we chose the notation $\freehyp{\Sigma, \mathcal E}$ for the analogous concept involving Frobenius theories and FROPs. This is  meant to indicate that we have formed the free \textit{hypergraph category} over $(\Sigma, \mathcal E)$, where a hypergraph category is a symmetric monoidal category where each object is equipped with a fixed choice of Frobenius algebra (see e.g.~\cite{KissingerHypergraph}).
\end{remark}


Syntactic rewriting in a FROP is defined just as it was for PROPs in Definition~\ref{defn:rewprop}, except when defining a rule $l \rewr r$, the equations
\[
\input{./figures/lhs-ctx.tikz} \qquad\qquad
\input{./figures/rhs-ctx.tikz}
\]
are taken modulo the Frobenius equations in addition to the SMC equations.


\subsection{Coloured PROPs and FROPs}
\label{sec:coproduct}

We will conclude our discussion on syntactic foundations by introducing multi-sorted versions of the constructions we have introduced before.  Aside from being of interest as a generalisation of multi-sorted algebraic structures, these concepts will come in handy in Section~\ref{sec:multi-frob} when it comes to formalising rewriting modulo multiple, interacting Frobenius algebras.

For a set $\col$, let $\col^{\star}$ denote the set of words over $\col$.
Then $\col^{\star}$ carries the structure of a monoid, where the unit is the empty word (written $\varepsilon$) and binary operation is word concatenation (written $vw$ for all words $v,w\in \col^\star$).

A multi-sorted SMT is a triple $(\col, \Sigma, \mathcal E)$ where $\col$ is a set of sorts, or \emph{colours}, $\Sigma$ a set of operations having arities and coarietes in $\col^{\star}$, and $\mathcal E$ is a set of equations between $\Sigma$-terms with matching arities and co-arities.

Just as single-sorted SMTs present PROPs, multi-sorted SMTs present \textit{coloured PROPs}.

\begin{definition}[Coloured PROP]
Given a finite set $\col$ of colours, a \emph{$\col$-coloured PROP} $\catA$ is a symmetric strict monoidal category where the set of objects is $\col^{\star}$ and the monoidal product on objects is word concatenation. A morphism from a $\col$-coloured PROP $\catA$ to a $\col'$-coloured PROP $\catA'$ is a symmetric strict monoidal functor $H \: \catA\rightarrow \catA'$ acting on objects as a monoid homomorphism induced by a function $\col \to \col'$. Coloured PROPs and their morphisms form a category $\CPROP$.
\end{definition}

It is worth to note that fixing a colour $\sbl$ and restricting to $\{\sbl\,\}$-coloured PROPs and morphisms between them yields $\PROP$ as a full sub-category  of $\CPROP$.

Both categories $\CPROP$ and $\PROP$ have coproducts, which will be useful for the constructions to follow. It is instructive to see how the coproduct in $\PROP$ differs from the one in $\CPROP$. In $\PROP$, coproducts consist of formal compositions of morphisms from the two consistituent components, on a single sort. For example, the coproduct of $\frob$ with itself will consist of diagrams made from two copies of the Frobenius algebra generators, which we will typically depict in two different colours
\begin{equation*}
\Bmult \quad \Bunit \quad \Bcomult \quad \Bcounit \quad
\BsRmult \quad \BsRunit \quad \BsRcomult \quad \BsRcounit,
\end{equation*}
modulo two copies of the Frobenius equations, one for each colour. On the other hand, the coproduct in $\CPROP$ will yield morphisms on two \textit{different} sorts
\begin{equation*}
\Bmult \quad \Bunit \quad \Bcomult \quad \Bcounit \quad
\Rmult \quad \Runit \quad \Rcomult \quad \Rcounit,
\end{equation*}
again modulo two copies of the Frobenius equations. In particular, the generators of the $\sbl$ Frobenius algebra cannot be composed with the generators of the $\sr$ Frobenius algebra. To emphasise this difference, we write the coproduct in $\CPROP$ as $\CFrob{\sbl} + \CFrob{\sr}$. More generally for a set of colours $\col$, we will write $\CFrob{\col}$ for  $\sum_{c\in \col} \CFrob{}$.

The category $\CFrob{\col}$ will play an important role in the coming sections. Much like for $\frob$, we can make a combinatoric version of $\CFrob{\col}$ using cospans. However, rather than cospans of sets, we should use cospans of sets whose elements are labelled by colours in $\col$. Formally, we can express $\col$-coloured sets as objects of the slice category $\mathbb F \downarrow \col$, whose objects are pairs of a finite cardinal $[n]$ and a labelling function $w : [n] \to \col$ assigning a colour to each element in $[n]$ and whose morphisms are functions respecting the labelling.

\begin{proposition}\label{pro:FCcoloured}
  $\mathbb F \downarrow \col$ is a coloured PROP.
\end{proposition}

\begin{proof}
  The pair $([n], w : [n] \to \col)$ can equivalently be given as a word in $\col^\star$ of length $n$ whose $i$-th letter is $w(i)$. As a slice category, $\mathbb F \downarrow \col$ inherits coproducts from $\mathbb F$ that make it a strict symmetric monoidal category. It is straightforward to check the inherited coproduct acts on objects by concatenation of words.
\end{proof}

\begin{theorem}\label{thm:frobcol}
  For a finite cardinal $\col \in \mathbb F$, $\CFrob{\col} \cong \Cospan{\mathbb F \downarrow \col}$ is an isomorphism of coloured PROPs.
\end{theorem}

\begin{proof}
Using a simple inductive argument, Lack showed in~\cite{Lack2004a} that $\CFrob{} \cong \Cospan{\mathbb F}$. This generalises easily to the multi-sorted case since
\[
\CFrob{C} = \sum_{c\in \col} \CFrob{}
          \cong \sum_{c\in \col} \Cospan{\mathbb F}
          \cong \sum_{c\in \col} \Cospan{\mathbb F \downarrow 1}
          \cong^\star \Cospan {\mathbb F \downarrow \sum_{c\in \col} 1}
          \cong \Cospan {\mathbb F \downarrow \col}
\]
The first equality is just the definition of $\CFrob{C}$, the first isomorphism follows from Lack's result, the next isomorphism follows from the fact that $\mathbb{F}\cong \mathbb{F}\downarrow 1$, and the last isomorphism is again obvious. Thus we need to justify only the isomorphism marked $\star$. The reason for this is, essentially, the fact that coproducts and pushouts commute, and indeed \emph{as categories} $\sum_{c\in \col} \Cospan{\Set_f \downarrow 1}
\cong^\dagger \Cospan {\Set_f \downarrow \sum_{c\in \col} 1}$, where $\Set_f$ is the category of finite sets and functions. As coloured PROPs, one additionally needs to keep in mind the objects are not mere sets but words. This boils down to the argument given in the proof of Proposition~\ref{pro:FCcoloured}, but let us elaborate. The categories $\sum_{c\in \col} \Cospan{\Set_f \downarrow 1}$ and $\Cospan {\Set_f \downarrow \sum_{c\in \col} 1}$ can both be considered to have the objects of $\Set_f\downarrow \col$, i.e. functions $X\rightarrow \col$ where $X$ is a finite set. Choosing an ordering of the elements of $X$, as in the proof of Proposition~\ref{pro:FCcoloured}, is a uniform way of passing from categories to coloured PROPSs in all three cases, meaning that the $\dagger$ isomorphism of categories above implies the $\star$ isomorphism of coloured PROPs.
\end{proof}

Morphisms and composition look much like they did in the example at the end of Section~\ref{sec:frob}, but with sets replaced by coloured sets. Taking $\col = [2] \cong  \{ \sbl, \sr \}$, an example of composition in $\Cospan{\mathbb F \downarrow \col}$ is the following
\[
\scalebox{0.75}{\input{./figures/cfrob-morphism-cospan.tikz}}
\ \ \textrm{\Huge $\poi$} \ \
\scalebox{0.75}{\input{./figures/cfrob-morphism-cospan-2.tikz}}
\ \ =\ \
\scalebox{0.75}{\input{./figures/cfrob-morphism-cospan-3.tikz}}
\]



Even though coproducts in $\CPROP$ yield disjoint colours, the colours in two $\col$-coloured PROPs can be identified by using a pushout, as we see in the following example.

\begin{example}\label{ex:perm}
The free $\col$-coloured PROP on the theory $(\col, \varnothing)$ with an empty signature is written $\perm{\col}$ and has arrows $w \to v$ the permutations of $w$ into $v$ (thus arrows exist only when the word $v$ is an anagram of the word $w$). Given $\col$-coloured PROPs $\catA$ and $\catA'$, we use notation $\catA +_{\scriptscriptstyle \col} \catA'$ for the \emph{pushout} in $\CPROP$ of the span of the inclusions $\catA \tl{} \perm{\col} \tr{} \catA'$: in a nutshell, $\catA +_{\scriptscriptstyle \col} \catA'$ is the coproduct $\catA + \catA'$ where we have identified the copy from $\catA$ and from $\catA'$ of each $c \in C$. Thus $\catA +_{\scriptscriptstyle \col} \catA$ is also a $\col$-coloured PROP.
\end{example}

\begin{remark}
\label{rem:perm}
For reasons that will become clear later, it will often be more valuable if, rather than identifying the two colours, we formally introduce an isomorphism between them. That is, we introduce two new \textit{colour change} generators $\{ \BRchange, \RBchange \}$ and impose the equations
\[
\input{./figures/brb-iso.tikz} \qquad\qquad \input{./figures/rbr-iso.tikz}
\]
In this case, we will obtain a coloured PROP that is equivalent (but not isomorphic) to the one we described in Example~\ref{ex:perm}.
\end{remark}

The notions of Frobenius theory and FROP extend in the obvious way to
$\col$-coloured Frobenius theories $(\col, \Sigma, \mathcal{E})$ and
FROPs $\freehyp{\col, \Sigma, \mathcal{E}}$.
Namely, each colour $c \in \mathcal C$ is equipped with a distinct Frobenius algebra.
Then, similar to equations~\eqref{eq:n-frob}, these induce a Frobenius algebra on
any word $w \in \col^\star$.

\begin{example}\label{ex:bipartite}
Fix a set $\col = \{ \sbl, \sr \}$ of colours and a signature $\Gamma$ consisting of the two colour-change operations, $\{ \BRchange, \RBchange \}$. We may construct the free coloured FROP $\freehyp{\col, \Gamma, \varnothing}$. Here is an example of a string diagram
in this category
\begin{equation}\label{eq:bipartite}
\input{./figures/two-col-ex.tikz}
\end{equation}
We claim that $\freehyp{\col, \Gamma, \varnothing}$ is the same as the category of finite directed \emph{bipartite graphs} (with interfaces). This will become clear in Example \ref{ex:bipartite2}, after the characterisation provided by Corollary \ref{cor:freehyp-fterm}.
\end{example}

With the addition of colours, we now have all of the tools we need to understand string diagram rewriting from the purely syntactic point of view, where we are rewriting $\Sigma$-terms modulo a set of structural equations.

\if\ismain0 

\bibliographystyle{plain}
\bibliography{catBib3Rev}

\fi 

\if\ismain0 

\setcounter{section}{3} 
\tableofcontents

\fi 

\section{Combinatorial foundations of string diagram rewriting} \label{sec:isosyntaxgraphs}

While one can get quite some mileage out of treating string diagrams purely syntactically, we argued in Section~\ref{sec:intro} that this point of view is often unwieldy. This comes from the fact that we are never doing rewriting on $\Sigma$-terms themselves, but rather $\Sigma$-terms modulo the SMC (and Frobenius) equations. This phenomenon is not unique to string diagrams: complications inevitably arise whenever one considers rewriting modulo a set of equations~\cite{huet1980confluent,peterson1981complete,bachmair1989completion}.

One way to avoid the complexity of rewriting modulo equations is to choose a better representation for string diagrams that `absorbs' the SMC and Frobenius equations directly into the representation. This is analogous to the way, in term rewriting systems, it can be fruitful to consider multisets of expressions directly, rather than terms modulo associativity and commutativity. In this section, we will see that

\begin{quote}
  \it Hypergraphs give a canonical, combinatorial representation for string diagrams.
\end{quote}

A hypergraph is a generalisation of a graph, which comes in a few different variations. In all of these variations, edges, which connect precisely two nodes, are replaced by \textit{hyperedges}, which connect arbitrary numbers of nodes together. The particular variation we will focus on have hyperedges that are both directed and ordered. That is, each hyperedge has an ordered list of source nodes and an ordered list of target nodes. We will depict these hyperedges using a symbol that looks rather like a box in a string diagram, with source nodes connecting to the left and target nodes connecting to the right
\ctikzfig{hyperedge}
This is by no means a coincidence: these will indeed play the role of the boxes that represent morphisms in string diagrams.


Once we represent string diagrams using hypergraphs, we will show that rewriting of string diagrams can be accomplished using double pushout (DPO) rewriting. This is a standard technique for performing rewrites on graphs and graph-like structures, where the lefthand-side of a rule is first `cut out' of the target graph using an operation called the \textit{pushout complement}, then the righthand-side is `glued in' using a pushout. The whole process results in a diagram of two pushout squares side-by-side (cf. equation~\eqref{eq:dpo2}), hence the name \textit{double} pushout.

The only extra complexity we need to handle when applying the DPO approach to string diagrams is to account for the inputs and outputs of a string diagram, i.e. the wires left dangling to the left and the right. These form an \textit{interface} to a possibly larger diagram (e.g. one consisting of multiple string diagrams plugged together), and this interface should be respected by rewriting. To account for this, we introduce double-pushout rewriting with interfaces (DPOI) in Section~\ref{sec:dpoi}. This will give us the right tool for establishing a formal correspondence between the syntactic notion of string diagram rewriting in the previous section and the combinatorial one developed in this section.

\subsection{The category of labelled hypergraphs}\label{sec:hypergraphs}

DPO rewriting makes sense in any category with pushouts, but it is often considered in a category where those pushouts obey certain well-behavedness conditions, such as adhesive categories~\cite{Lack2005}. For our purposes, we will skip an abstract overview of DPO rewriting in an arbitrary adhesive category and focus on the specific category $\Hyp{}$ of directed hypergraphs.

An object $G$ of $\Hyp{}$ is a hypergraph, which consists of a set of \emph{nodes} $G_\star$
and for each $k,l\in\N$ a (possibly empty) set of \emph{hyperedges} $G_{k,l}$ with $k$ (ordered) sources and $l$ (ordered) targets. That is, for each $0\leq i < k$ we have the $i$\textsuperscript{th} source map $s_i\colon G_{k,l}\to G_\star$,
and for each $0\leq j < l$, the $j$\textsuperscript{th} target map $t_j \colon G_{k,l}\to G_{\star}$. The arrows of $\Hyp{}$ are hypergraph homomorphisms: functions $G_\star \to H_\star$
such that, for each $k,l$, $G_{k,l}\to H_{k,l}$, they respect the source and target maps in the obvious way. The following definition characterises $\Hyp{}$ as a presheaf topos, and as such, it is adhesive~\cite{Lack2005}.

\begin{definition}[Hypergraphs]
\label{defn:hyp}
The category of finite directed hypergraphs $\Hyp{}$
is the functor category ${\mathbb F}^{\mathbf{I}}$ where
${\mathbf{I}}$ has as objects pairs of natural numbers $(k,l) \in \N \times \N$ together with one extra object $\star$. 
For each $k,l\in\N$, there are $k+l$ arrows from $(k,l)$ to $\star$.
\end{definition}

Nodes will be drawn as dots and a $(k,l)$ hyperedge $h$ will be drawn as a rounded box, whose connections on the left represent the list $[s_1(h), \ldots, s_k(h)]$, ordered from top to bottom, and whose connections on the right give $[t_1(h), \ldots, t_l(h)]$.

\begin{example}
Let $G$ be the hypergraph with nodes $\{ v_1, \ldots, v_8 \}$, a $(3,3)$-hyperedge $h_1$, a $(2,1)$-hyperedge $h_2$, and a $(1,0)$-hyperedge $h_3$, and the following source and target maps
\[
\begin{array}{l}
s_1(h_1) := v_1 \\
s_2(h_1) := v_2 \\
s_3(h_1) := v_3 \\
\end{array}
\quad
\begin{array}{l}
t_1(h_1) := v_5 \\
t_2(h_1) := v_6 \\
t_3(h_1) := v_6
\end{array}
\qquad , \qquad
\begin{array}{l}
s_1(h_2) := v_3 \\
s_2(h_2) := v_4 \\
t_1(h_2) := v_8 \\
\end{array}
\qquad , \qquad
\begin{array}{l}
s_1(h_3) := v_6 \\
\end{array}
\]
Then $G$ is drawn as follows
\begin{equation}\label{eq:hypergraph-ex}
  \input{./figures/hypergraph-ex.tikz}
\end{equation}
\end{example}

There are often many ways one can generalise concepts from graphs to hypergraphs. In order to fix conventions, we will define some basic concepts for hypergraphs that will be useful for later.

\begin{definition}\label{def:connection}
  For a hyperedge $h$, node $v$ and $\alpha \in \{ s, t \}$, a \textit{connection} for $v$ is a triple $(\alpha, h, i)$ such that $\alpha_i(h) = v$. The \textit{degree} $\textrm{deg}(v)$ of a node $v$ is the number of distinct connections in $G$.
\end{definition}

Connections are sometimes called \textit{tentacles} in the hypergraph literature. For us, their main utility is obtaining the correct notion of degree of a node when it is connected multiple times to the same hyperedge. For example, $v_6$ in \eqref{eq:hypergraph-ex} has degree 3, even though it is only connected to 2 distinct hyperedges.

\begin{definition}\label{def:path}
  A \textit{path} in a hypergraph is an alternating list $p = [p_1, \ldots, p_n]$ of hyperedges and nodes such that for all hyperedges $p_i$, the nodes $p_{i-1}$ and $p_{i+1}$ are a source and target for $p_i$, when they are defined (i.e. when $i > 1$ and $i < n$, respectively). A hypergraph is said to be \textit{acyclic} if it has no path containing the same node twice.
\end{definition}

A monoidal signature $\Sigma$ can be regarded as a directed hypergraph $G_{\Sigma}$ with a single node. Each symbol in the signature is depicted as a hyperedge with a number of sources equal to the arity and a number of targets equal to the coarity, all connected to the single node. For example, $\Sigma = \{ o_1 \: 2 \to 2, o_2 \: 1 \to 0 \}$ yields
\begin{equation}\label{eq:sig-example}
  G_\Sigma \ :=\
  \input{./figures/signature-single-sort.tikz}
\end{equation}

This extends naturally to multi-sorted signatures $(\col, \Sigma)$ by letting the nodes of $G_{(\col,\Sigma)}$ correspond to colours in $C$ and for a symbol $o\: u \to v$ in $\Sigma$, adding a corresponding hyperedge where the $i$\textsuperscript{th} colour in $u$ is the  $i$\textsuperscript{th} source node and the $j$\textsuperscript{th} colour in $v$ is the  $j$\textsuperscript{th} target node. For instance the hypergraph for $(\col, \Sigma)$ where $\col = \{ c_1, c_2\}$ and $\Sigma = \{ o_1 \: c_1 c_2 \to c_2 c_2, o_2 \: c_2 \to \epsilon \}$ is depicted as follows
\begin{equation}\label{eq:col-sig-example}
\input{./figures/signature-unlabeled.tikz}
\end{equation}



The utility of writing a monoidal signature $(\col, \Sigma)$ as a hypergraph is that we can now define hypergraphs labelled by $(\col, \Sigma)$ as the slice category $\Hyp{\col, \Sigma} := \Hyp{} \downarrow G_{\col,\Sigma}$. That is to say, an object of $\Hyp{\col, \Sigma}$ consists of a hypergraph $G$ together with a graph homomorphism $l\: G \to G_{\col, \Sigma}$. Intuitively $l$ \emph{labels} each node of $G$ with a colour in $\col$ and each hyperedge with an operation in $\Sigma$.  Observe that this definition ensures that a $\Sigma$-operation $o \: u \to v$ labels a hyperedge only when the label of its input (resp. output) nodes forms the word $u$ (resp. $v$). We call such objects $(\col, \Sigma)$-hypergraphs and we visualise them as hypergraphs whose nodes $n$ are coloured by $l(n)$ and whose hyperedges $h$ are labelled by $l(h)$.

\begin{example}
Considered the following (unlabelled) hypergraph
\ctikzfig{hypergraph-unlab-ex}
labelled by the signature \eqref{eq:sig-example} as follows
\[
l :=
\begin{cases}
v_1, v_5 \mapsto \sbl \\
v_2, v_3, v_4, v_6, v_7 \mapsto \sr \\
h_1, h_2 \mapsto f \\
h_3 \mapsto g
\end{cases}
\]
with $c_1 := \sbl$ and $c_2 := \sr$.
This is depicted as
\ctikzfig{hypergraph-lab-ex}
\end{example}


\subsection{String diagrams as cospans of hypergraphs}\label{sec:hyp-inform}

We have already written hypergraphs in a way that is suggestive of how, in the next section, we will relate them to string diagrams. Namely, hyperedges play the role of the `boxes' in a string diagram and nodes play the role of `wires', insofar as they allow us to connect boxes to each other. So, it should seem plausible how one could interpret a string diagram as a hypergraph and vice-versa. For example
\ctikzfig{sd-to-graph}

We will formalise this translation in Section~\ref{sec:equiv}, but before we do so, we need to answer a couple of open questions. First, in translating from a string diagram to a hypergraph, we seem to have lost some data. Namely, we no longer know which nodes should be treated as inputs and which as ouputs and in what order. We can solve this by replacing hypergraphs with cospans of hypergraphs $M \rightarrow G \leftarrow N$ where $M$ and $N$ are discrete hypergraphs embedding the inputs and outputs, respectively
\ctikzfig{sd-cospan}
Here, the labels are used to indicate how the two graphs $M$ and $N$ are embedded into $G$.

For generic $M, N$, this does not impose an ordering on the inputs and outputs, but we always take $M$ and $N$ to be finite cardinals, i.e. sets of the form $[n] = \{0, 1, \ldots, n-1\}$, considered as hypergraphs. More formally, there is a faithful, coproduct-preserving functor $D : \mathbb F \to \Hyp{\Sigma}$ sending each $i \in \textrm{ob}{\mathbb F} = \mathbb N$ to a hypergraph whose set of nodes is $[i]$ and sending each function to the induced homomorphism of discrete hypergraphs. Then, we can let $M := Dm, N := Dn$.

By introducing a functor that `picks out' the objects, we get a more refined notion of a cospan category than the one encountered in section~\ref{sec:frob}.

\begin{theorem}\label{thm:cspF}
Let $\mathbb X$ be a PROP whose monoidal product is a coproduct, $\cat C$ a category with finite colimits, and $F: \mathbb X \rightarrow \cat C$ a coproduct-preserving functor. Then there exists a PROP $\DCospan{F}{\cat C}$ whose arrows $m$ to $n$ are isomorphism classes of $\cat C$ cospans $Fm \ra C \la Fn$.
\end{theorem}

\begin{proof}
  Composition in $\DCospan{F}{\cat C}$ is given by pushout as in Definition~\ref{defn:cospan}.
Given $Fm \ra C \la Fm' \in \DCospan{F}{\cat C}(m,m')$ and $Fn \ra D \la Fn' \in \DCospan{F}{\cat C}(n,n')$
their monoidal product is the cospan
\[
F(m+n) \xrightarrow{\sim} Fm+Fn \ra C + D \la Fm' + Fn' \xleftarrow{\sim} F(m'+n')
\]
where the leftmost and rightmost maps are iso since $F$ preserves the monoidal product (given by the coproduct). It follows that this data defines a strict monoidal category.

The symmetries are inherited from $\mathbb X$, being the following cospans
\[
F(m+n) \xrightarrow{F\sigma_{m,n}} F(n + m) \la F(n + m)
\]
To see that the symmetries are natural, it suffices to note that the symmetry structure in $\mathbb X$
is determined by the universal property of the coproduct, which is preserved
by $F$.
\end{proof}

Our main example is $\DCospan{D}{\Hyp{\Sigma}}$, which we sometimes refer to as the category of \textit{discrete cospans of hypergraphs}. Composition and monoidal product in this category correspond exactly to the intuitive notions of plugging string diagrams together and putting diagrams side-by-side.

\begin{example}
  Consider the following morphisms $G : 2 \to 3, H: 3 \to 1$ in $\DCospan{D}{\Hyp{\Sigma}}$
  \[ G :=\ \ \input{./figures/cospan-G.tikz} \]
  \[ H :=\ \ \input{./figures/cospan-H.tikz} \]
  We can form the composition and monoidal product as follows
  \[ G \poi H = \ \ \input{./figures/cospan-GpoiH.tikz} \]
  \[ G \tns H = \ \ \input{./figures/cospan-GtnsH.tikz} \]
\end{example}

The composition and tensor product in $\DCospan{D}{\Hyp{\Sigma}}$ now give us a way to interpret string diagrams (or more precisely, $\Sigma$-terms) as hypergraphs. For each of the generators $o : m \to n$ in $\Sigma$, we can form a cospan $m \rightarrow O \leftarrow n$ where $O$ consists of a single $o$-labelled hyperedge with $m$ input nodes and $n$ output nodes
\[
\input{./figures/hypergraph-gen.tikz}
\]
Thus, any $\Sigma$-term is interpreted by taking compositions and monoidal products of such cospans.

This gives us a recipe for interpreting any string diagram as a cospan of hypergraphs, but what about interpreting a generic cospan of hypergraphs as a string diagram? In particular, note that all of the examples above have involved hypergraphs where each node has at most one in-hyperedge and one out-hyperedge. We made no such restriction when we defined hypergraphs, so how can we make sense of more general hypergraphs where a single node is connected to many hyperedges?


Thankfully, we already answered this question without realising it back in Section \ref{sec:frob}, when we showed that $\Cospan{\mathbb F}$ is isomorphic to the PROP of Frobenius algebras. The only missing ingredient is to find a relationship between $\Cospan{\mathbb F}$ and $\DCospan{D}{\Hyp{\Sigma}}$. It turns out that the first embeds faithfully in the latter, a fact that arises as a special case of the following theorem.

\begin{theorem}\label{thm:adjunctionchain}
  Let $\mathbb X$ be a PROP whose monoidal product is a coproduct and $\cat C$ a category such that $\mathbb X$ and $\cat C$ have finite colimits, and $F : \mathbb X \to \cat C$ a colimit-preserving functor.
Then there is a homomorphism of PROPs
$\widetilde F : \Cospan{\mathbb X} \to \DCospan{F}{\cat{C}}$ that sends
 $m \xrightarrow{f} X \xleftarrow{g} n$ to $Fm \xrightarrow{Ff} FX \xleftarrow{Fg} Fn$. If $F$ is full and faithful, then $\widetilde F$ is faithful.
\end{theorem}
\begin{proof}
Since $F$ preserves finite colimits, $\widetilde F$ preserves composition (since it preserves pushouts) and monoidal product (since it preserves coproducts). Symmetries, which are inherited from $\mathbb X$, are clearly preserved. Finally, suppose that $\widehat{F}(m \xra{f} X \xla{g} n) = \widehat{F}(m \xra{f'} Y \xla{g'} n)$. Then we have a commutative diagram in $\cat C$
\[
\xymatrix@R=1pt{
& {FX} \ar[dd]^\psi \\
{Fm} \ar[ur]^{Ff} \ar[dr]_{Ff'} & & {Fn} \ar[ul]_{Fg} \ar[dl]^{Fg'} \\
& {FY}
}
\]
where $\psi$ is an iso. Since $F$ is full there exists $\varphi:X\to Y$ with $F\varphi=\psi$. Since $F$ is faithful, $\varphi$ is an isomorphism. Hence, the cospans $m\xra{f} X \xla{g} n$ and $m\xra{f'} X \xla{g'} n$ are equal in $\Cospan{\mathbb X}$, so $\widehat F$ is faithful.
\end{proof}

From this, we get the following corollary.

\begin{corollary}\label{cor:frob-into-csphyp}
  There is a faithful PROP homomorphism $\widetilde D : \Cospan{\mathbb F} \to \DCospan{D}{\Hyp{\Sigma}}$.
\end{corollary}

That is, we get an embedding of the PROP $\cat{Frob} \cong \Cospan{\mathbb F}$ into our PROP $\DCospan{D}{\Hyp{\Sigma}}$ of `combinatorial string diagrams'.

\begin{definition}\label{frobto}
  Let $\frobTosem{\cdot}\from\frob \to \DCospan{D}{\Hyp{\Sigma}}$ be the homomorphism obtained by
  composing the isomorphism of Proposition~\ref{prop:cospanfrob} with the homomorphism of
  Corollary~\ref{cor:frob-into-csphyp}.
\end{definition}

Following the recipe from Section~\ref{sec:frob} relating Frobenius algebra diagrams to cospans, the four basic generators of a Frobenius algebra map to cospans as follows
\[
\begin{array}{ccccccc}
\Bmult & \quad\mapsto\quad & \input{./figures/csp-mult.tikz}
&\qquad&
\Bcomult & \quad\mapsto\quad & \input{./figures/csp-comult.tikz} \\[1cm]
\Bunit & \quad\mapsto\quad & \input{./figures/csp-unit.tikz}
&\qquad&
\Bcounit & \quad\mapsto\quad & \input{./figures/csp-counit.tikz}
\end{array}
\]
where all of the cospan maps are given by the unique function into the one-element set. These capture in a compositional way all the ways in which a node in a hypergraph could have different numbers of in- and out-hyperedges.

\begin{example}
  Consider the following composition of string diagrams
  \[ \input{./figures/ab.tikz} \ \ \textrm{\Large $\poi$}\ \  \Bmult \ \ =\ \   \input{./figures/ab-mu.tikz} \]
  The corresponding composition in $\DCospan{D}{\Hyp{\Sigma}}$ looks like this
  \[ \input{./figures/cospan-ab.tikz} \ \ \textrm{\Large $\poi$}\ \  \input{./figures/csp-mult.tikz} \]
  This is computed by taking the pushout, which identifies the two nodes connected to the outputs of the $a$ and $b$-labelled hyperedges. Hence, we obtain
  \[ \input{./figures/ab-mu.tikz}\ \ \mapsto\ \ \input{./figures/cospan-ab-mu.tikz} \]
  By composing with the Frobenius algebra generators, we can obtain even more general cospans. For example
  \ctikzfig{abc-mu}
  Note how, like in Section~\ref{sec:frob}, connected components of Frobenius algebra generators become single nodes in the combinatorial picture. In the next section, we will see that any cospan in $\DCospan{D}{\Hyp{\Sigma}}$ arises from a string diagram, possibly with some Frobenius algebra generators, in this way.
\end{example}

\begin{remark}
  For coloured PROPs, we have a similar situation, except that we get a different Frobenius algebra for each colour and we end up with hypergraphs whose nodes are labelled by the colours associated to those Frobenius algebras.

  Fix a set of colours $\col \in \mathbb F$. As shown in Theorem~\ref{thm:frobcol}, $\Cospan{\mathbb F \downarrow \col}$ is isomorphic to the coproduct $\sum_{c\in \col} \frob_c$ consisting of one copy of \frob for each of the colours $c \in \col$. We define a coproduct preserving, full and faithful functor $D_C$ from $\mathbb F \downarrow \col$ to the category of finite hypergraphs $\Hyp{\col, \Sigma}$, which sends a coloured set to its corresponding discrete (node-labelled) hypergraph.

  This extends to a faithful coloured PROP homomorphism $\widetilde{D_C}$ in the same way as in Corollary~\ref{cor:frob-into-csphyp}. Hence, following Definition~\ref{frobto}, we get an interpretation $\frobTosem{.}_C : \sum_{c\in \col} \frob \to \DCospan{D_C}{\Hyp{\col, \Sigma}}$ from coloured Frobenius algebras into cospans of hypergraphs with coloured nodes.
\end{remark}

\subsection{DPO rewriting for hypergraphs}\label{sec:backgroundDPO}

We now introduce the basics of DPO rewriting, applied to $\Sigma$-hypergraphs.
We recall the DPO approach to rewriting applied to a category $\catC$ with pushouts.
A \emph{DPO rule} is a span $L \xleftarrow{} K \xrightarrow{} R$ in $\Hyp{\Sigma}$. $L$ gives the LHS of the rule, $R$ gives the RHS, whereas $K$ gives the \textit{invariant subgraph} associated with the rule. For our purposes, $K$ will always be a discrete hypergraph consisting of the inputs and the outputs of the two sides of the rule.

\begin{example}\label{ex:dpo-rule}
  Consider the following hypergraph DPO rule
  \begin{equation}\label{eq:dpo-rule-ex}
    \input{./figures/fg-rule-ex-dpo.tikz}
  \end{equation}
  where the numbering is used to indicate how $K$ embeds into $L$ and $R$, respectively. While we will not establish a formal correspondence between string diagram equations and DPO rules until Section~\ref{sec:equiv}, one can see intuitively how this corresponds to the following equation between string diagrams
  \ctikzfig{fg-rule-ex}
  Namely, hyperedges play the role of boxes and vertices track how the boxes are connected to each other. The embeddings of $K$ into $L$ and $R$ play an important role in the rule above: they maintain the correspondence between inputs/outputs on the LHS and inputs/outputs on the RHS. If we change these embeddings, we will change the rule. For example, if $K \to R$ in \eqref{eq:dpo-rule-ex} is mapped $0 \mapsto 1, 1 \mapsto 0, 2 \mapsto 2$, the rule would become
  \ctikzfig{fg-rule-swap-ex}
\end{example}

\begin{remark}\label{rem:coproduct-boundary}
  Note that if we considered $L$ and $R$ as cospans $2 \xrightarrow{i} L \xleftarrow{o} 1$ and $2 \xrightarrow{i'} R \xleftarrow{o'} 1$ following Section~\ref{sec:hyp-inform}, $K \cong 2 + 1$, the coproduct of the input and the output sets. The maps from $K$ to $L$ and $R$ are given by the induced copairings, $[i,o]: K \to L$ and $[i',o']: K \to R$. This will play a role in the next section when we establish a formal correspondence between syntax and combinatorics.
\end{remark}

A \emph{DPO rewriting system} $\mathcal{R}$ is a set of DPO rules.
Given hypergraphs $G$ and $H$, we say that $G$ \emph{rewrites} into $H$ ---notation $G \DPOstep{\mathcal{R}} H$---
if there exists $L \tl{} K  \tr{} R$ in $\mathcal{R}$, object $C$ and morphisms
such that the squares below are pushouts
\begin{equation}\label{eq:dpo2}
\raise12pt\hbox{$
\xymatrix@R=10pt@C=20pt{
L \ar[d]_m   &  K \ar[d]
\ar@{}[dl]|(.8){\text{\large $\urcorner$}}
\ar@{}[dr]|(.8){\text{\large $\ulcorner$}}
\ar[l] \ar[r]  & R \ar[d] \\
G &  C \ar[l] \ar[r]  & H }$}
\end{equation}
Typically the object $C$ and the arrows $K \tr{} C \tr{} G$ are computed from
$K \tr{} L \tr{m} G$ in such a way that the left square above forms a pushout.
In this case, the arrows $K \tr{} C \tr{} G$ are called a \textit{pushout complement}.
Hence, DPO rewriting can be seen as two distinct steps: first computing the pushout
 complement $K \tr{} C \tr{} G$, then pushing out the span $C \tl{} K \tr{} R$
 to produce the rewritten object $H$. A \emph{derivation} from $G$ into $H$
is a sequence of such rewriting steps.

\begin{example}\label{ex:simple-rewrite}
Consider a rewriting system with the following rule
\[
\input{./figures/brule-ex.tikz}\ \
\]
The rule  has a matching into the following hypergraph
\begin{equation}\label{eq:graph-ex}
\input{./figures/graph-ex.tikz}
\end{equation}
yielding the following DPO rewriting diagram
\[ \scalebox{0.9}{\input{./figures/rewrite-ex.tikz}} \]
The above rewrite is the hypergraph equivalent of applying the rule $f;g \Rew{} g;f$ to the term $f;g;f$ to obtain $g;f;f$.
\end{example}

Pushout complements do not necessarily have to exist or be unique. For a fixed rule $L \tl{} K \tr{} R$, a \textit{matching} $m : L \to G$ is a morphism with the additional property that the pushout complement of $K \tr{} L \tr{m} G$ exists. For the example $\catC := \Hyp{\col, \Sigma}$, we can give the precise conditions under which $m$ is a matching,
drawing e.g. from~\cite{HandbookDPO}.

\begin{definition}
Morphisms $K \tr{i} L \tr{m} G$ satisfy the \textit{no-dangling} condition if, for every hyperedge not in the image of $m$, every node of its source and target is either (i) not in the image of $m$ or (ii) in the image of $i \poi m$.
 They satisfy the \emph{no-identification} condition if any two nodes merged by $m$ are in the image of $i$.
 \end{definition}

\begin{proposition}
A pushout complement of $K \tr{i} L \tr{m} G$ exists if and only if it satisfies the no-dangling and no-identification conditions.
\end{proposition}


Uniqueness is a simpler story, as it only relies on $i$ to be mono.

\begin{proposition}
If $i : K \to L$ is mono, the pushout complement of $K \tr{i} L \tr{m} G$ is unique (up to commuting isomorphism), when it exists. That is, when $C, C'$ are two pushout complements, there exists an isomorphism $C \tr{\cong} C'$ such that the following diagram commutes
\[
\raise12pt\hbox{$
\xymatrix@R=10pt@C=20pt{
L \ar[d]_m   &
K \ar[d]
\ar[l]
\ar[rdd]  & \\
G &  C \ar[l] \ar[rd]|-{\cong} & \\
  &  & \ar[ull] C'
}$}
\]
\end{proposition}

If $i$ is not mono, the pushout may not be unique, see Section~\ref{sec:contexts} below. In fact, the property holds for any adhesive category, of which labelled directed hypergraphs are an example~\cite{Lack2005}. The case where $K\to L$ is mono is important, because the matching $m$ fully determines the resulting hypergraph $H$, due to the uniqueness of pushout complements. A rule $L \tl{} K  \tr{} R$ is said to be \emph{left-linear} if the morphism $K\to L$ is mono.

On the other hand, we will occasionally study rules that are \textit{not} left-linear, in which case the pushout complement does not need to be unique. In the case of hypergraphs, all of the distinct pushout complements can still be effectively enumerated~\cite{heumuller2011construction}, which in our setting is closely related to enumeration of contexts when rewriting modulo the Frobenius equations. This will be covered in detail in Section~\ref{sec:contexts}.

\subsection{Double-Pushout Rewriting with Interfaces}\label{sec:dpoi}
Our first observation is that an extension of the traditional DPO rewriting, acting on
hypergraphs \emph{with interfaces}, actually fits best our purposes.
This is for two main reasons. First, taking interfaces into account is essential to
adequately map syntactic rewriting into DPO rewriting. Indeed, string diagrams are
interpretable as \emph{cospans} of hypergraphs, and the source and target of such
cospans constitute the interface of the diagram. Second, DPO rewriting with interfaces
(henceforth, DPOI) is of independent interest in the context of confluence.
It is well-known~\cite{Plump10} that local confluence for DPO is undecidable.
As shown in the sequel to
this paper (after~\cite{BGKSZ-esop17,BGKSZ-partthree}), DPOI enjoys the Knuth-Bendix property:
joinability of critical pairs is the same as local confluence. Hence, DPOI in this case
closely matches standard term rewriting, whereas DPO is analogous to restricting rewriting to ground terms, where confluence is undecidable.

\begin{remark}\label{rem:dpoi-related}
DPOI has appeared in different guises in the graph rewriting literature, such as rewriting with \textit{borrowed contexts}~\cite{Ehrig2004}, the graphical encodings of process calculi~\cite{Gadducci07,BonchiGK09}, and some foundational studies connecting DPO rewriting with computads in cospans~\cite{Gadducci1998,Sassone2005a}.
\end{remark}

Fix a category $\catC$ with pushouts. We provide a definition of rewriting for morphisms $G \tl{}J$ in $\catC$. When $\catC$ is $\Hyp{\Sigma}$, we call them \emph{(hyper)graphs with interface}. The intuition is that $G$ is a hypergraph and $J$ is an interface that allows $G$ to be ``glued'' to a context.

Given $G \leftarrow J$ and $H \leftarrow J$ in $\catC$, \emph{$G$ rewrites into $H$ with interface $J$} --- notation $(G \tl{} J) \DPOstep{\mathcal{R}} (H \tl{} J)$ --- if there exist rule $L \tl{} K  \tr{} R$ in $\mathcal{R}$ and object $C$ with suitable morphisms
in $\catC$
such that the diagram below commutes, where the marked squares are pushouts

\begin{equation}\label{eq:dpo2a}
\raise25pt\hbox{$
\xymatrix@R=10pt@C=20pt{
L \ar[d]_m   &  K \ar[d]
\ar@{}[dl]|(.8){\text{\large $\urcorner$}}
\ar@{}[dr]|(.8){\text{\large $\ulcorner$}}
\ar[l] \ar[r]  & R \ar[d] \\
 G &  C \ar[l] \ar[r]  & H \\
&  J \ar[u] \ar[ur]  \ar[ul]
}$}
\end{equation}
Hence, the interface $J$ is preserved by individual rewriting steps.

\begin{remark}\label{rem:interface-cospan}
  Similar to the case of rewriting rules (cf. Remark~\ref{rem:coproduct-boundary}) if $G$ comes from a cospan representing a string diagram, the interface $J$ is the coproduct of the inputs and the outputs.
\end{remark}

When $\catC$ has a (strict) initial object $0$ (e.g. in $\Hyp{\Sigma}$, $0$ is the empty hypergraph), ordinary DPO rewriting can be considered as a special case, by taking $J$ to be $0$. Like for traditional DPO, rewriting steps are considered up to isomorphism: $G_1 \leftarrow J: f_1$ and $G_2 \leftarrow J: f_2$ are isomorphic if there is an isomorphism $\varphi \colon G_1 \to G_2$ with $f_1\poi \varphi = f_2$.

In Section~\ref{sec:backgroundDPO} we said that a morphism is called a \textit{match} if a suitable pushout complement exists. In DPOI, the condition for being a match is strictly stronger, as this pushout complement must furthermore respect the interface in~\eqref{eq:dpo2a}.

\begin{example}
  We now extend Example~\ref{ex:simple-rewrite} to DPOI rewriting. We consider the same rewriting rule as before
\[
\input{./figures/brule-ex.tikz}\ \
\]
but now the target graph also has an interface
\begin{equation}\label{eq:bgraph-ex}
\input{./figures/bgraph-ex.tikz}
\end{equation}
Hence, rewriting produces the following DPOI diagram, where the interface of the target graph is tracked on the bottom row
\[ \scalebox{0.9}{\input{./figures/brewrite-ex.tikz}} \]
Note that, while the top two pushout squares in this diagram are identical to the ones in Example~\ref{ex:simple-rewrite}, the presence of an interface has an effect on whether a rewriting rule applies. For example, if the hypergraph in \eqref{eq:bgraph-ex} is given the following, different interface
\begin{equation}\label{eq:bgraph-ex2}
\input{./figures/bgraph-ex2.tikz}
\end{equation}
the LHS now has no match, as the new interface no longer factors through the pushout complement.

While the interface in~\eqref{eq:bgraph-ex2} would not arise from a morphism in a generic monoidal category, it can arise when we include Frobenius algebra structure. Namely, it corresponds to one of the following diagrams\ctikzfig{funny-interface}
Whether it corresponds to the left or right diagram above depends on whether the node `$1$' came from an input or an output in the associated cospan (cf. Remark~\ref{rem:interface-cospan}).
\end{example}

\if\ismain0 

\bibliographystyle{plain}
\bibliography{catBib3Rev}

\fi 

\if\ismain0 

\setcounter{section}{4} 
\tableofcontents

\fi 

\section{Equivalence of syntax and combinatorics}\label{sec:equiv}

\subsection{Terms as cospans}

In Section~\ref{sec:background}, we defined $\syntax{\Sigma}$ as the free PROP generated by a signature $\Sigma$. By freeness, we can completely define the PROP homomorphism
\[
  \synTosem{\cdot}\from \syntax{\Sigma} \to \DCospan{D}{\Hyp{\Sigma}}
\]
by specifying how it acts on the generators in $\Sigma$. We do this by sending each generator $o : k \to l$ to a cospan $k \rightarrow O \leftarrow l$ where $k, l$ are discrete hypergraphs and $O$ is hypergraph containing a single $o$-labelled hyperedge connected to a single distinct node for each input and output. The cospan maps embed the inputs and outputs in the obvious way
\[
\input{./figures/hypergraph-gen.tikz}
\]
This mapping of generators extends by universal property to a PROP homomorphism
$\synTosem{\cdot} \: \syntax{\Sigma} \to \DCospan{D}{\Hyp{\Sigma}}$ sending any $\Sigma$-term in $\syntax{\Sigma}$ to its associated composition of cospans.

\subsection{Characterisation theorem for string diagrams}

We now have all the ingredients needed to prove the equivalence of the free (i.e. syntactic) category $\freehyp{\Sigma}$ of string diagrams modulo Frobenius structure and its combinatoric representation $\FTerm{\Sigma}$. We begin by noting that $\freehyp{\Sigma}$ is isomorphic to the coproduct of PROPs $\syntax{\Sigma} + \frob$, and in the following we will use them
interchangeably, depending on what is most convenient.
%
Then, we define $\allTosem{\cdot}$ as the copairing in  $\PROP$ of the faithful functors
$\synTosem{\cdot} \: \syntax{\Sigma} \to \FTerm{\Sigma}$
and
$\frobTosem{\cdot} \from \frob \to \FTerm{\Sigma}$.

\begin{theorem}\label{th:characterisation} $\allTosem{\cdot} \: \syntax{\Sigma} + \frob \to \DCospan{D}{\Hyp{\Sigma}}$ is an isomorphism of PROPs.
\end{theorem}

\begin{proof}
It suffices to verify that $\FTerm{\Sigma}$ satisfies the universal property of the coproduct $\syntax{\Sigma} + \frob$ in $\PROP$.

\begin{equation} \label{eq:univcoproductgen}
\vcenter{\xymatrix@R=5pt@C=45pt{
{\syntax{\Sigma}} \ar[ddrr]_\alpha \ar[rr]^-{\synTosem{\cdot}} &&
{\FTerm{\Sigma}} \ar@{.>}[dd]^\gamma && \ar[ll]_{\frobTosem{\cdot}} {\frob} \ar[ddll]^{\beta} \\
\\
&& {\catA} &&
}}
\end{equation}
 Given $\alpha$, $\beta$, and a PROP $\catA$ as in \eqref{eq:univcoproductgen}, we need to show the existence of a unique $\gamma$ making the diagram commute. Now, because all morphisms in \eqref{eq:univcoproductgen} are identity-on-objects, it suffices to show that any arrow of $\FTerm{\Sigma}$ can be decomposed in an essentially unique way into an expression where all the basic constituents lie in the image of $\synTosem{\cdot}$ or $\frobTosem{\cdot}$.

To this aim, fix a cospan $n \tr{f} G \tl{g} m$ in $\FTerm{\Sigma , \col}$, where $G$ has set of nodes $N$, set of hyperedges $E$ and a labelling function $\chi \: E \to \Sigma$. We pick an order $e_1,\dots,e_j$ on the hyperedges in $E$. Let $\tilde{n} \tr{i} \tilde{E} \tl{o} \tilde{m}$ be the cospan defined as $\bigoplus_{1 \leq i \leq j} \synTosem{\chi(e_i)}$. Intuitively, $\tilde{E}$ piles up all the hyperedges of $G$, but disconnected from each other. $\tilde{n}$ and $\tilde{m}$ are the concatenations of all the inputs, respectively outputs of these hyperedges.

There are obvious functions $f \colon n \to N$, $g \colon m \to N$, $j \colon \tilde{n} \to N$ and $p \colon \tilde{m} \to N$ mapping nodes to their occurrence in the set $N$ of all nodes of $G$. All this information is now gathered in the following composition of cospans \footnote{
Note in \eqref{eq:decomp} $N$ is seen both as a discrete hypergraph --- when appearing in the carrier --- and as an object of $\FTerm{\Sigma}$ --- when appearing in the domain or codomain of a cospan. We did not emphasise this distinction to not overload notation.
}
\begin{equation}\label{eq:decomp}
(n \tr{f}  N \tl{\copair{id, j}} N \tns \tilde{n}) \ \poi \ (N \tns \tilde{n} \tr{id \tns i} N \tns \tilde{E} \tl{id \tns o} N \tns \tilde{m}) \ \poi \ (N \tns \tilde{m} \tr{\copair{id, p}} N \tl{g} m)
\end{equation}
Copairing maps $\copair{id, j}$ and $\copair{id, p}$ are well-defined as $\tns$ is also a coproduct in $\Hyp{\Sigma}$. One can compute that the result of composing \eqref{eq:decomp} (by pushout) is indeed isomorphic to $n \tr{f} G \tl{g} m$.

Towards a definition of $\gamma$, we need to check that every component of \eqref{eq:decomp} is in the image of either $\synTosem{\cdot}$ or $\frobTosem{\cdot}$. First, the middle cospan is clearly in the image of $\synTosem{\cdot}$, as it is the monoidal product of the identity cospan $w \tr{id} w \tl{id} w$ with $\tilde{n} \tr{i} \tilde{E} \tl{o} \tilde{m}$, which itself is the monoidal product of cospans each in the $\synTosem{\cdot}$-image of some generator in $\Sigma$. Second, we have that the two outmost cospans are in the image of $\frobTosem{\cdot}$: this is because, by Definition~\ref{frobto} and Proposition~\ref{prop:cospanfrob}, any morphism $u_1 \tr{} u_3 \tl{} u_2 $ of $\FTerm{\Sigma}$, with $u_1$, $u_2$, $u_3$ discrete, is in the image of $\frobTosem{\cdot}$.

Therefore, $\gamma$ can be defined on $n \tr{f} G \tl{g} m$ by the values of $\synTosem{\cdot}$ and $\frobTosem{\cdot}$ on its decomposition as in \eqref{eq:decomp}. This is a correctly and uniquely defined assignment: in the construction of the decomposition \eqref{eq:decomp}, the only variable parts are the different orderings that are picked for nodes and for hyperedges in $\tilde{E}$, but these are immaterial since all the involved categories are symmetric monoidal.
\end{proof}

We now observe two interesting consequences of this theorem. Recall from Section~\ref{sec:frop} that $\freehyp{\Sigma}$ is the free FROP over the signature $\Sigma$.

\begin{corollary}\label{cor:freehyp-fterm}
  $\freehyp{\Sigma} \cong \DCospan{D}{\Hyp{\Sigma}}$
\end{corollary}

\begin{proof}
  $\frobTosem{\cdot} \: \frob \to \DCospan{D}{\Hyp{\Sigma}}$ defines a FROP structure on $\DCospan{D}{\Hyp{\Sigma}}$ and the isomorphism of Theorem \ref{th:characterisation} extends to one of FROPs. Then the result follows by Proposition~\ref{prop:cospanfrob} and Theorem~\ref{th:characterisation}.
\end{proof}

The next corollary states that there is no `information loss' in passing from the free PROP to the free FROP on $\Sigma$.

\begin{corollary} \label{cor:faithful}
  $\synTosem{\cdot}  \: \syntax{\Sigma} \to \DCospan{D}{\Hyp{\Sigma}}$ is faithful.
\end{corollary}
\begin{proof}
We use the fact that coproducts of PROPs can be computed as pushouts in the category $\SMC$ of small symmetric monoidal categories. In particular, $\syntax{\Sigma} + \frob$ in $\PROP$ arises as
 \begin{equation}\label{eq:pushout}
 \vcenter{
\xymatrix@R=20pt@C=40pt{
 \ar[d]_{!_1} \perm{}  \ar[r]^-{!_2} \drcorner & \frob \ar[d]^-{\frobTosem{\cdot}}\\
 \ar[r]_-{\synTosem{\cdot}} \syntax{\Sigma} & {\syntax{\Sigma} + \frob \cong \DCospan{D}{\Hyp{\Sigma}}}
}
}
\end{equation}
in $\SMC$, where $\perm{}$ is the category of finite sets and bijections (the initial object in the category $\PROP$, see Example \ref{ex:perm}) and the maps $!_1$ and $!_2$ are given by initiality of $\perm{}$. Intuitively, in \eqref{eq:pushout} $\syntax{\Sigma} + \frob$ is built as the disjoint union of the categories $\syntax{\Sigma}$ and $\frob$ where one identifies the two copies of each object $n \in \mathbb N$
and the symmetric monoidal structure of the two categories, i.e. the morphisms in the image of $\perm{}$.

Now, in order to prove that $\synTosem{\cdot}$ is faithful, we can use a result~\cite[Th.~3.3]{MacDonald2009} about amalgamation in $\CAT$ (which transfers to $\SMC$). As all the functors in \eqref{eq:pushout} are identity-on-objects and $!_1$, $!_2$ are faithful, it just requires to show that $!_1$ and $!_2$ satisfy the so-called 3-for-2 property: for $!_1$, this means that, given $h= f \poi g$ in $\syntax{\Sigma}$, if any two of $f,g,h$ are in the image of $!_1$, then so is the third. This trivially holds as every arrow of $\perm{}$ is an isomorphism. The argument for $!_2$ is identical.
\end{proof}

\subsection{Characterisation theorem for coloured PROPs}
Of crucial importance for the results in the latter half of this paper is that all of the results from the previous section extend in the obvious way to coloured PROPs. Notably, we can state a version of Theorem~\ref{th:characterisation} for $\col$-coloured PROPs, with
$\allTosem{\cdot}_\col$ the copairing in $\CPROP$, modulo the identification of the colours, of the faithful functors
$\synTosem{\cdot}_\col  \: \syntax{\col, \Sigma} \to \DCospan{D_\col}{\Hyp{\col, \Sigma}}$
and
$\frobTosem{\cdot}_\col \from \frob_\col \to \DCospan{D_\col }{\Hyp{\col,\Sigma}}$,
the $\col$-coloured versions of $\synTosem{\cdot}$ and $\frobTosem{\cdot}$,
 respectively.

\begin{proposition}\label{th:characterisation-col}
$\allTosem{\cdot}_\col \: \syntax{\col,\Sigma} +_\col \CFrob{\col} \to \DCospan{D_\col}{\Hyp{\col,\Sigma}}$ is an isomorphism of
$\col$-coloured PROPs.
\end{proposition}

It is also worth noting that the Frobenius structure identifies the hypergraphs with no hyperedges, i.e. the sets of $\col$-labelled nodes.

\begin{corollary}
There is an isomorphism of $\col$-coloured PROPS
$\frob_{\col} \cong \DCospan{D_\col}{\Hyp{\col,\emptyset}}$.
\end{corollary}

As arrows of $\DCospan{D_\col}{\Hyp{\col,\Sigma}}$ are the same thing as cospans in the slice category $\FINSET \downarrow \col$, this corollary is essentially a restatement of Theorem~\ref{thm:frobcol}.

\begin{example}\label{ex:bipartite2}
Consider the free FROP $\freehyp{\col, \Sigma}$ for colours $\col = \{\sbl,\sr\}$ introduced in Example \ref{ex:bipartite}. By Corollary~\ref{cor:freehyp-fterm}, $\freehyp{\col, \Sigma} \cong \DCospan{D_\col}{\Hyp{\col,\Sigma}}$.  In hypergraphs of $\DCospan{D_\col}{\Hyp{\col,\Sigma}}$, hyperedges correspond to switches \BRchange or \RBchange.  Since these hyperedges are in fact edges (i.e. they have one input and one output node), we can draw them as such. Since they connect any two nodes only when these have a different colour, what we obtain are exactly finite directed bipartite graphs.  For example, the combinatorial presentation of the diagram \eqref{eq:bipartite} from Example~\ref{ex:bipartite} is
\[ \input{./figures/two-col-ex.tikz} \qquad \xmapsto{\allTosem{\cdot}}\qquad \input{./figures/two-col-ex-graph.tikz}\]
This shows that Theorem \ref{th:characterisation} not only provides a combinatorial representation for algebraic structures, but conversely it allows us to give a purely algebraic representation for bipartite graphs.
\end{example}

\subsection{Characterisation theorem for rewriting}

DPOI rewriting deals with hypergraphs $(G \tl{} I)$ with a \emph{single} interface. These may be seen as a particular class of morphisms in $\FTerm{\Sigma}$, namely those of the form $0 \tr{} G \tl{} I$, where $0$ is the empty hypergraph (which also serves as the initial object of $\FTerm{\Sigma}$).
This means that DPOI rewriting can be meaningfully defined on morphisms of $\FTerm{\Sigma}$ with source $0$.
However, our semantics $\allTosem{\cdot}$ maps diagrams of $\syntax{\Sigma}+\frob$ to cospans with \emph{any} source. This gap can be overcome: in order to interpret syntactic rewriting as DPOI rewriting, we need an intermediate step in which we `fold' the two interfaces $m, n$ of a string diagram $a \: m \to n$ into one $m + n$.

For cospans $m \xrightarrow{i} G \xleftarrow{o} n$, we already noted in Remark~\ref{rem:interface-cospan} that we can do this simply by using the copairing $[i,o] : m + n \to G$. We will now show that there is an equivalent syntactic operation to the passage from $m \xrightarrow{i} G \xleftarrow{o} n$ to $0 \to G \xleftarrow{[i,o]} m + n$. First, we note that we can use the Frobenius algebra structure in $\freehyp{\Sigma}$ to build `cup' and `cap' morphisms $\cup : 0 \to 2$ and $\cap : 2 \to 0$ as follows
\begin{equation}\label{def:cupcap-frob-def}
  \input{./figures/cup-frob-def.tikz} \qquad\qquad
  \input{./figures/cap-frob-def.tikz}
\end{equation}
It follows from the Frobenius equations that the cup and cap maps satisfy the following equations
\begin{equation}\label{eq:cupcap-frob-eqs}
  \input{./figures/cupcap-frob-eqs.tikz}
\end{equation}
These equations capture the fact that the object $1$ is equal to its own \textit{dual}. This structure extends to cups and caps $\cup_n : 0 \to n + n$, $\cap_n : n + n \to 0$ for an arbitrary object $n$ as follows
\[
  \input{./figures/cup-many-def.tikz} \qquad\qquad
  \input{./figures/cap-many-def.tikz}
\]
This collection of cups and caps for every object $n$ endows a FROP with the structure of a compact closed category, which is furthermore coherently self-dual in the sense of~\cite{selinger2010selfdual}.

\begin{remark}
  The fact that hypergraph categories (i.e. categories where every object is equipped with a chosen Frobenius algebra structure) are always compact closed was noted in~\cite{Carboni1987}, motivating the original name of \textit{well-supported compact closed categories}.
\end{remark}

We can use the `cup' part of the compact structure to define an operation $\rewiring{\cdot}$ that bends all of the input wires around to become outputs. That is, for a map $a : m \to n$, we can form $\rewiring{a} : 0 \to m + n$ as $\cup_m \poi (1_m \tns a)$. Or, as a diagram
\begin{equation}\label{eq:name-of-a}
  \input{./figures/name-of-a.tikz}
\end{equation}
This operation is sometimes called forming the `name' of a morphism~\cite{Abramsky2004} and can be inverted by post-composing with $\cap_m \tns 1_n$ and applying the first equation in~\eqref{eq:cupcap-frob-eqs}.

\begin{proposition}\label{prop:rewriting-is-copairing}
  Suppose $\allTosem{a}$ yields a cospan $m \xrightarrow{i} A \xleftarrow{o} n$, then $\allTosem{\rewiring{a}}$ yields a cospan isomorphic to $0 \to A \xleftarrow{[i,o]} m + n$.
\end{proposition}

\begin{proof}
  From the definition of the cup map in terms of Frobenius structure, it follows that
  \[
    \allTosem{\cup} \ =\ \input{./figures/cup-cospan.tikz}
  \]
  and hence
  \[
    \allTosem{\cup_m} \ =\ \input{./figures/cup-many-cospan.tikz}
  \]
  The result then follows from interpreting $\rewiring{a} := \cup_m \poi (1_m \oplus a)$ as a composition of cospans and computing the result by pushout.
\end{proof}

Using $\rewiring{\cdot}$ will enable us to pass freely between syntactic rewriting of $\Sigma$-terms with non-trivial inputs and outputs to rewriting $\Sigma$-terms that only have outputs. We can do this both for the $\Sigma$-term being rewritten and for the rewriting rule itself. Namely, a rewriting rule $\rrule{l}{r}$ can be replaced by an equivalent (modulo Frobenius structure) rule $\rrule{\rewiring{l}}{\rewiring{r}}$. Such terms and rules having only a single interface (the outputs) will in turn correspond directly to DPOI rewriting under the interpretation functor $\allTosem{\cdot}$. Putting these ingredients together, we are ready to state our main equivalence theorem for string diagram rewriting.


\begin{theorem}\label{thm:frobeniusrewriting}
Let $\rrule{l}{r}$ be any rewriting rule on $\syntax{\Sigma}+\frob$. Then
\[
a \Rightarrow_{\rrule{l}{r}} b  \quad \text{ iff } \quad \allTosem{\rewiring{a}} \DPOstep{\allTosem{\rrule{\rewiring{l}}{\rewiring{r}}}}  \allTosem{\rewiring{b}}\text{ .}
\]
\end{theorem}
\begin{proof}
On the direction from left to right, suppose that $a \Rightarrow_{\rrule{l}{r}} b$. Thus, by definition
\begin{equation}\label{eq:proof1}
\input{./figures/syn-lhs.tikz} \qquad \input{./figures/syn-rhs.tikz}
\end{equation}
Using the Frobenius structure of $\syntax{\Sigma}+\frob$ we can put $\rewiring{a}$ in the following shape
\begin{equation}\label{eq:rw-proof1}
\input{./figures/rw-proof1.tikz}
\end{equation}
where
\ctikzfig{a1star}
The dashed line in \eqref{eq:rw-proof1} decomposes the rightmost diagram into $\rewiring{l} \: 0 \to i+j$ followed by a diagram of type $i + j \to m + n$, which we name $\tilde{a}$. With analogous reasoning
\begin{gather}\label{eq:rewiringcomp}
\input{./figures/rw-proof2.tikz}
\quad\text{meaning that}\quad \rewiring{a} = \rewiring{l} \poi \tilde{a} \text{ and }\rewiring{b} = \rewiring{r}\poi\tilde{a}.
\end{gather}
Next, we introduce cospans giving semantics to the various diagrams
\begin{equation}\label{eq:defscospans}
\begin{aligned}
\allTosem{\rewiring{l}} = 0 \tr{} L \tl{} i + j \qquad \allTosem{\tilde{a}} = i + j \tr{} C \tl{} m + n \qquad \allTosem{\rewiring{r}} = 0 \tr{} R \tl{} i + j \\
\allTosem{\rewiring{a}} = 0 \tr{} G \tl{} m + n \qquad \allTosem{\rewiring{b}} = 0 \tr{} H \tl{} m + n.
\end{aligned}
\end{equation}
Equation \eqref{eq:rewiringcomp} tells that the cospan giving semantics to $\rewiring{a}$ (respectively, $\rewiring{b}$) is the composite of cospans giving semantics to $\rewiring{l}$ (respectively, $\rewiring{r}$) and $\tilde{a}$. As composition of cospans is by pushout, we obtain a double-pushout diagram as in \eqref{eq:dpo2} with $J = i + j$ and $I = m + n$, meaning that $\allTosem{\rewiring{a}} \DPOstep{\allTosem{\rrule{\rewiring{l}}{\rewiring{r}}}}  \allTosem{\rewiring{b}}$.

We now conclude the proof by showing the right to left direction of the statement. Suppose that $\allTosem{\rewiring{a}} \DPOstep{\allTosem{\rrule{\rewiring{l}}{\rewiring{r}}}}  \allTosem{\rewiring{b}}$. Naming cospans $\allTosem{\rewiring{a}}$, $\allTosem{\rewiring{b}}$, $\allTosem{\rewiring{l}}$ and $\allTosem{\rewiring{r}}$ as in \eqref{eq:defscospans}, this implies by definition the existence of a pushout complement $C$ yielding a DPOI diagram as \eqref{eq:dpo2} with $J = v_1v_2$ and $I = w_1w_2$. Now, pick $\hat{a} \: v_1 v_2 \to w_1 w_2$ such that $\allTosem{\hat{a}} = v_1 v_2 \tr{} C \tl{} w_1 w_2$, which exists by fullness of $\allTosem{\cdot}$. Because composition in $\FTerm{\Sigma}$ is by pushout, the existence of such a DPOI diagram yields
\begin{equation}\label{eq:secondproof}
\begin{aligned}
\allTosem{\rewiring{a}} = (0 \tr{} G \tl{} w_1 w_2) = (0 \tr{} L \tl{} v_1 v_2)\poi(v_1 v_2 \tr{} C \tl{} w_1 w_2) =   \allTosem{\rewiring{l}}\poi\allTosem{\hat{a}} \\
\allTosem{\rewiring{b}} = (0 \tr{} H \tl{} w_1 w_2) = (0 \tr{} R \tl{} v_1 v_2)\poi(v_1 v_2 \tr{} C \tl{} w_1 w_2) =   \allTosem{\rewiring{r}}\poi\allTosem{\hat{a}}.
\end{aligned}
\end{equation}
Because $\allTosem{\cdot}$ is faithful, \eqref{eq:secondproof} yields decompositions $\rewiring{a} = \rewiring{l}\poi\hat{a}$ and $\rewiring{b} = \rewiring{r}\poi\hat{a}$ also on the syntactic side. This allows for a rewriting step $a \Rew{\rrule{l}{r}} b$ as below, where the dashed lines show how the syntactic matching (\emph{cf.} the shape \eqref{eq:rewpropmatch}) is performed
\[
\input{./figures/rw-proof3.tikz}
\]
\end{proof}

Furthermore, we can prove an analogous result for the rewriting of coloured PROPs,
which thus subsumes and generalises the one for PROPs.
Now, syntactic rewriting occurs in $\syntax{\col,\Sigma} +_\col \CFrob{\col}$,
DPOI rewriting is in $\DCospan{D_\col}{\Hyp{\col,\Sigma}}$, and the
folding operation $\rewiring{\cdot}_\col$ works as expected in the coloured setting.
The result is stated below without proof.

\begin{proposition}\label{pr:colfrobeniusrewriting}
Let $\rrule{l}{r}$ be any rewriting rule on $\syntax{\col,\Sigma} +_\col \CFrob{\col}$.
Then
\[
a \Rightarrow_{\rrule{l}{r}} b  \quad \text{ iff } \quad
\allTosem{\rewiring{a}_\col}_\col
\DPOstep{\allTosem{\rrule{\rewiring{l}_\col}{\rewiring{r}_\col}}_\col}
\allTosem{\rewiring{b}_\col}_\col\text{ .}
\]
\end{proposition}


\begin{example}
We give an illustration of the correspondence of
Proposition~\ref{pr:colfrobeniusrewriting}.
Fix $\col = \{\sbl,\sr\}$ and $\Sigma = \{ \cgr[height=.4cm]{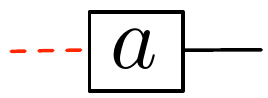}\}$,
and take the $\col$-coloured PROP
$\syntax{\Sigma,\col} +_{\scriptscriptstyle \col} (\CFrob{\sbl} + \CFrob{\sr})$.
We consider a rule $\alpha$ on such a PROP, together with its interpretation in
$\DCospan{D_\col}{\Hyp{\col,\Sigma}}$
\[
\input{./figures/col-rule-ex.tikz}
\]
We 
use numbers to indicate how the morphisms in the cospan are defined. Notice that these legs may be non-injective.
Also, notice how the interpretation ``absorbs'' the Frobenius component. With the above rule, one can perform the syntactic rewriting step $\cgr[height=.7cm]{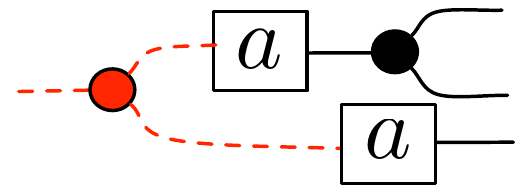} \quad \Rew{\alpha} \quad \cgr[height=.7cm]{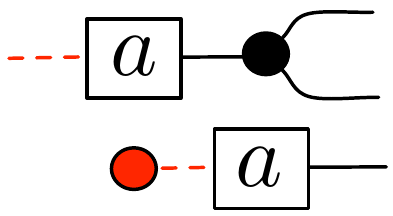}$. It is implemented in $\DCospan{D_\col}{\Hyp{\col,\Sigma}}$ via the DPOI rewriting step below
\[
\input{./figures/col-rewrite-ex.tikz}
\]
Note that the node labelled $2$ in the resulting graph is not in the image of the interface (corresponding to a $\Runit$) and the node labelled by $1$ and $3$ is in the image of two nodes in the interface (corresponding to a $\Bcomult$). Hence, the outcomes of syntactic and DPOI rewriting coincide.
\end{example}

\subsection{Pushout complements and rewriting modulo Frobenius}\label{sec:contexts}

We conclude our discussion of the connection between syntactic and combinatoric rewriting by studying rule applications for which more than one pushout complement may exist. We will see that, rather than a ``bad'' feature of combinatoric rewriting in this approach, this naturally subsumes the extra freedom one obtains in performing rewrites modulo Frobenius structure.


As discussed in Section~\ref{sec:backgroundDPO}, in adhesive categories such as
$\Hyp{\Sigma}$, pushout complements for $K \to L \to G$ are guaranteed to be unique
only when the morphism $K \to L$ is mono. If we have a look again
at a DPOI rewrite from Section~\ref{sec:dpoi}
\begin{equation}\label{eq:dpo2b}
\raise25pt\hbox{$
\xymatrix@R=10pt@C=20pt{
L \ar[d]_m   &  K \ar[d]
\ar@{}[dl]|(.8){\text{\large $\urcorner$}}
\ar@{}[dr]|(.8){\text{\large $\ulcorner$}}
\ar[l] \ar[r]  & R \ar[d] \\
 G &  C \ar[l] \ar[r]  & H \\
&  J \ar[u] \ar[ur]  \ar[ul]
}$}
\end{equation}
it follows that the rewritten graph $H$ is totally determined by the map $L \to G$ only in the case where the mapping of $K$ into the LHS $L$ of the rule is mono. However, when rewriting modulo Frobenius structure, there are cases where this is not true. For example, consider the relatively harmless-looking equation that captures the fact that $f$ is a one-sided inverse of $g$
\begin{equation}\label{eq:fg-section}
  \input{./figures/fg-section.tikz}
\end{equation}
If we translate this into a hypergraph rewriting rule from the LHS to the RHS, we have no problem, because the embedding of $K$ into $L$ is mono
\begin{equation}
\label{eq:fg-section-LR}
\input{./figures/fg-section-LR.tikz}
\end{equation}
Hence, a rewriting step involving this rule will be uniquely fixed by the matching $m$.
Consider applying the rule above to this diagram
\[
\input{./figures/fg-section-ex.tikz}
\]
which corresponds to this cospan of hypergraphs
\begin{equation}
\label{eq:fg-section-ex-h}
\input{./figures/fg-section-ex-h.tikz}
\end{equation}
Then, we see the LHS of the rewriting rule \eqref{eq:fg-section-LR} has exactly two matchings, corresponding to the upper and lower copies of $f$ and $g$, respectively. Since the embedding of the boundary into the LHS of \eqref{eq:fg-section-LR} is mono, each of these matchings corresponds to a unique way to rewrite the diagram
\bigskip
\[\input{./figures/fg-section-rw1.tikz}\]
\bigskip
\[\input{./figures/fg-section-rw2.tikz}\]
\bigskip

However, the identity wire in equation~\eqref{eq:fg-section} is captured as a single node in the image of two nodes in the boundary of the rewriting rule: one corresponding to the input of the wire and one corresponding to the output. Hence, if we consider equation~\eqref{eq:fg-section} instead as a rewriting rule from the RHS to the LHS, we obtain the following span of hypergraphs
\begin{equation}
\label{eq:fg-section-RL}
\scalebox{0.9}{\input{./figures/fg-section-RL.tikz}}
\end{equation}
Now, there is a matching of the LHS of rule \eqref{eq:fg-section-RL} for each of the nodes in \eqref{eq:fg-section-ex-h}. Unlike before, a single matching corresponds to multiple ways to rewrite the diagram, modulo Frobenius. For the matching $m$ whose image is the node $a_1$, one possible rewrite corresponds to the obvious one, where we expand the wire between $f$ and $g$ to contain another copy of $f$ and $g$
\begin{equation}
\label{eq:fg-section-norm-rw}
\scalebox{0.9}{\input{./figures/fg-section-norm-rw.tikz}}
\end{equation}
But this is not the only rewrite we can perform at this location. We can also apply some Frobenius algebra equations to create another, distinct rewrite here
\begin{equation}
\label{eq:fg-section-alt-rw}
\scalebox{0.9}{\input{./figures/fg-section-alt-rw.tikz}}
\end{equation}

Each of these corresponds to a DPO rewrite with the same matching $m$, but with a different \textit{context}. That is, a different graph $C$ in the DPO diagram~\eqref{eq:dpo2b} is chosen to complete the left pushout square. It is not clear \textit{a priori} how many such contexts exist, or even if the collection of possible contexts is even finite, up to isomorphism.

Thankfully, this question has already been answered for hypergraphs
by~\cite{heumuller2011construction}, which shows that the set of pushout
complements for a given pair of morphisms $K \to L \to G$ is finite whenever
$K, L$ and $G$ are so, and gives a method for enumerating them.
We now will briefly sketch how this method works and apply it to the example above,
referring the interested reader to~\cite{heumuller2011construction} for
the relevant proofs.

We begin by constructing an ``exploded'' context $K + \widetilde G$, which is the disjoint union of $K$ and a new graph $\widetilde G$ defined as follows
\begin{enumerate}
  \item add one vertex to $\widetilde G$ for every vertex $v \in G$ not in the image of $m$,
  \item add one hyperedge to $\widetilde G$ for every hyperedge $h \in G$ not in the image of $m$, and
  \item for each hyperedge $h \in \widetilde G$, let the $i$-th source $\widetilde s_i(h)$ be $s_i(h)$ if $s_i(h) \in \widetilde G$, and otherwise let it be a new, fresh vertex; define the targets similarly.
\end{enumerate}
Then, \cite{heumuller2011construction} showed that any pushout complement $C$ arises as a quotient of this exploded context. That is, we obtain a morphism $n : K \to C$ by composing the embedding of $K$ into $K + \widetilde G$ with the quotient map. If pushing out $n$ along $K \to L$ gives our matching $m : L \to G$, we have a valid pushout complement.

In principle, there are exponentially many possible quotients of this graph to consider. However thanks to the fact that the quotient of $K + \widetilde G$ must yield a valid pushout complement, we typically only need to consider a small fraction of the possible quotients. By construction of the exploded context, we have an induced map $q: K + \widetilde G \to G$ that sends nodes and hyperedges in $\widetilde G$ which came from $G$ to themselves and nodes in $K$ to their image under $K \to L \to G$. It was shown in \cite{heumuller2011construction} that all of the pushout complements arise from $q$ by identifying nodes in the fibres of $q$, i.e. sets of the form $q^{-1}(v)$ for nodes $v \in G$. We can therefore enumerate contexts by looking at all the possible quotients of non-trivial fibres of $q$ and seeing which of them push out to the correct graph. Hence, if $q$ only has a small number of non-trivial fibres, and the fibres themselves are also relatively small, it is practical to enumerate all of the possible contexts.

To see how this works on our example, we first compute the exploded context for the matching $m$ whose image is the vertex $a_1$ in~\eqref{eq:fg-section-ex-h}
\bigskip
\begin{equation}\label{eq:fg-section-exploded}
\input{./figures/fg-section-exploded.tikz}
\end{equation}
\bigskip
The vertices $\{k_0, k_1\}$ come from $K$, and the vertices $\{a_1', a_1''\}$ are new, fresh copies of the vertex $a_1$, which was in the image of $m$.

We can form candidate contexts for $m$ as quotients of the graph above. By the discussion above, we can conclude that the only allowed quotients of \eqref{eq:fg-section-exploded} are the ones that identify nodes in the only non-trivial fibre of $q : K + \widetilde G \to G$, which is $q^{-1}(a_1) = \{a_1', a_1'', k_0, k_1\}$. Of the 15 possible partitions for the set $q^{-1}(a_1)$, five will yield a map $n : K \to C$ that pushes out to give $m: L \to G$, as required. The rest will result in a graph $G'$ with strictly more nodes that $G$. The five good candidates for $C$ are the following
\bigskip
\begin{equation}
\label{eq:fg-section-contexts}
\scalebox{0.9}{$\begin{array}{c}
\input{./figures/fg-section-q1.tikz} \ \input{./figures/fg-section-q2.tikz} \\[1cm]
\input{./figures/fg-section-q3.tikz} \ \input{./figures/fg-section-q4.tikz}
\ \input{./figures/fg-section-q5.tikz}
\end{array}$}
\end{equation}
\bigskip

\noindent
Note that, even though we see the same context graph multiple times, they differ in the embedding $n : K \to C$, as indicated by the labels. The first context yields the ``obvious'' rewrite~\eqref{eq:fg-section-norm-rw}, whereas if we compute the DPOI diagram for the third context, we obtain~\eqref{eq:fg-section-alt-rw}
\[ \scalebox{0.8}{\input{./figures/fg-section-dpo.tikz}} \]

If we perform this rewriting step for each of the five different contexts in~\eqref{eq:fg-section-contexts}, we obtain five different results
\[ \scalebox{0.9}{\input{./figures/fg-section-rewrites1.tikz}} \]
\[ \scalebox{0.9}{\input{./figures/fg-section-rewrites2.tikz}} \]
Thanks to Theorem~\ref{thm:frobeniusrewriting}, we can find all of the distinct ways of rewriting a diagram with a given rule, modulo Frobenius, by enumerating all of the matchings, then enumerating all of the valid contexts.

Finally, note that even though the mapping $K \to L$ in rule~\eqref{eq:fg-section-RL} is not mono, there is only one context in~\eqref{eq:fg-section-contexts} that yields a sound rewrite for matching $m$ in the absence of Frobenius structure. This is a general feature of string diagram rewriting \textit{without} Frobenius structure, which we will revisit in the second paper in this series~\cite{BGKSZ-parttwo}.




\section{Example: Group Algebras}\label{sec:example-group-alg}

In order to showcase the techniques we have introduced, we now study a string diagram rewriting system relevant to the representation theory of finite groups. We introduce boxes $m : 2 \to 1, i : 1 \to 1, u: 0 \to 1$ satisfying the following associativity, inverse, and unit laws
\begin{equation}\label{eq:hopf-aui}
  \begin{array}{ccc}
    \input{./figures/ga-assoc.tikz} & \qquad & \input{./figures/ga-inverse.tikz} \\[8mm]
    \input{./figures/ga-left-unit.tikz} & \qquad &
    \input{./figures/ga-right-unit.tikz}
  \end{array}
\end{equation}
which are furthermore ``natural'' with respect to $(\Bcomult, \Bcounit)$ in the following sense
\begin{equation}\label{eq:hopf-natural}
  \begin{array}{ccc}
    \input{./figures/ga-m-natural.tikz} & \qquad & \input{./figures/ga-m-natural2.tikz} \\[15mm]
    \input{./figures/ga-i-natural.tikz} & \qquad & \input{./figures/ga-i-natural2.tikz} \\[15mm]
    \input{./figures/ga-u-natural.tikz} & \qquad & \input{./figures/ga-u-natural2.tikz}
  \end{array}
\end{equation}

\begin{remark}
  Note that the Frobenius structure is used to allow expressions that refer to inputs in a non-linear fashion. For instance, if one thinks of the input wire of the top-right equation in~\eqref{eq:hopf-aui} as a variable $x$, this roughly corresponds to the term equation $m(x, i(x)) = u$. Notably, the lefthand-side refers to $x$ twice (indicated by ``copying'' the input with \Bcomult) and the rightside-side refers to $x$ zero times (indicated by ``deleting'' the input with \Bcounit).
\end{remark}

Let $(\Vect_k, \otimes)$ be the symmetric monoidal category of finite-dimensional vector spaces and linear maps with the tensor product. The models of the SMT defined above are an important structure in the study of finite groups.


\begin{proposition}
  For $\Sigma = \{ m, i, u \}$ and $\mathcal E$ given by the equations \eqref{eq:hopf-aui} and \eqref{eq:hopf-natural}, the models of $(\Sigma, \mathcal E)$ in $(\Vect_k, \otimes)$ are group algebras. That is, a model consists of vector space $V$ spanned by the elements of a finite group $G$ and (bi)linear maps $m : V \otimes V \to V$, $i : V \to V$, $u : k \to V$ defined on the basis of group elements as follows
\begin{equation}\label{eq:group-algebra}
  m(g \otimes h) = gh \qquad\qquad
  i(g) = g^{-1} \qquad\quad
  u(1) = e
\end{equation}
\end{proposition}

\begin{proof} \textit{(Sketch)}
  This follows from the fact that fixing a (special commutative) Frobenius algebra over a finite-dimensional vector space is equivalent to fixing a basis of vectors ``copied'' by $\delta := \Bcomult$, i.e. satisfying $\delta(v) = v \otimes v$ (see~\cite{CPV}). Then, equations \eqref{eq:hopf-natural} imply that $m, u, i$ send basis elements to basis elements, and the equations \eqref{eq:hopf-aui} imply the usual associativity, unit, and inverse laws of a group. For details, see e.g.~\cite{Kassel-quantumgroups2012} or \cite{CoeckeKissinger_book}.
\end{proof}

Now, applying the techniques introduced in the previous section, we can translate the 10 equations \eqref{eq:hopf-aui} and \eqref{eq:hopf-natural} into combinatoric form. For example, the associativity, inverse, and first naturality equation translate as follows
\begin{equation}\label{eq:hopf-comb}
  \begin{array}{c}
    \input{./figures/ga-assoc-graph.tikz} \\[8mm]
    \input{./figures/ga-inverse-graph.tikz} \\[8mm]
    \input{./figures/ga-m-natural-graph.tikz}
  \end{array}
\end{equation}
The remaining 7 equations are similar.

We can use these rules to obtain a terminating rewriting strategy for normalising expressions over $\Sigma = \{ m, u, i \}$, at least in the case of acyclic hypergraphs. Intuitively, we eliminate as many $u$ and $i$ boxes as possible, associate the $m$ boxes to the right, and push $m, u$ and $i$ boxes as far toward the outputs as possible.

First, we note that the rules \eqref{eq:hopf-natural} can have bad behaviour if they are applied in a completely na\"ive fashion modulo Frobenius. For example, applying the naturality rule for $i$, we can get infinite chains like this
\begin{equation}\label{eq:ga-bad-rw}
  \scalebox{0.8}{\input{./figures/ga-bad-rw.tikz}}
\end{equation}

However, such rewrites are not making any progress in ``pushing $i$ towards the outputs''. We can fix this problem by constraining the rewriting with using a total order than measures progress and only accept rewrites that are decreasing with respect to that order.

Whereas this is difficult to do in a purely syntactic fashion, it is fairly straightforward to define ``progress'' using the combinatoric structure. First, we will introduce a number that measures how far a given hyperedge is from an output.

\begin{definition}\label{def:right-depth}
  For a hyperedge $h$ in an acyclic hypergraph with interface $G \xleftarrow{b} K$, let $p$ be the shortest path connecting $h$ to a node that is not in the source of any hyperedge (which by acyclicity always exists). The \textit{branching depth} of $h$ is the sum of the \textit{branching degree} of every node in $p$, where the branching degree of node $v$ is $\textrm{max}(0, \textrm{deg}(v) + |b^{-1}(v)| - 2)$.
\end{definition}

Intuitively, the branching degree of a node captures how many \Bmult and \Bcomult maps a single node represents, and the branching depth captures a notion of distance from an output (or a \Bcounit), measured in terms of those maps.

The applications of the rules obtained from \eqref{eq:hopf-natural} that ``make progress'' do so by replacing a single hyperedge $h$ with (possibly) multiple hyperedges whose branching depth is strictly smaller than $h$. Conversely, the ``bad'' sequences we want to rule out, such as \eqref{eq:ga-bad-rw}, generate more hyperedges without decreasing the branching depth of the original one.

Let $\mathcal D(G) \in \mathbb N^\star$ be a word where $\mathcal D(G)_k$ is the number of hyperedges with branching depth $k$. By comparing words of the form $\mathcal D(G)$ lexicographically, we obtain a total order on hypergraphs that measures ``progress'' in the sense we are after.

\begin{definition}
  For a totally ordered set $A$, the reverse lexicographic order $\prec_\ell$ is defined recursively on the set $A^\star$ of words as follows: $u \prec_\ell v$ if
  \begin{itemize}
    \item $|u| < |v|$, or
    \item $|u| = |v| = k$ and $u_k < v_k$, or
    \item $|u| = |v| = k$, $u_k = v_k$, and $u_1...u_{k-1} \prec_\ell v_1...v_{k-1}$.
  \end{itemize}
\end{definition}

\noindent \textbf{Reduction Strategy}
\textit{We begin with an acyclic hypergraph with interface $G \leftarrow J$. The graph is normalised with respect to rules obtained from \eqref{eq:hopf-aui} and \eqref{eq:hopf-natural} as follows}
\begin{enumerate}
  \item pick a rule from \eqref{eq:hopf-aui} and apply it if possible;
  \item pick a rule from \eqref{eq:hopf-natural} and apply it if possible, subject to the condition that $\mathcal D(G)$ is strictly reduced with respect to $\prec_\ell$;
  \item repeat until no rule can be applied in steps 1 and 2.
\end{enumerate}

\begin{theorem}
  The \textbf{Reduction Strategy} above terminates.
\end{theorem}

\begin{proof}
  The unit and inverse laws from \eqref{eq:hopf-aui} strictly decrease the number of hypereges in $G$, which will strictly decrease $\mathcal D(G)$. By construction, applying the rules in \eqref{eq:hopf-natural} must strictly decrease $\mathcal D(G)$. The associativity law strictly decreases the number of $m$-labelled hyperedges connected to the left (i.e. `upper') input of another $m$-labelled hyperedge, while leaving $\mathcal D(G)$ fixed. Since all of these quantities are finite and bounded below, the reduction strategy must terminate.
\end{proof}

\if\ismain0 

\bibliographystyle{plain}
\bibliography{catBib3Rev}

\fi 

\if\ismain0 

\setcounter{section}{5} 
\tableofcontents

\fi 

\section{Rewriting modulo multiple Frobenius algebras}\label{sec:multi-frob}

The previous sections have shown that the Frobenius equations are an intrinsic 
part of the structural rules of diagrammatic theories, distinguished from the 
domain-specific equations in a rewriting system. 
However, the hypergraph representation can absorb the structure of a Frobenius 
algebra, but only one per sort. This is in contrast with many applications of
Frobenius algebra rewriting -- e.g. in quantum theory~\cite{Coecke2008,CoeckeKissinger_book}, graphical
linear algebra~\cite{GLA}, concurrency theory~\cite{BHPSZ-popl19}, circuits~\cite{BPSZ-lics19},
and control theory~\cite{Bonchi2014b,BaezErbele-CategoriesInControl} -- which typically deal with multiple, interacting
Frobenius algebras on a single sort.


%
%
In order to deal elegantly with multiple Frobenius algebras on a single sort,
we need an intermediate step, moving from a single-sorted to 
a coloured setting. Concretely, our starting domain will be a PROP
$\catC$, with $n$ Frobenius algebras. From this, we build a 
$n$-coloured PROP $\catDE$, where each colour carries 
one of the $n$ different Frobenius structures. We then make each of the colours
formally isomorphic by introducing pairs of ``colour-switch'' maps (e.g.
$\RBoutput$ and $\BRinput$) and imposing equations making
them inverses to each other.


We shall then prove that $\catC \to \catDE$ is actually an \emph{equivalence} of coloured
PROPs, meaning that $\catDE$ is a faithful representation of the information carried 
by $\catC$. By working in the multi-sorted setting provided by $\catDE$, we can now 
exploit the correspondence established in Proposition~\ref{pr:colfrobeniusrewriting}.
However, we will need a further step: the elimination of redundant 
colour-switch maps after the application of a rule. This is obtained by normalising
graphs with respect to the confluent and terminating rewriting system $\Upsilon$ that
removes pairs of inverse colour-switch maps. Thus our implementation 
establishes a new correspondence stating that ``Rewriting PROPs with $\col$ 
Frobenius algebras''
corresponds to ``DPO rewriting of hypergraphs with $\col$-sorted nodes, 
in $\Upsilon$-normal form.''
%

\subsection{The Polychromatic Interpretation}\label{sec:polychrome}

Throughout this and the next section we fix a PROP
$$\catC \df \syntax{\Sigma} + \frob + \frob$$
 freely generated by a signature $\Sigma$ and two Frobenius algebras (\emph{cf.} Section~\ref{sec:coproduct}), together with a rewriting system $\RS$ on $\catC$. Our goal is to provide a DPOI implementation for $\RS$-rewriting in $\catC$.

\begin{remark}
Even though our exposition deals with rewriting modulo two Frobenius algebras, this is just for simplicity. The theory works for an arbitrary number of Frobenius algebras, via a straightforward generalisation of the developments presented in this section.
\end{remark}

Towards this goal, this section provides the intermediate step of representing $\catC$ 
in terms of a coloured PROP $\catDE$; this setup will make our diagrammatic theory 
adapted to DPOI  rewriting, via Proposition~\ref{pr:colfrobeniusrewriting}. 
$\catDE$ is defined as follows. Consider a signature of ``colour conversion'' 
operations $\Gamma$ (\emph{cf.} Remark~\ref{rem:perm}
and Example~\ref{ex:bipartite}), which we denote graphically as
$$\Gamma = \{ \BRinput \colon \sbl \to \sr, \RBoutput \: \sr\to\sbl \}$$
together with  equations
$$\Upsilon = \{ \RBRio = \Rid , \BRBio = \Bid \}$$
Then $\catD$ is defined as the $\{\sbl,\sr\}$-coloured PROP
\begin{equation}\label{eq:D} 
\catD \df \syntax{\{\sbl,\sr\},\Sigma \uplus \Gamma} 
+_{\scriptscriptstyle \{\sblS,\srS\}} (\CFrob{\sbl}+\CFrob{\sr})
\end{equation}
and $\catDE$ as $\catD$ quotiented by $\Upsilon$. Notice that $\catDE$ is generated by the same signature $\Sigma$ as $\catC$, including operations on sort $\sbl$, but also from the colour conversion operations in $\Gamma$. Whereas the two Frobenius structures in $\catC$ were on the same sort, the two in $\catDE$ are on two different sorts: $\sbl$ and $\sr$. We use $+_{\scriptscriptstyle \{\sblS,\srS\}}$ (\emph{cf.} Example \ref{ex:perm}) to identify the sorts of $\CFrob{\sbl} + \CFrob{\sr}$ with those of 
$\syntax{\{\sbl,\sr\},\Sigma \uplus \Gamma}$.

We now define the ``polychromatic interpretation'' $\chrome{(\cdot)} \: \catC \to \catDE$. Intuitively, $\chrome{(\cdot)}$ will ``shift'' one of the two Frobenius structures of $\catC$ from sort $\sbl$ to sort $\sr$, so that each sort hosts a single Frobenius algebra. Formally, $\chrome{(\cdot)} \: \catC \to \catDE$ is a morphism of coloured props, where $\catC$ is here seen as a $\{\sbl\,\}$-coloured PROP. It suffices to define $\chrome{(\cdot)}$ on the generating objects and arrows of $\catC$. For objects, the single sort $\sbl$ of $\catC$ is mapped to $\sbl$. For arrows, $\chrome{(\cdot)}$ acts as the identity with the exception of the generators of the second Frobenius algebra
\begin{align*}
\BsRmult \ \mapsto \ \RsRmultIO &&
\BsRunit \ \mapsto \ \RsRunitIO \\
\BsRcomult \ \mapsto \ \RsRcomultIO &&
\BsRcounit \ \mapsto \ \RsRcounitIO
\end{align*}
Notice that equations $\Upsilon$ are needed in order for this functor to be well-defined. For instance, they ensure preservation of the separability law for the ``red'' Frobenius algebra
\begin{eqnarray*}
\chrome{\left(\input{./figures/Rsep.tikz}\right)} &=& \input{./figures/IntPresSep1.tikz} \\
&=& \input{./figures/IntPresSep2.tikz} \\
&=& \BRBio = \Bid = \chrome{\left(\Bid\right)}
\end{eqnarray*}

\begin{remark}\label{rmk:quotient}
As $\chrome{(\cdot)}$ has been defined in terms of the generators of $\catC$ and $\catD$, it will sometimes be useful to regard, by abuse of notation, $\chrome{(\cdot)}$ as a mapping from formal string diagrams of the generators of $\catC$ to morphisms in $\catD$. It is worth noting that this mapping would not extend to a well-defined functor from $\catC$ to $\catD$, since the latter is missing the equations of $\Upsilon$.
\end{remark}

It is essential for our developments that the polychromatic interpretation is without loss (or gain) of information. This is guaranteed by the following result.

\begin{proposition}\label{lemma:equivPolyInt} $\chrome{(\cdot)}$ induces an equivalence $\catC \simeq \catDE$ in $\CPROP$. \end{proposition}

\begin{proof}
We have already shown that $\chrome{(\cdot)}$ gives a strict monoidal functor from $\catC$ to $\catDE$. We define another functor $K : \catDE \to \catC$ on objects by letting $K(\sr) = K(\sbl) = \sbl$. Since $\catDE$ is presented by generators and equations, it suffices to say what it does on generators of $\catDE$. It sends the generators $\Sigma$ and the two Frobenius algebras to their monochromatic versions, whereas it sends each of the two colour-changers in $\Gamma$ to $\id_{\sbl}$. One can straightforwardly check that this gives a well-defined, strict monoidal functor and that $K(\chrome{(\cdot)}) = \textit{Id}_{\catC}$. So, it remains only to give a natural isomorphism $\kappa : \textit{Id}_{\catDE} \cong \chrome{(K(\cdot))}$.

For a word $w$ in $\{\sbl, \sr\}$, $\chrome{(K(\cdot))} = \sbl^{|w|}$. So, let $\kappa_w : w \to \sbl^{|w|}$ be the unique monoidal product of $\id_{\sbl}$ and $\RBoutput$ morphisms of the correct type. This is an isomorphism by construction. Naturality then follows from the definition of $\chrome{(\cdot)}$ and the equations $\Upsilon$.
\end{proof}

\begin{remark}\label{rmk:truncation}
 The construction in this section `splits the difference' between the two coproducts 
 discussed in Section~\ref{sec:coproduct}. As noted there, the embedding 
 $U : \PROP \to \CPROP$ does not preserve coproducts. 
 However, we can consider the introduction of the colour changers and equations 
 $\Upsilon$ as a weak truncation operation on coloured PROPs $(\cdot)^\bullet$, 
 which forces all of the colours to be isomorphic to $\sbl$. Then, we do indeed have 
 an equivalence of coloured PROPs 
 $U(\catA + \catA') \simeq (U(\catA) + U(\catA'))^\bullet$. 
 Proposition~\ref{lemma:equivPolyInt} is then the instantiation of this fact for 
 $\catA := \syntax{\Sigma} + \frob$ and $\catA' := \frob$.
\end{remark}

\subsection{Interpreting the Rewriting}\label{sec:interp}

Now that we have represented $\catC$ as a coloured PROP $\catDE$, we can pass to hypergraphs by instantiating Proposition~\ref{th:characterisation-col}.

\begin{corollary}\label{cor:isoSynHypColoured}
There is an isomorphism of $\{\sbl,\sr\}$-coloured PROPs  between 
$\DCospan{F_{\{\sblS,\srS\}} I_{\{\sblS,\srS\}}}
{\Hyp{{\{\sblS,\srS\}},\Sigma\uplus\Gamma}}$
quotiented by $\isosem{\Upsilon}_{\{\sblS,\srS\}}$ and $\catDE$.
\end{corollary}

With respect to rewriting, Corollary \ref{cor:isoSynHypColoured} is still unsatisfactory: 
rewriting in $\catDE$ (and thus in $\catC$) corresponds to DPOI rewriting in 
$\DCospan{F_{\{\sblS,\srS\}} I_{\{\sblS,\srS\}}}
{\Hyp{{\{\sblS,\srS\}},\Sigma\uplus\Gamma}}$
only modulo the equations $\Upsilon$. 
Clearly, it would be computationally obnoxious to reason about rewriting of 
$\isosem{\Upsilon}_{\{\sblS,\srS\}}$-equivalence 
classes of graphs. We now proceed in steps towards a solution to the problem. 
First, henceforth we shall treat the two equations in $\Upsilon$ as rewriting rules 
on $\catD$, with a left-to-right orientation. For simplicity we denote by $\Upsilon$ also the 
resulting rewriting system. We then observe the result below.

\begin{lemma}\label{lemma:Estronglynormalising} $\Upsilon$ is terminating and confluent on $\catD$. \end{lemma}

Our next goal is then to show that rewriting modulo $\Upsilon$ can be simulated without loss of generality by putting a graph in $\Upsilon$-normal form and then applying the rewriting rule. 
However, a naive application of this approach immediately poses problems, as it is shown by the following example.

\begin{example}\label{ex:counterexE} Suppose $\Sigma$ contains an operation $o : 0 \to 1$ and consider a rewriting rule $\alpha$ defined as
  \ctikzfig{col-rw-break}
  Under the interpretation $\chrome{(\cdot)}$ it yields a rule $\chrome{\alpha}$ in $\catD$ defined as
  \ctikzfig{col-rw-break2}
  The translated rule conflicts with $\Upsilon$, meaning that $\Upsilon$ can erase $\chrome{\alpha}$-redexes. For instance
\begin{equation}\label{eq:counterex}
  \input{./figures/col-rw-break3.tikz}
\end{equation}
\end{example}
This kind of problematic example motivates the following transformation on the rewriting rules of $\catD$. As a preparatory step, we record the following lemma, where $\nf{c}$ is the unique $\Upsilon$-normal form of a morphism $c$ of $\catD$, guaranteed by Lemma \ref{lemma:Estronglynormalising}.



\begin{lemma}\label{lem:factRB} Let $l$ be a morphism of $\catD$ in the image of $\chrome{(\cdot)}$. Then, there is a morphism $l'$ not containing any $\Upsilon$-redex such that (in $\catD$)
\begin{equation}\label{eq:lhs}\cgr{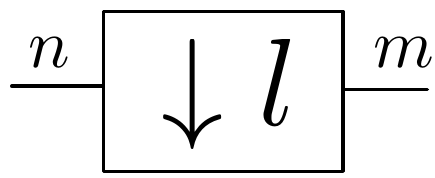} = \cgr[height=1.5cm]{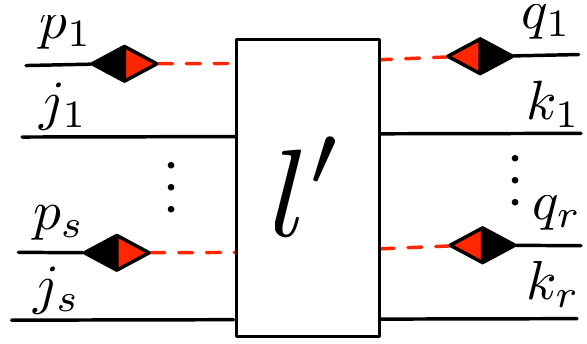}\end{equation}
where $p_1 + j_1 + \dots + p_s + j_s = n$ and $q_1 + k_1 + \dots + q_r + k_r = m$, with $p_i, j_i, q_i,k_i \in \N$. Moreover, there is a unique such $l'$ in $\catD$ for each $l$.
\end{lemma}
\begin{proof} Given $l$, by Lemma \ref{lemma:Estronglynormalising} there is a unique $\nf{l}$ in $\Upsilon$-normal form. Because $l$ is in the image of $\chrome{(\cdot)}$, besides $\BRinput$ and $\RBoutput$ it can only contain $\sbl$-sorted operators, and external dangling wires are also of sort $\sbl$: thus every wire of sort $\sr$ inside $\nf{l}$ can only be connected to the left boundary via a $\BRinput$ and to the right boundary via a $\RBoutput$. Using the laws of SMCs to ``pull out'' all such connecting operators towards the corresponding boundary, we obtain the LHS of \eqref{eq:lhs}.
\end{proof}

We are now ready to introduce the transformation that will remove the conflicts between a rewriting rule and $\Upsilon$.

\begin{definition}\label{def:ruletrans} Let $\alpha$ be
a rewriting rule
of type $\sbl^n \to \sbl^m$ on $\catD$
\[\cgr[height=.6cm]{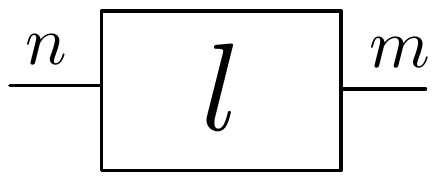} \Rew{\alpha} \cgr[height=.6cm]{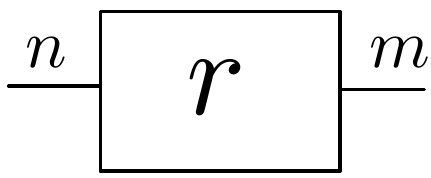}\]
We obtain $l'$ through Lemma \ref{lem:factRB}
\[\cgr[height=.6cm]{lhs.pdf} = \cgr[height=1.5cm]{lhsprimeIODots.pdf}\]
The rule $\itf{\alpha}$ is defined as
\[\cgr[height=1.5cm]{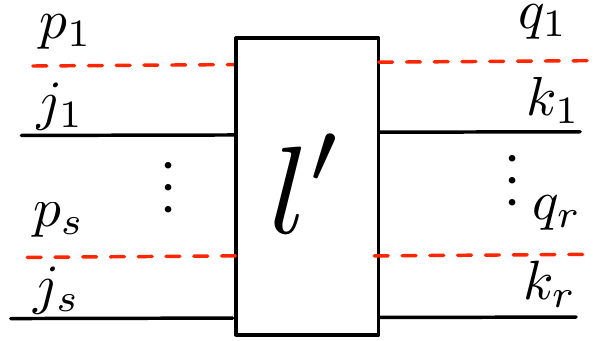} \Rew{} \cgr[height=1.5cm]{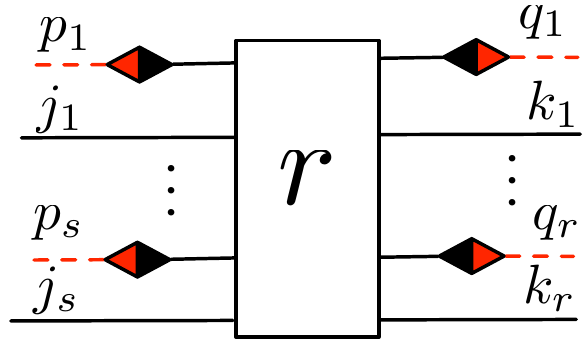} \]
Given a rewriting system $\RS$, we write $\itf{\RS} = \{\itf{\alpha} \mid \alpha \in \RS\}$.
\end{definition}

It is instructive to show how this transformation neutralises the problem of~\eqref{eq:counterex}.
\begin{example}
The rule $\chrome{\alpha}$ from Example \ref{ex:counterexE} is transformed into $\itf{(\chrome{\alpha})}$, defined as $\cgr[height=.45cm]{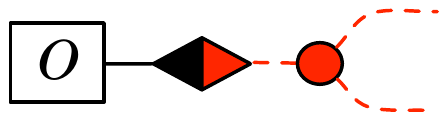} \Rew{} \cgr[height=.5cm]{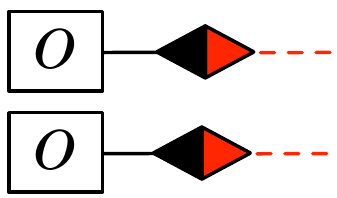}$
 Observe that coexistence with $\Upsilon$ is not problematic anymore, as $\Upsilon$ cannot erase $\itf{(\chrome{\alpha})}$-redexes. 
 
 For instance, in computation \eqref{eq:counterex} we are not stuck anymore
\begin{equation*}
  \input{./figures/col-rw-break4.tikz}
\end{equation*}
 \end{example}

%

 We claim that this ``improved'' rewriting system removes conflicts with $\Upsilon$. Or, to put it another way: a single syntactic rewriting step is possible if and only if it can be obtained as a combinatoric step on an $\Upsilon$-normal form, followed by renormalising using $\Upsilon$. Symbolically, we can write this as follows
\[
a \Rightarrow_{\RS} b  \quad \text{ iff } \quad \nf{\isosem{\chrome a}_{\{\sblS,\srS\}}}\rigidDPOstep{\isosem{\TRSC}_{\{\sblS,\srS\}}} \Trans{\rigidDPOstep{\isosem{\Upsilon}_{\{\sblS,\srS\}}}} \ \nf{\isosem{\chrome b}_{\{\sblS,\srS\}}}
\]

Proving the formal statement above will be the main task for the remainder of this section, and it appears as Theorem~\ref{th:mainth}.

Contrary to the situation depicted at the beginning of the section, in 
Theorem~\ref{th:mainth} DPOI  rewriting in the combinatorial domain 
is ``on-the-nose'': instead of dealing with 
$\isosem{\Upsilon}_{\{\sblS,\srS\}}$-equivalence classes of hypergraphs, 
we can now deal exclusively with 
$\isosem{\Upsilon}_{\{\sblS,\srS\}}$-normal forms, which 
thanks to Lemma~\ref{lemma:Estronglynormalising} are straightforward to compute.

The proof of Theorem~\ref{th:mainth} will go in steps. The theory so far ensures a correspondence between
 \begin{itemize}
 \item rewriting in $\catC$ and rewriting in $\catDE$, thanks to 
 Proposition~\ref{lemma:equivPolyInt};
 \item rewriting in $\catD$ and rewriting in 
 $\DCospan{F_{\{\sblS,\srS\}} I_{\{\sblS,\srS\}}}
{\Hyp{{\{\sblS,\srS\}},\Sigma\uplus\Gamma}}$,
 thanks to 
 Proposition~\ref{pr:colfrobeniusrewriting}.
 \end{itemize}
Thus the only missing link to complete the correspondence in Theorem \ref{th:mainth} is to adequately represent rewriting in $\catDE$ as rewriting in $\catD$. This is the remit of the next section.

\subsection{Adequacy of the Implementation}\label{sec:adequacy}

For the purposes of this section, let $\RS$ be a rewriting system on $\catD$. We focus on the missing piece of the proof of Theorem~\ref{th:mainth}: showing that the rule transformation of Definition \ref{def:ruletrans} is an adequate implementation for $\RS$-rewriting modulo $\Upsilon$ in $\catD$.

\begin{theorem}[Adequacy]\label{prop:adequacy} Let $c$ and $d$ be arrows in $\catD$. Then
\[ c \RLTrans{\RLRew{\Upsilon}}  \Rew{\RS}  \RLTrans{\RLRew{\Upsilon}} d \ \text{ iff } \nf{c} \ \Rew{\TRS} \Trans{\Rew{\Upsilon}} \ \nf{d}. \]
\end{theorem}

The proof 
will follow from Propositions~\ref{prop:completeness} and~\ref{prop:soundness}. For the right-to-left direction (completeness), we can prove a stronger statement.

\begin{proposition}\label{prop:completeness} $c \Rew{\TRS} d $ implies $c \RLTrans{\RLRew{\Upsilon}}  \Rew{\RS} d$.
\end{proposition}

\begin{proof} For the sake of readability, all the diagrams in the proofs of this section are depicted with unlabelled wires: it is intended that each wire stands for a number of parallel wires of the same type, arbitrary but compatible with its position in the diagram.

We know $c$ has a redex for $\itf{\alpha}$ for some rule $\alpha \in \RS$. If $\alpha$ is given by $l \Rew{} r$, then $\itf{\alpha}$ rewrites $c$ as
\begin{eqnarray}\label{eq:complenetessStep1}
\cgr{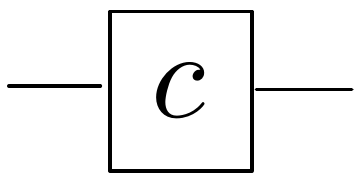} &=& \cgr[height=1.3cm]{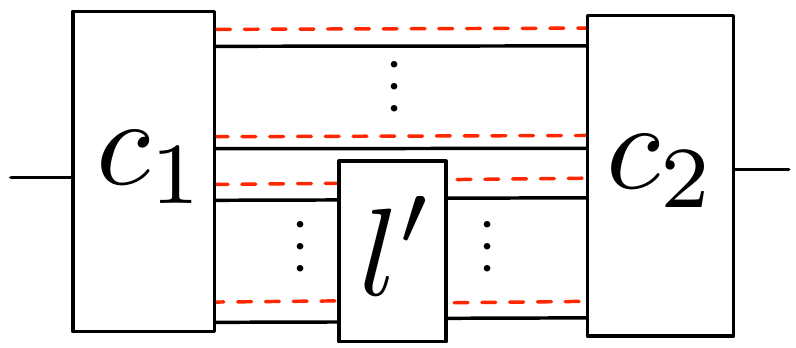} \\
&\Rew{\itf{\alpha}}& \cgr[height=1.3cm]{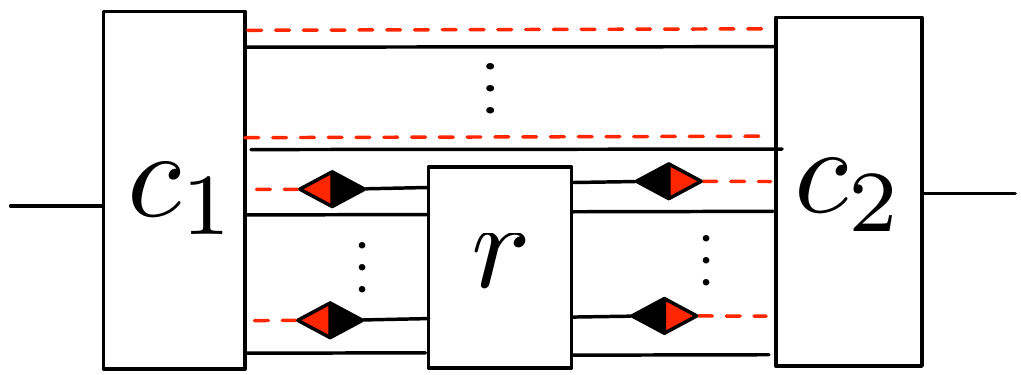} 
\end{eqnarray}
In light of \eqref{eq:complenetessStep1}, $c$ modulo $\Upsilon$ contains a redex for $\alpha$ as well
\begin{eqnarray*}
\cgr[height=1.3cm]{c1lprimec2Dots.pdf} & \RLTrans{\RLRew{\Upsilon}} &  \cgr[height=1.3cm]{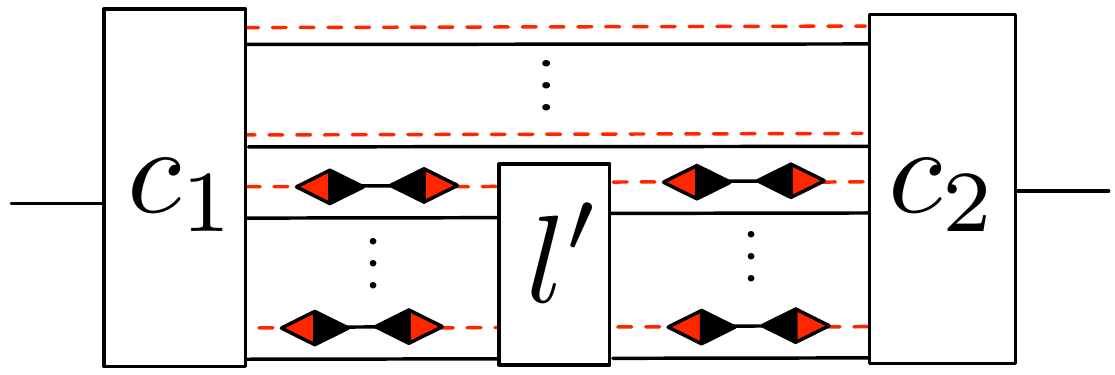} \\
 & \RLTrans{\RLRew{\Upsilon}} & \cgr[height=1.3cm]{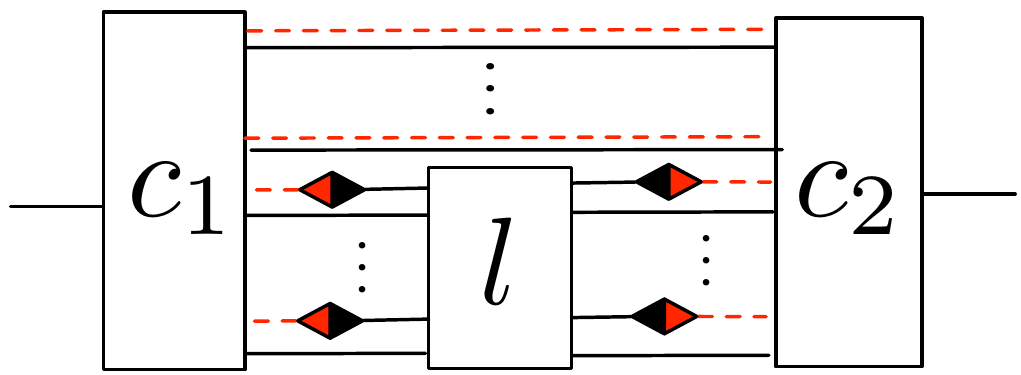} \\
 & \Rew{\alpha} &   \cgr[height=1.3cm]{c1oric2Dots.pdf}
\end{eqnarray*}
where the second step is justified by the definition of $l'$ as in \eqref{eq:lhs}. This proves the statement.
\end{proof}

The left-to-right direction (soundness) of Thorem~\ref{prop:adequacy} requires more work. First, we have that $\TRS$ is as powerful as $\RS$, modulo $\Upsilon$.

\begin{lemma}\label{lem:equiv} $c \RLTrans{\RLRew{\Upsilon}}  \Rew{\RS}  \RLTrans{ \RLRew{\Upsilon}} d$ iff $c \RLTrans{\RLRew{\Upsilon}} \Rew{\TRS} \RLTrans{\RLRew{\Upsilon}} d$.
\end{lemma}

\begin{proof}
It suffices to show that $c \Rew{\RS} d$ implies $c \RLTrans{\RLRew{\Upsilon}} \Rew{\TRS} \RLTrans{\RLRew{\Upsilon}} d$ and $c \Rew{\TRS} d$ implies $c \RLTrans{\RLRew{\Upsilon}}  \Rew{\RS}  \RLTrans{ \RLRew{\Upsilon}} d$.
For the left-to-right implication, the assumption is that $c$ contains a redex for a rule $\alpha \in \RS$, say of the form $l \Rew{} r$
\begin{eqnarray}
\cgr{cDots.pdf} &=& \cgr[height=1.3cm]{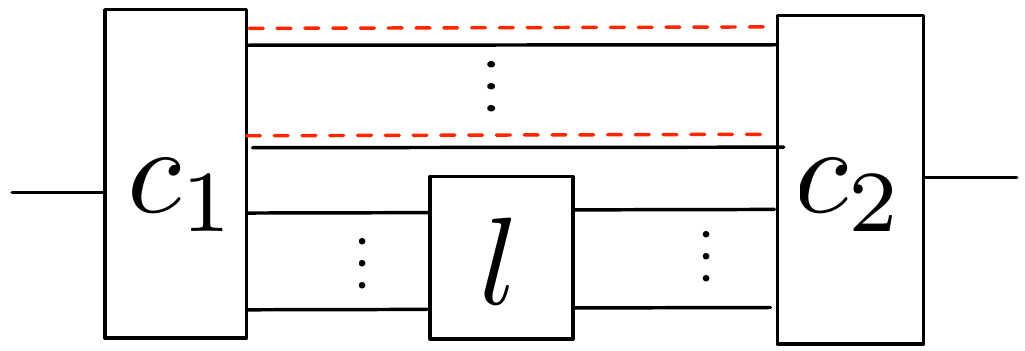} \nonumber \\
&\Rew{\alpha}& \cgr[height=1.3cm]{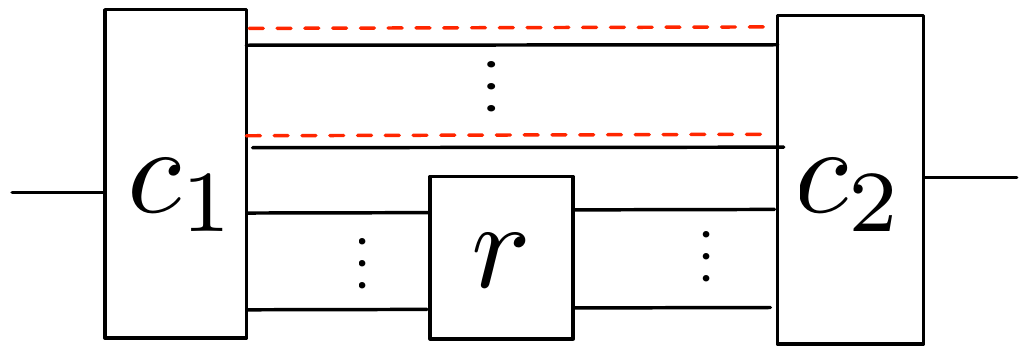} \label{eq:soundnessEquivalence}
\end{eqnarray}
Then, modulo $\Upsilon$, $c$ also contains a redex for $\itf{(l \Rew{} r)}$. Applying this rule yields the same outcome, modulo-$\Upsilon$, as~\eqref{eq:soundnessEquivalence}
\begin{eqnarray*}
\cgr[height=1.3cm]{c1lc2Dots.pdf} &\RLTrans{\RLRew{\Upsilon}}& \cgr[height=1.3cm]{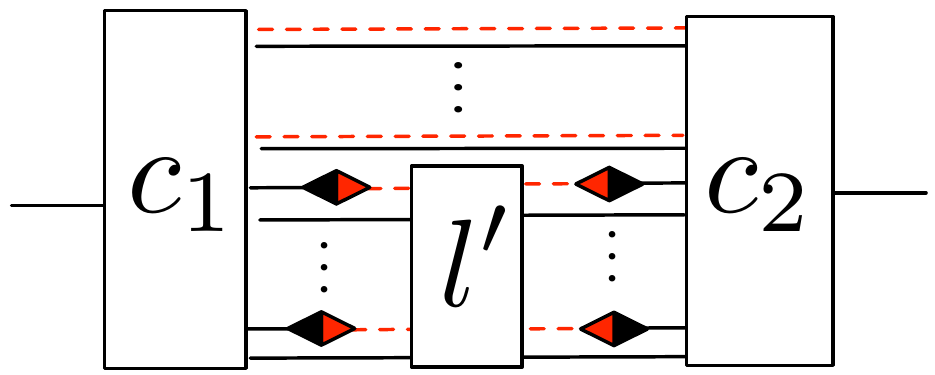}\\
 &\Rew{\itf{\alpha}}& \cgr[height=1.3cm]{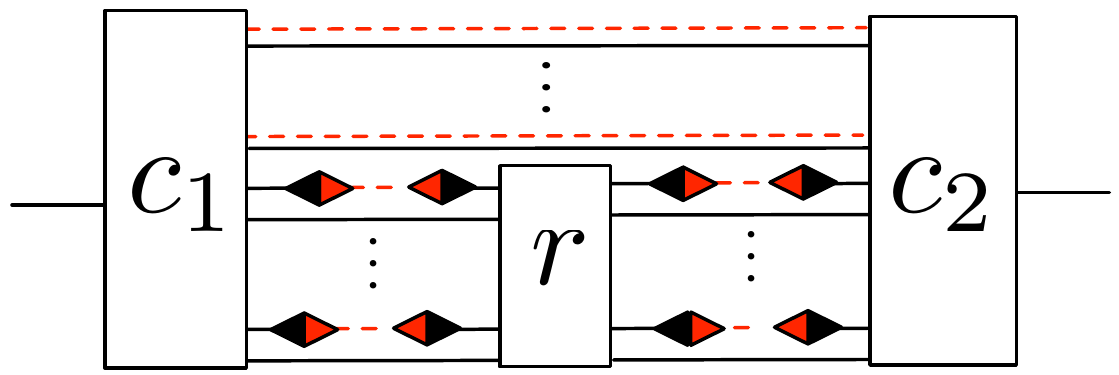} \\
  &\RLTrans{\RLRew{\Upsilon}}& \cgr[height=1.3cm]{c1rc2Dots.pdf}
\end{eqnarray*}
This proves the left-to-right implication of the statement. The right-to-left implication is given by Proposition~\ref{prop:completeness}.
\end{proof}

The next step is to show that $\Rew{\TRS}$ satisfies a ``diamond property'' with respect to $\Upsilon$. This property implies that $\Upsilon$-rewriting does not interfere with $\TRS$-rewriting--- whence the latter can be assumed without loss of generality to work on arrows in $\Upsilon$-normal form, as in the desired implementation (Theorem \ref{prop:adequacy}). As shown in Example \ref{ex:counterexE}, the diamond property fails for arbitrary rewriting systems and justifies the introduction of the transformation $\itf{(\cdot)}$.

\begin{lemma}[Diamond Property] \label{lem:diamond} If $c \Rew{\TRS} d$ and $c \Rew{\Upsilon} e$ then there exists an $f$ such that $d \Rew{\Upsilon} f$ and $e \Rew{\TRS} f$
\begin{equation*}
\xymatrix@C=25pt@R=5pt{
& \ar@{=>}[dl]_>>>>{\TRS} c \ar@{=>}[dr]^>>>>{\Upsilon} & \\
d \ar@{=>}[dr]_>>>>{\Upsilon} & & e \ar@{=>}[dl]^>>>>{\TRS} \\
& f &
}
\end{equation*}
\end{lemma}
\begin{proof}
This is immediate from the fact that, by Definition \ref{def:ruletrans}, the application of $\TRS$ cannot introduce $\Upsilon$-redexes. Therefore, $\TRS$ and $\Upsilon$ are orthogonal rewriting systems (i.e. they have no critical pairs between each other).
\end{proof}

We are now ready to show soundness.

\begin{proposition}\label{prop:soundness}
$
c \RLTrans{\RLRew{\Upsilon}}  \Rew{\RS}  \RLTrans{\RLRew{\Upsilon}} d
\text{ implies }
\nf{c} \Rew{\TRS} \Trans{\Rew{\Upsilon}} {\nf{d}}.
$
\end{proposition}
\begin{proof}
Since $\Upsilon$ is confluent and terminating (Lemma \ref{lemma:Estronglynormalising}), the conclusion is equivalent to $\nf{c} \Rew{\TRS} \RLTrans{\RLRew{\Upsilon}} d$, so we focus on this statement.

Assume $c \RLTrans{\RLRew{\Upsilon}}  \Rew{\RS}  \RLTrans{\RLRew{\Upsilon}} d$. By Lemma \ref{lem:equiv}, this implies ${c \RLTrans{\RLRew{\Upsilon}}  \Rew{\TRS}  \RLTrans{\RLRew{\Upsilon}} d}$. Since $\Upsilon$ is confluent and terminating, this implies ${c \Trans{\Rew{\Upsilon}} \nf{c} \LTrans{\LRew{\Upsilon}} \Rew{\TRS}  \RLTrans{\RLRew{\Upsilon}} d}$.

We can now drop the first part of the rewrite sequence and focus on $\nf{c} \LTrans{\LRew{\Upsilon}} \Rew{\TRS}  \RLTrans{\RLRew{\Upsilon}} d$. By repeatedly applying Lemma \ref{lem:diamond}, we can commute $\Rew{\TRS}$ through $\LTrans{\LRew{\Upsilon}}$ to obtain $\nf{c} \Rew{\TRS} \ \LTrans{\LRew{\Upsilon}} \RLTrans{\RLRew{\Upsilon}} d$. Finally, merging $\LTrans{\LRew{\Upsilon}}$ and $\RLTrans{\RLRew{\Upsilon}}$ yields  $\nf{c} \Rew{\TRS} \RLTrans{\RLRew{\Upsilon}} d$ as required.
\end{proof}


We now have all the ingredients to prove the main theorem of this section: the DPOI  rewriting implementation of rewriting in $\catC$.

\begin{theorem}\label{th:mainth} Let $\RS$ be a rewriting system on $\catC$. Then
\[
a \Rightarrow_{\RS} b  \quad \text{ iff } \quad \nf{\isosem{\chrome a}_{\{\sblS,\srS\}}}\rigidDPOstep{\isosem{\TRSC}_{\{\sblS,\srS\}}} \Trans{\rigidDPOstep{\isosem{\Upsilon}_{\{\sblS,\srS\}}}} \ \nf{\isosem{\chrome b}_{\{\sblS,\srS\}}}
\]
\end{theorem}


\begin{proof} 
First, we have a correspondence at the level of syntactic rewriting (Definition \ref{defn:rewprop}) in the props $\catC$ and $\catDE$
\begin{equation}\label{eq:mainth1}
\boxed{
\begin{aligned} a \Rightarrow_{\RS} b  \\ \text{in } \catC \quad \end{aligned}
}
\quad \text{ iff }  \quad
\boxed{
\begin{aligned} \chrome{a} \Rightarrow_{\chrome{\RS}} \chrome{b}  \\ \text{in } \catDE \quad \end{aligned}
}
\end{equation}
This is ensured by the fact that $\chrome{(\cdot)}$ is a functorial and 
full and faithful mapping. Second, we interpret $\Upsilon$ as a set of 
rewriting rules instead of a set of equations. Then, rewriting in $\catDE$ 
is just the same as rewriting in $\catD$ modulo $\Upsilon$-rewriting. 
Starting from the right-hand side of \eqref{eq:mainth1}
\begin{equation}\label{eq:mainth2}
\boxed{
\begin{aligned} \chrome{a} \Rightarrow_{\chrome{\RS}} \chrome{b}  \\ \text{in } \catDE \quad \end{aligned}
}
\quad \text{ iff }  \quad
\boxed{
\begin{aligned}  \chrome{a} \RLTrans{\RLRew{\Upsilon}} \Rightarrow_{\chrome{\RS}}\RLTrans{\RLRew{\Upsilon}} \chrome{b}  \\ \text{ in } \catD \qquad \ \end{aligned}
}
\end{equation}
where $\chrome{a}$ and $\chrome{b}$ are understood on the right as arrows of $\catD$, \emph{cf.} Remark \ref{rmk:quotient}. Third, we use Theorem \ref{prop:adequacy} to give an implementation for rewriting modulo-$\Upsilon$. Starting from the right-hand side of \eqref{eq:mainth2}
 \begin{equation}\label{eq:mainth3}
 \boxed{
\begin{aligned}  \chrome{a} \RLTrans{\RLRew{\Upsilon}} \Rightarrow_{\chrome{\RS}}\RLTrans{\RLRew{\Upsilon}} \chrome{b}  \\ \text{ in } \catD \qquad \ \end{aligned}
}
\quad \text{ iff }  \quad
\boxed{
\begin{aligned}  \nf{\chrome{a}} \Rew{\TRSC} \Trans{\Rew{\Upsilon}} \nf{\chrome{b}}  \\ \text{ in } \catD \qquad \end{aligned}
}
\end{equation}
 Last, Corollary~\ref{cor:isoSynHypColoured} and 
 Proposition~\ref{pr:colfrobeniusrewriting}
 yield the correspondence between rewriting in $\catD$ and DPO-rewriting in 
 $\DCospan{F_{\{\sblS,\srS\}} I_{\{\sblS,\srS\}}}
{\Hyp{{\{\sblS,\srS\}},\Sigma\uplus\Gamma}}$. 
Starting from the right-hand side of \eqref{eq:mainth3}
\begin{equation}\label{eq:mainth4}
\begin{aligned}
\boxed{
\begin{aligned}  \nf{\chrome{a}} \Rew{\TRSC} \Trans{\Rew{\Upsilon}} \nf{\chrome{b}}  \\ \text{ in } \catD \qquad \end{aligned}
}
 \quad \text{ iff }  \quad
\boxed{
\begin{aligned} \isosem{ \nf{\chrome{a}} }_{\{\sblS,\srS\}} 
\rigidDPOstep{\isosem{\TRSC}_{\{\sblS,\srS\}}} 
\Trans{\rigidDPOstep{\isosem{\Upsilon}_{\{\sblS,\srS\}}}} 
\ \isosem{ \nf{\chrome{b} } }_{\{\sblS,\srS\}}  \\ \text{ in } 
\DCospan{F_{\{\sblS,\srS\}} I_{\{\sblS,\srS\}}}
{\Hyp{{\{\sblS,\srS\}},\Sigma\uplus\Gamma}}
 \qquad \end{aligned}
}
 \end{aligned}
\end{equation}
Note that 
$\isosem{ \nf{\chrome{a}}}_{\{\sblS,\srS\}} = \nf{\isosem{ \chrome{a}}_{\{\sblS,\srS\}}}$,
 where the normal form on the right is computed in the category
$\DCospan{D_{\{\sblS,\srS\}}}
{\Hyp{{\{\sblS,\srS\}},\Sigma\uplus\Gamma}}$ according to the rules 
$\isosem{\Upsilon}_{\{\sblS,\srS\}}$. 
To conclude, by chaining \eqref{eq:mainth1} to \eqref{eq:mainth4} 
we obtain the statement of the theorem.
\end{proof}

This theorem gives us an effective combinatorial method for rewriting modulo multiple Frobenius algebras.

\section{Example: Interacting Bialgebras}\label{sec:example}

We now turn to one of the main examples of multiple interacting Frobenius algebras: the case of two Frobenius algebras that together interact as a bialgebra. In some sense, this specialises the example from Section~\ref{sec:example-group-alg} that had a single Frobenius algebra $\sbl$ interacting with a group $(m : 2 \to 1, u: 0 \to 1, i: 1 \to 1)$. Here, we assume that $m$ and $u$ are themselves part of a second Frobenius algebra. In this case, the associativity and unit equations come for free, so it remains to state analogous equations to\eqref{eq:hopf-natural}, which make the two Frobenius structures ``natural'' with respect to each other
\begin{equation}\label{eq:bialg-rules}
\input{./figures/ib1.tikz}
\end{equation}
\bigskip
\begin{equation}\label{eq:bialg-rules1}
\input{./figures/ib2.tikz}
\end{equation}
For simplicity, we also require that the induced `cup' and `cap' maps coincide
\begin{equation}\label{eq:capcup}
  \input{./figures/capcup.tikz}
\end{equation}
which will entail the ``inverse'' equation from~\eqref{eq:hopf-aui} for $i = \textit{id}_1$ (see rule $(h)$ in the derived rules \eqref{eq:derived-rules} below). Hence, it is a strict specialisation of the group algebra structure introduced in Section~\ref{sec:example-group-alg}.

This system, referred to as $\mathbb{IB}$~\cite{BialgAreFrob14} (`Interacting Bialgebras'), has appeared ubiquitously in the study of component-based systems across different research areas. It forms the core of the \textit{ZX-calculus}~\cite{Coecke2008}, which has recently been extended to give sound and complete equational theories for approximately~\cite{LoriaCompleteness} and fully~\cite{OxfordCompleteness} universal families of quantum circuits. Also, it has been employed to reason about signal processing circuits in control theory~\cite{Bonchi2014b,BaezErbele-CategoriesInControl}, electrical circuits~\cite{BPSZ-lics19}, and Petri nets~\cite{BHPSZ-popl19}.

\begin{remark}
  Recall that the models in $(\Vect_k, \otimes)$ of the structure from Section~\ref{sec:example-group-alg} correspond exactly to representations of finite groups (i.e. group algebras). The models of a pair of Frobenius algebras $\sbl, \sr$ satisfying equations~\eqref{eq:bialg-rules} and \eqref{eq:bialg-rules1} specialise this fact: they correspond exactly to the representations of finite \textit{Abelian} groups. If we additionally impose~\eqref{eq:capcup}, these are representations of finite Abelian groups whose elements are all self-inverse. See \cite{CoeckeKissinger_book}, Section 9.6.1 for details.
\end{remark}

\subsection{A Representation Theorem}
This section will propose a rewriting strategy that exploits the DPO implementation  presented in the previous sections and allows for turning any diagram of such an 
"interacting bialgebra" into a suitable normal form.

The first step is to note that from the rules above one can derive (see e.g.~\cite{Coecke2008}) the following two rules, which will soon be useful
\begin{equation}\label{eq:derived-rules}
  \input{./figures/derived-rules.tikz}
\end{equation}


A generic diagram composed of generators from these two Frobenius algebras consists of arbitrarily many alternating layers of $\sbl$ and $\sr$ generators
\[ \cgr{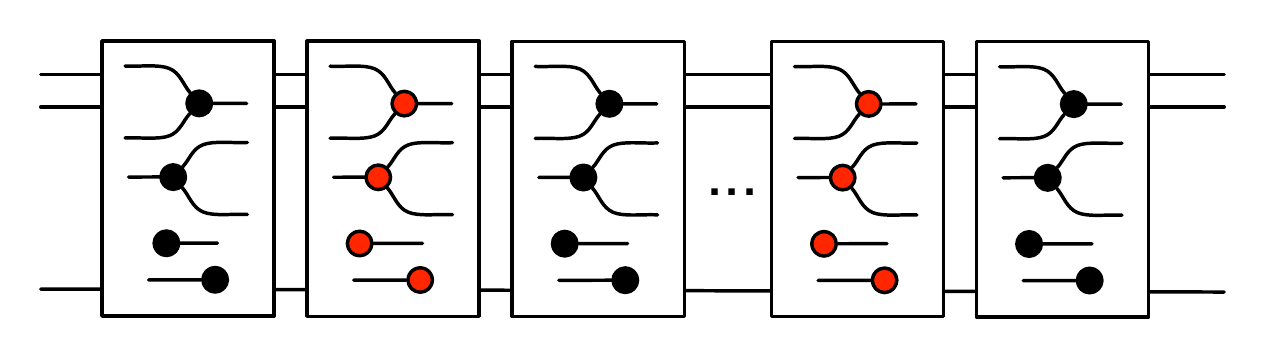} \]

Any such diagram can be rewritten into a \textit{$\sbl$-reduced form} that consists of just four layers: an initial layer of $\sbl$-comonoid structure, followed by
$\sr$-monoid structure, followed by $\sr$-comonoid structure, followed by a final layer of $\sbl$-monoid structure
\begin{equation}\label{eq:reduced-form}
  \cgr{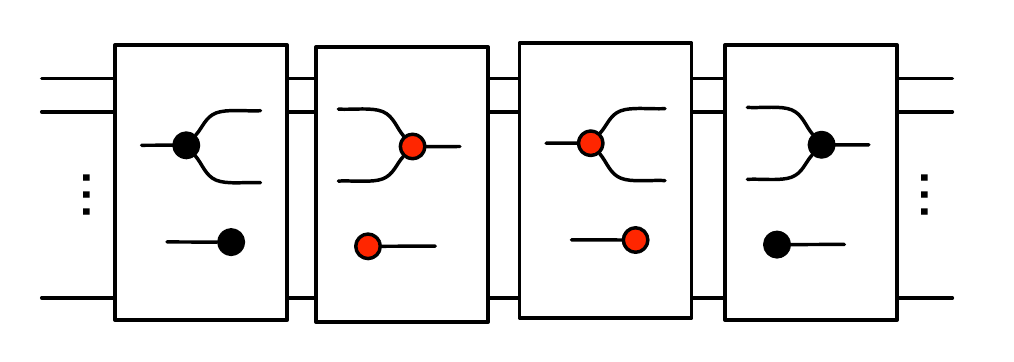}
\end{equation}
We call this the \textit{$\sbl$-reduced form} because there is no internal layer of $\sbl$ generators. We now characterise these forms in terms of their associated hypergraphs with interfaces. To express hypergraphs with interfaces compactly and unambiguously, we adopt the following notational conventions
\begin{enumerate}
  \item As we did in Example~\ref{ex:bipartite2}, hyperedges corresponding to $\BRinput$ and $\RBoutput$ are depicted as unlabelled, directed edges between nodes of appropriate colour, hence
  \[ \input{./figures/BRgraph.tikz} = \isosem{\BRinput} \quad
     \input{./figures/RBgraph.tikz} = \isosem{\RBoutput} \]
  \item 
  To avoid writing interfaces explicitly, we will indicate these by consistently placing labels above nodes for inputs and below nodes for outputs.
  For example, the following hypergraph with interfaces is abbreviated as
  \[ \input{./figures/hypergraph-abbrev.tikz} \]
\end{enumerate}
Using these conventions, the rules in the system $\Upsilon$ can be written as the following two DPOI rules
\[ \input{./figures/multicol-rules.tikz} \]
Hence, normalising with respect to $\Upsilon$ contracts away any node with precisely one in-edge and one out-edge. We are now ready to characterise $\sbl$-reduced forms.

\begin{proposition}\label{prop:reduced-form}
  A string diagram generated by two Frobenius algebras $\sbl$ and $\sr$ is in $\sbl$-reduced form as in \eqref{eq:reduced-form} (modulo Frobenius equations) if and only if its associated hypergraph with interfaces $I_1 \rightarrow G \leftarrow I_2$ satisfies the following conditions. $G$ is directed acyclic and every $\sbl$ node in $G$ is either
  \begin{enumerate}
    \item[(I)] in the image of a single node in $I_1$ and has no in-edges,
    \item[(O)] in the image of a single node in $I_2$ and has no out-edges, or
    \item[(IO)] in the image of one or more nodes in \textbf{both} $I_1$ and $I_2$.
  \end{enumerate}
\end{proposition}

\begin{proof}
  When hypergraph nodes representing Frobenius algebra generators are composed, they fuse together. Hence, the nodes in $G$ correspond to maximal connected components of Frobenius algebra generators of the same colour.

  First, suppose a string diagram is in the form of \eqref{eq:reduced-form}. Then, each $\sbl$ node in the hypergraph $G$ corresponds to a maximal connected component of $\sbl$ Frobenius generators in \eqref{eq:reduced-form}. We first note that any (co)unit connected to a (co)multiplication can be reduced away. Hence, we need to consider only 5 cases for connected components of $\sbl$ generators: (1) a counit applied to an input wire, (2) a unit applied to an output wire, (3) a tree of comultiplications applied to an input wire, (4) a tree of multiplications connected to an output wire, or (5) a connected component of cases (3) and (4). Cases (1) and (3) yield a $\sbl$ node of type (I). Cases (2) and (4) yield a node of type (O), and case (5) yields a node of type (IO).

  Conversely, we can interpret each of the $\sbl$ nodes of types (I) and (O) as cases (1)-(4) described above. The only difficult case is nodes of type (IO). These can be interpreted as a `zig-zag' of $\sbl$ comultiplications in the first layer of \eqref{eq:reduced-form} and multiplications in the last layer, with no $\sr$ generators in between
  \[ \input{./figures/zigzag-boundary.tikz} \qquad \ \raisebox{-1mm}{$\mapsto$} \qquad \  \cgr{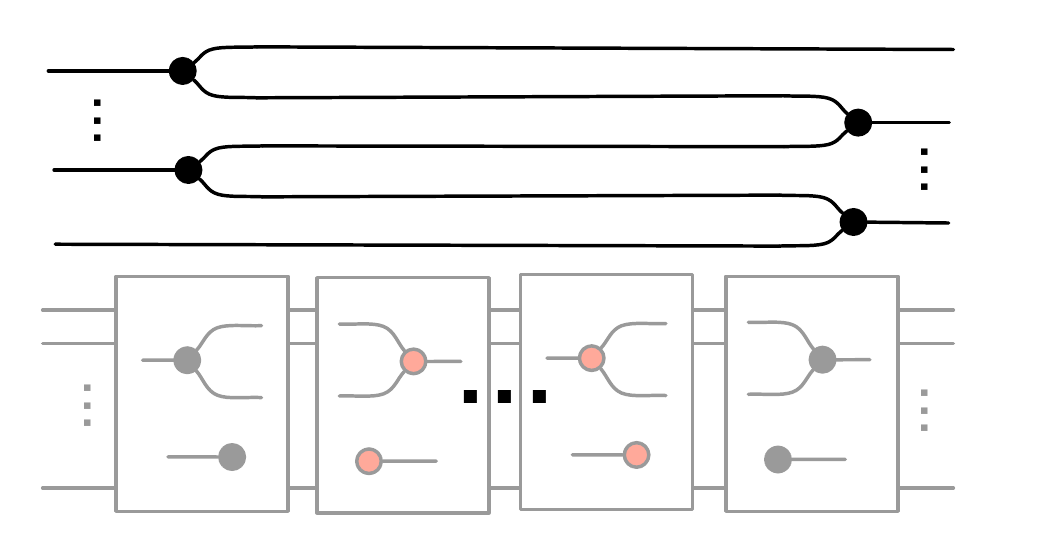} \]
\end{proof}

Crucially, a hypergraph with interfaces that satisfies the conditions above contains no \textit{interior $\sbl$ nodes}, i.e. nodes not in the image of $I_1$ or $I_2$. Eliminating these nodes will form the main component of the strategy below. In order to obtain a hypergraph with interfaces satisfying these conditions, we first perform the transformation of the interacting bialgebra rules into a DPOI rewriting system. This is a  mechanical procedure, but for clarity, we will show it explicitly for the rule $(b)$. Following the recipe of Theorem \ref{th:mainth}, we first use $\chrome{(\cdot)}$ to get the polychromatic interpretation ---\eqref{eq:bialg-transformed2} below--- then apply $\itf{(\cdot)}$ to shift the colour change maps on inputs/outputs to the right-hand side ---\eqref{eq:bialg-transformed}--- and finally apply $\isosem{\cdot}_{\{\sblS,\srS\}}$ to interpret 
\begin{align}
 &\input{./figures/bialgProcess1.tikz}& \label{eq:bialg-transformed2} \\[5mm]
 & \input{./figures/bialgProcess2.tikz} \label{eq:bialg-transformed}& \\[5mm]
 & \input{./figures/bialg-graph.tikz}\label{eq:bialg-transformed1}&
\end{align}

We can give a similar treatment to $(cp1)$, $(cp2)$, and $(u)$ rules. Equivalently, we can introduce a family of rules $K_{m,n}$ for $m, n \geq 0$
\begin{equation}\label{eq:bialg-graph-Kmn}
 \input{./figures/bialg-graph-Kmn.tikz}
\end{equation}
where the righthand-side contains the fully connected bipartite graph from $m$ red nodes to $n$ black nodes. Note when $m=0$ (resp. $n=0$), we interpret the range $0...m-1$ (resp. $0...n-1$) as empty.  This family of rules is implied by the Frobenius equations and $(b)$, $(cp1)$, $(cp2)$, and $(u)$. Conversely, it implies these 4 rules as special cases (see e.g.~\cite{CoeckeKissinger_book}).


We have mentioned the derived equations in box~\eqref{eq:derived-rules} because, once we translate them into DPOI rules, we see that rule $(d)$ allows to reverse the direction of an arbitrary edge, 
rule $(h)$ to delete parallel edges, and rules $(u1)$ and $(u2)$ to delete single, isolated nodes
\[ \input{./figures/derived-graphs.tikz} \]

Note the rule $(D)$ (and its converse) allow us to essentially work with undirected graphs, as we can always reverse an edge directions if necessary to create a match. This renders the additional ``primed'' equations in box \eqref{eq:bialg-rules1} redundant, since they are the same as the rules above, but with the directions reversed.

Hence, the only other rule we need is a rule for introducing red caps
\ctikzfig{bialg-graph-cap}




\smallskip

\noindent \textbf{Reduction Strategy}
%
%
\textit{We begin with a hypergraph with interfaces $I_1 \rightarrow G \leftarrow I_2$, whose interfaces $I_1, I_2$ are all of the $\sbl$-sort. It should be understood that after every rewriting step, the graph is normalised with respect to rules $(\Upsilon{}1)$ and $(\Upsilon{}2)$. The strategy proceeds as follows}

\begin{enumerate}
  \item Reduce as much as possible using rules $(U1)$, $(U2)$, and $(H)$, using the rule $(D)$ to  reverse edge directions as necessary.
  \item If there are no interior $\sbl$ nodes, go to step 5. Otherwise, apply the rule $K_{mn}$ to an interior $\sbl$ node $v$ and one neighbouring $\sr$ $w$ node to remove it as follows:
    \ctikzfig{remove-interior-bl-simp}
    where we again use the $(D)$ rule to reverse edge directions as necessary.
  \item If there are remaining interior $\sbl$ nodes, go to step 1.
  \item If a $\sbl$ node is in the image of multiple nodes in $I_1$ and of no nodes in $I_2$, apply the converse of rule $(CA)$ to split it into multiple $\sbl$ nodes connected by $\sr$ nodes. For example
  \[ \input{./figures/split-inputs.tikz} \]
  We split nodes only in the image of $I_2$ similarly.
  \item Apply $(D)$ or its converse to direct the remaining edges from the image of $I_1$, to the $\sr$ nodes, then to the image of $I_2$.
\end{enumerate}

The ``essential trick'' in this strategy is step (2), which removes pairs of adjacent nodes $(v, w)$ at the expense of introducing some additional edges. Since we are always removing nodes and parallel edges in the ``main loop'' (i.e. steps 1-3), the number of nodes in the graph always goes down, and the number of edges is bounded above by the number of nodes. Steps (4) and (5) are then just a finite amount of post-processing in order to get the exact form in Proposition~\ref{prop:reduced-form}.

\begin{theorem}
  The \textbf{Reduction Strategy} above terminates and yields a graph in reduced form.
\end{theorem}

\begin{proof}
  Each iteration of steps 1-3 reduces the number of interior $\sbl$ nodes by 1. Hence it terminates after $n$ iterations for $n$ interior $\sbl$ nodes, with no interior $\sbl$ nodes. Step 4 guarantees all remaining, non-interior $\sbl$ nodes are of the form (I), (O), or (IO) as in Proposition~\ref{prop:reduced-form}, and step 5 guarantees the directed acyclicity conditions.
\end{proof}



\begin{remark}
  The quantum circuit optimisation tool PyZX~\cite{pyzx} uses a version of the \textbf{Reduction Strategy} above to simplify phase-free diagrams using the ZX-calculus.
\end{remark}

\subsection{Rewriting as quantifier elimination}
We close this section with a brief discussion about the $\sbl$-reduced form, and its relationship to the semantics of $\mathbb{IB}$. It was shown in \cite{BialgAreFrob14} that the PROP for $\mathbb{IB}$ is isomorphic to the PROP $\textbf{LinRel}(\mathbb Z_2)$ of $\mathbb Z_2$-linear relations. That is, morphisms $S : m \to n$ are linear sub-spaces $S \subseteq Z_2^m \times Z_2^n \cong Z_2^{m+n}$, $\oplus$ is given by direct product, and composition is done relation-style
\begin{equation}\label{eq:relation-style}
  (v,w) \in (S ; T) \iff \exists u . (v,u) \in S, (u,w) \in T
\end{equation}

As explained in~\cite{ZanasiThesis}, the $\sbl$-reduced form (called the \textit{cospan form} therein) enables us to `read off' $S$ as a homogeneous system of equations (or equivalently, as a basis for $S^\perp$). In this form, $\sbl$ nodes correspond to variables, and $\sr$ nodes to equations, whose LHS and RHS consist of those variables connected by in-edges and out-edges, respectively. For example, the diagram below represents the space of solutions to the following system of equations
\begin{equation}\label{eq:homog-system}
  \input{./figures/cospan-ex1.tikz} \qquad\mapsto\qquad
\left(\begin{array}{rcl}
  \textrm{\color{blue}$x_0$} + \textrm{\color{blue}$x_1$} \!\!&\!\!
  \overset{\textrm{\color{red}$e_0$}}{=} \!\!&\!\!
  \textrm{\color{blue}$y_0$} \\
  0 \!\!&\!\!
  \overset{\textrm{\color{red}$e_1$}}{=} \!\!&\!\!
  \textrm{\color{blue}$y_0$} \\
  \textrm{\color{blue}$x_0$} + \textrm{\color{blue}$x_1$} \!\!&\!\!
  \overset{\textrm{\color{red}$e_2$}}{=} \!\!&\!\!
  0
\end{array}\right)
\end{equation}

This interpretation gives a semantical view of the \textbf{Reduction Strategy} as a quantifier elimination procedure. 
The main purpose of the procedure is to eliminate interior $\sbl$ nodes. Since these nodes arise from sequential compositions
 in $\textbf{LinRel}(\mathbb Z_2)$, equation~\eqref{eq:relation-style} tells that they correspond to existentially quantified variables 
\[ \!\!\input{./figures/existential.tikz}
\ \mapsto\  \exists \textrm{\color{blue}$z_0$} . \left(\!\!
\begin{array}{rcl}
  \textrm{\color{blue}$x_0$} + \textrm{\color{blue}$x_1$} \!\!\!&\!\!\!
  \overset{\textrm{\color{red}$d_0$}}{=} \!\!\!&\!\!\!
  \textrm{\color{blue}$z_0$} \\
  \textrm{\color{blue}$z_0$} \!\!\!&\!\!\!
  \overset{\textrm{\color{red}$e_0$}}{=} \!\!\!&\!\!\!
  \textrm{\color{blue}$y_0$} \\
  0 \!\!\!&\!\!\!
  \overset{\textrm{\color{red}$e_1$}}{=} \!\!\!&\!\!\!
  \textrm{\color{blue}$y_0$} \\
  \textrm{\color{blue}$z_0$} \!\!\!&\!\!\!
  \overset{\textrm{\color{red}$e_2$}}{=} \!\!\!&\!\!\!
  0
\end{array}\!\!
\right) \]
The core of the \textbf{Reduction Strategy} are steps 2 and 3. The former isolates an existentially quantified variable $z$ on the LHS of an equation $e$, and step 3 substitutes any occurrence for that variable with its RHS, simultaneously eliminating $z$ and $e$. Applying this procedure to the diagram above yields the $\sbl$-reduced form in~\eqref{eq:homog-system}
\[ \input{./figures/existential-reduce.tikz} \]

Since everything in $\mathbb{IB}$ and the \textbf{Reduction Strategy} is colour-symmetric, we can use the same strategy to compute the analogous $\sr$ reduced forms. To do so, we first pre- and post-compose with colour changers to obtain a graph with an interface consisting entirely of $\sr$ nodes, then apply the \textbf{Reduction Strategy} with the colours reversed. 
Applying this to example~\eqref{eq:homog-system} yields 

%
\[ 
\input{./figures/reverse-colours-abbrev.tikz} 
\]
This again gives a canonical representation of a sub-space $S$ (called the \textit{span form}), but this time as a basis for $S$ itself, rather than $S^\perp$ \cite{ZanasiThesis}. $\sbl$ nodes correspond to basis vectors, where the presence of an edge indicates a $1$ in the corresponding position. The final diagram in the rewrite sequence above represents $S$ as $\textbf{span}\{((1,1),(0))\}\subseteq \mathbb Z_2^{2} \times \mathbb Z_2$, which is indeed the space of solutions to the system given in~\eqref{eq:homog-system}.

Note that we have focused on the case of $\mathbb{IB}$ and $\mathbb Z_2$-linear equations because it is the simplest. It was shown in~\cite{interactinghopf} that this system generalises straightforwardly to a system $\mathbb{IB}_F$ that has as its PROP $\textbf{LinRel}(F)$ for an arbitrary field $F$. In that case, we introduce a family of generators for each $a \in F \backslash{\{0,1\}}$ that give weights to edges. By modifying the \textbf{Reduction Strategy} to account for these weights, we can still obtain a (slightly more elaborate) procedure for removing internal nodes. This then gives the graphical analogue to quantifier elimination over an arbitrary field $F$.

Interestingly, this graphical version of quantifier elimination is \textit{inherently compositional}. It is possible to introduce generators and relations, breaking the semantic connection with $\textbf{LinRel}(F)$, while using \textbf{Reduction Strategy} on sub-diagrams in the $\mathbb{IB}_F$ fragment. This technique can exploit the fact that the ZX-calculus contains $\mathbb{IB}$ to perform peephole optimisations on quantum circuits, even though the latter have a more complex semantics than $\textbf{LinRel}(\mathbb Z_2)$.

For yet another perspective, recall that $\mathbb{IB}$ enjoys a modular characterisation in terms of distributive laws of PROPs \cite{BialgAreFrob14}, which prescribes that each diagram can be turned into \emph{cospan form} and \emph{span form}. As observed, these correspond to $\sbl$-reduced and $\sr$-reduced forms respectively: thus our result provides algorithmic means to reach them, which were lacking in the abstract picture. It also fills the main gap in formulating the realisability procedure for signal flow graphs \cite{Bonchi2015,BonchiSZ17} entirely as a diagram rewriting procedure.






\if\ismain0 

\bibliographystyle{plain}
\bibliography{catBib3Rev}

\fi 

\if\ismain0 

\setcounter{section}{6} 
\tableofcontents

\fi 

\section{Conclusions}

Increasingly, string diagrams are establishing themselves as the yardstick formalism to
reason compositionally about graphical models of computation across different fields.
These developments demand a mathematical foundation of how to \emph{compute} with
equational theories of string diagrams, seen as rewriting systems. In this work, we laid out
such foundations, in the form of a sound and complete interpretation of string diagram
rewriting as double-pushout rewriting on a suitable domain of hypergraphs. One
fundamental aspect of this modelling is the presence of \emph{interfaces}: the
compositional nature that is intrinsic to string diagrams has to be adequately mirrored in the
hypergraph interpretation, which we achieved by resorting to the theory of cospans, graphs
with interfaces and the associated notion of rewriting.

From the viewpoint of string diagrams, the key advantage of working under the
hypergraph interpretation is that the structural laws usually imposed on the
diagrammatic syntax get absorbed in the combinatorial representation. In this way,
instead of having to deal with the subtleties of rewriting modulo those structural laws,
one may use double-pushout rewriting ``on-the-nose'' on the corresponding hypergraphs.

Furthermore, we saw how absorbing all the structural laws becomes way subtler once
there are more than one Frobenius structure at stake. In that situation, we were able to
come up with an adequate interpretation by imposing extra structure to distinguish the
different Frobenius algebras, which was suitably normalised during the rewriting
procedure.

We concluded the paper with an extended example, in which we proposed
a terminating rewriting strategy for the theory of Interacting Bialgebras. This is a
well-studied diagrammatic calculus with cross-disciplinary applications. The richness
of the calculus, featuring two bialgebras and two Frobenius algebras, had previously made
it very difficult and unwieldy to study its rewriting properties. Using our
interpretation, the two Frobenius algebra structures get absorbed, leaving only
the bialgebra rules as non-trivial rewrites, which
drastically simplifies its study. This led us to a representation theorem, a reduction
strategy, and a semantical view of such strategy as quantifier elimination.

\smallskip

The work presented in this paper is only the first part of the story: we have presented a
characterisation of string diagram rewriting modulo Frobenius algebras, and showed that
it matches naturally double-pushout rewriting of hypergraphs with interfaces. But what
about equational theories that do not feature any Frobenius algebra? Building on the framework
developed so far, in the sequel of this paper we show how also these theories, in which
rewriting just happens modulo the laws of symmetric monoidal categories, without any
additional structure, can be characterised in terms of double-pushout rewriting. Next, we
complete our framework by investigating confluence in the context of string diagram rewriting.

\paragraph{Acknowledgements} We are thankful to the anonymous referees for their helpful comments; Fabio Gadducci acknowledges
support from Italian MIUR PRIN  2017FTXR7S IT-MATTERS (Methods and Tools for Trustworthy Smart Systems);
Fabio Zanasi acknowledges support from EPSRC grant EP/V002376/1.
Pawe{\l} Soboci\'{n}ski was supported by the ESF funded Estonian IT Academy research measure (project 2014-2020.4.05.19-0001) and the Estonian Research Council grant PRG1210.
\if\ismain0 

\bibliographystyle{plain}
\bibliography{catBib3Rev}

\fi 

\bibliographystyle{alpha}
\bibliography{catBib3Rev}

\end{document}